\documentclass[11pt]{iopart}
\usepackage{iopams}

\expandafter\let\csname equation*\endcsname\relax
\expandafter\let\csname endequation*\endcsname\relax

\usepackage{graphicx,graphics,epsfig,subfigure,bm,bbm,amssymb,amsmath,amsfonts,mathrsfs}
\usepackage{times}
\usepackage[breaklinks=true,colorlinks=true,linkcolor=blue,urlcolor=blue,citecolor=blue]{hyperref}

\usepackage[matrix,frame,arrow]{xypic}

\makeatletter

\@ifundefined{textcolor}{}
{%
 \definecolor{BLACK}{gray}{0}
 \definecolor{WHITE}{gray}{1}
 \definecolor{RED}{rgb}{1,0,0}
 \definecolor{GREEN}{rgb}{0,1,0}
 \definecolor{BLUE}{rgb}{0,0,1}
 \definecolor{CYAN}{cmyk}{1,0,0,0}
 \definecolor{MAGENTA}{cmyk}{0,1,0,0}
 \definecolor{YELLOW}{cmyk}{0,0,1,0}
}

\newcommand{\abs}[1]{\left\vert#1\right\vert}

\newcommand{\ket}[1]{\left\vert#1\right\rangle}
\newcommand{\bra}[1]{\left\langle#1\right\vert}

\newcommand{\ignore}[1]{}

\newcommand{\beq}{\begin{equation}}
\newcommand{\eneq}{\end{equation}}
\newcommand{\beqnn}{\begin{equation*}}
\newcommand{\eneqnn}{\end{equation*}}
\newcommand{\beqy}{\begin{eqnarray}}
\newcommand{\eneqy}{\end{eqnarray}}
\newcommand{\beqynn}{\begin{eqnarray*}}
\newcommand{\eneqynn}{\end{eqnarray*}}

\begin{document}

\title[Long-time multi-qubit memory]{Dynamical decoupling sequences for
multi-qubit \\ dephasing suppression and long-time quantum memory}

\author{Gerardo A. Paz-Silva$^{1,2}$, Seung-Woo Lee$^{1,3}$, Todd J. Green$^{1,4}$, \\
and Lorenza Viola$^1$}

\address{$^1$ Department of Physics and Astronomy, Dartmouth College,
6127 Wilder Laboratory, Hanover, New Hampshire 03755, USA}

\address{$^2$ Centre for Quantum Dynamics \& Centre for Quantum Computation and 
Communication Technology, Griffith University, Brisbane, Queensland 4111, 
Australia} 

\address{$^3$ Quantum Universe Center, Korea Institute for Advanced Study,  
Seoul 02455, Korea}

\address{$^4$ Centre for Engineered Quantum Systems, School
of Physics, The University of Sydney, New South Wales 2006,
Australia}

\ead{\rm{lorenza.viola@dartmouth.edu}}

\date{\today}

\begin{abstract}
We consider a class of multi-qubit dephasing models that combine classical noise 
sources and linear coupling to a bosonic environment, and are controlled by arbitrary sequences 
of dynamical decoupling pulses. Building on a general transfer filter-function 
framework for open-loop control, 
we provide an exact representation of the controlled dynamics for arbitrary stationary 
{\em non-Gaussian} classical and quantum noise statistics, with analytical expressions emerging 
when all dephasing sources are Gaussian.  
This exact characterization is used to establish two main results.
First, we construct multi-qubit sequences that ensure maximum high-order error suppression 
in {\em both} the time and frequency domain and that can be {\em exponentially more efficient} 
than existing ones in terms of total pulse number.
Next, we show how long-time multi-qubit storage may be achieved 
by meeting appropriate conditions for the emergence of a {\em fidelity plateau} under sequence 
repetition, thereby generalizing recent results for single-qubit memory under Gaussian dephasing.
In both scenarios, the key step is to endow multi-qubit sequences with a suitable 
{\em displacement anti-symmetry} property, which is of independent interest for applications ranging 
from environment-assisted entanglement generation to multi-qubit noise spectroscopy protocols. 
\end{abstract}


\renewcommand*\contentsname{}

\tableofcontents

\section{Introduction}

Characterizing and counteracting decoherence from noise environments which may in general 
exhibit both {\em temporal and spatial correlations} is a central challenge for realizing high-fidelity quantum 
information processing (QIP) and fault-tolerant quantum computation \cite{qec,Martinis}. Of particular 
interest are purely dephasing environments, which provide an accurate physical description 
whenever relaxation processes associated with energy exchange occur over a characteristic time scale 
that is substantially longer than that associated with dephasing dynamics. 
Provided that the noise arises predominantly from low-frequency components, and that external control is 
available over time scales that are short compared to the resulting temporal correlations, open-loop 
techniques based on dynamical decoupling and dynamically error-corrected gates 
\cite{qec,Viola1998,Viola1999Dec,CDD,UDD,Kaveh2009,Kaveh2010} 
provide a powerful tool for boosting operational fidelities -- potentially eliminating ``coherent'' (highly 
``non-Markovian'') errors that dominate worst-case error estimates in rigorous threshold analyses 
\cite{Steve,Harrison}. 

Our focus in this work is dynamical decoupling (DD) for a class of purely dephasing models, 
that describe {\em stationary but not necessarily Gaussian correlated noise} from combined classical 
and quantum sources on a multi-qubit system.  
More precisely, as detailed in Sec. \ref{sub:models}, we consider arbitrary single- and 
two-qubit classical time-dependent phase noise, along with quantum noise from a dephasing ``free field'' 
Hamiltonian -- namely, a linear diagonal spin-boson model where non-Gaussian statistics may arise solely 
from the bath initial state \cite{Leggett1987,Hanggi,Hui,NGpaper}.    
Our motivation is twofold: on the one hand, while general-purpose multi-qubit DD sequences based on 
concatenation and nesting are known, which may in principle ensure the desired 
error suppression up to arbitrary high order \cite{CDD,Wang2011,WanPaz} (Sec. \ref{control}), it may 
be possible to design less resource-intensive DD protocols by tailoring construction to 
dephasing models with particular characteristics; on the other hand, while an approach to practical long-time 
high-fidelity quantum memory has been recently proposed based on the idea of engineering a ``coherent 
plateau'' \cite{Memory} by sequence repetition, the existing analysis is only applicable to a single qubit 
exposed to Gaussian phase noise.   

We tackle the above issues under the simplifying assumption that the required DD pulses may be 
instantaneously effected on selected (subsets of) qubits -- subject, however, to a realistic constraint on their minimum 
separation, or ``switching time'' \cite{Memory,limits}.  We leverage a general filter-transfer 
function formalism for characterizing open-loop error-suppression capabilities in both the time and the 
frequency domain using {\em fundamental filter functions} (FFs) as building blocks \cite{PazViola}.  
Our first result, and starting point for subsequent analysis, is an exact characterization for the 
multi-qubit controlled dynamics in terms of a suitably defined time-ordered cumulant expansion, given in Sec. 
\ref{sub:exact}.  While it is known that an exact solution exists for the free evolution of a system subject 
to Gaussian bosonic phase noise \cite{Palma1996,Duan1998,Reina2002},  
and is in fact equivalent to one resulting from a second-order Magnus perturbative treatment \cite{Hanggi}, 
our analysis extends this equivalence to DD-controlled evolution -- recovering in the process a number of 
partial results in the literature, limited to two qubits under special symmetry assumptions 
\cite{Kim2000,Hu2007,Su2012,Pan2012,Song2013}.  
For generic {\em non-Gaussian} dephasing noise, our representation shows how arbitrary high-order cumulants 
may still be expressed in terms of {\em only two} FFs associated with the control modulation of single- and 
two-qubit terms. 

Sections \ref{Decou} and \ref{longterm} contain our core results on two control tasks of increasing complexity: 
(i) Suppressing errors as effectively as possible, so that quantum information is preserved over a 
short-time regime of interest (e.g., a single gating period);  
(ii) Ensuring that quantum information is preserved with high fidelity for arbitrarily 
long storage times, in principle, with on-demand access.  Provided that selective control over 
individual qubits is available, we show that new multi-qubit DD sequences may be constructed, so that 
a {\em displacement anti-symmetry} property is obeyed by the control switching functions, both in the 
simplest case of $N=2$ qubits (Sec. \ref{sub:selective}) and, by appropriately orchestrating the sign pattern 
for every qubit pair, for general $N$ (Sec. \ref{sub:multisel}).  Such sequences ensure the 
same order of error suppression as the best known nested or concatenated protocols, while {\em
also} maximizing their ``filtering order'' in the frequency domain \cite{PazViola}.  Further to that, for a fixed system 
size $N$, DD sequences incorporating displacement anti-symmetry are {\em exponentially more efficient} in 
terms of total pulse number, as long as any direct coupling between qubits is time-independent.  The possibility 
of achieving multi-qubit long-time storage by meeting appropriate conditions for the emergence of a {\em fidelity plateau} 
under sequence repetition is first established for the important case of arbitrary classical stationary phase noise 
in Sec. \ref{p1}. Generalization of this result to dephasing scenarios that also include bosonic noise sources again 
relies crucially on incorporating displacement anti-symmetry in the DD sequences used for repetition (Sec. \ref{p2}). 

Provided that the relevant conditions for the emergence of a fidelity plateau can be satisfied, 
we argue in Sec. \ref{sec:ent} that the combined use of multi-qubit DD sequences that do not (or, respectively, 
do) obey displacement anti-symmetry provides a venue for controlled generation and storage of 
{\em multi-partite entanglement} mediated by a common quantum environment, extending known schemes
for bipartite entanglement generation \cite{Braun2002,Oh2006}.  A summary of our main results is presented 
in Sec. \ref{conclusion}, along with an outlook to future research.  By way of concrete illustration of our general 
FF formalism, additional detail on the simple yet practically important case of a DD-controlled two-qubit system 
under Gaussian dephasing from combined classical and quantum bosonic sources is included in the Appendix.

\section{Multi-qubit Controlled Dephasing Dynamics}
\label{sec:dynamics}

\subsection{Gaussian versus non-Gaussian dephasing models}
\label{sub:models}

We consider a class of purely-dephasing noise models on $N$ qubits in which no energy exchange 
takes place between the system and an environment (``bath'') modeled by a continuum of classical 
and/or quantum modes -- see Sec. \ref{sec:realistic} for further discussion on their physical relevance.
The general form of the relevant open-system Hamiltonian reads
\begin{equation}
\label{gralH}
H(t) = H_S (t) \otimes {\mathbb I}_B + {\mathbb I}_S \otimes H_B + H_{SB},
\end{equation}
where $H_S(t) $ and $H_B$ denote, respectively, the internal Hamiltonian for the system and bath, 
and $H_{SB}$ describes their interaction. For a purely dephasing model, 
$[H_{SB}, H_S]=0$, implying the existence of a preferred (energy or 
computational) basis \cite{qec}. 
Without loss of generality, we take this to be the ${z}$ basis, and use $Z_\ell$ to denote the Pauli 
operator $\sigma_z$ acting on qubit $\ell$.  A general multi-qubit dephasing Hamiltonian that includes up to 
two-body noisy qubit interactions and linear system-bath coupling may then be written as
\beqy
H_S (t) &=&\sum_{\ell=1}^N  Z_{{\ell}} \left( d_{{\ell}}+  \zeta_{{\ell}}(t) \right) + \hspace*{-2mm}
\sum_{\ell\neq \ell' =1}^N  \hspace*{-1mm}Z_{{\ell}} \otimes Z_{{\ell'}} \left( d_{\ell,\ell'}+ \eta_{\ell,\ell'}(t) \right), 
\label{gralHs} \\
H_{SB}&=&\sum_{\ell=1}^{N}  Z_{{\ell}} \otimes {B}_{{\ell}} ,
\label{gralHsb}
\eneqy
where we allow for dephasing due to both {\em classical and 
quantum noise sources}, as represented by the fluctuating system Hamiltonian $H_S(t)$ and the system-bath 
interaction Hamiltonian $H_{SB}$, for suitable (Hermitian) operators ${B}_{{\ell}}$ acting on $B$.  
Specifically, 
$\zeta_\ell(t)$ and $\eta_{\ell, \ell'}(t)$ are classical stochastic processes describing random fluctuations of the local 
energy splittings and two-local (Ising) coupling strengths, $d_\ell$ and $d_{\ell,\ell'}$.

A formally similar treatment of classical and quantum noise sources is possible upon moving to the 
interaction picture associated with the bath Hamiltonian $H_B$. The total Hamiltonian becomes 
$H (t) \mapsto \tilde{H}(t) = \tilde{H}_S (t) \otimes {\mathbb I}_B + \tilde{H}_{SB}(t)$, with 
$ \tilde{H}_S (t) \equiv H_S (t) ,$ 
$\tilde{H}_{SB}(t) \equiv \sum_{\ell=1}^{N} Z_{{\ell}}  \otimes {B}_{{\ell}}(t) ,$ and 
where, for notational convenience, we simply use ${B}_{{\ell}}(t)$ to denote the interaction-picture representation 
of ${B}_{{\ell}}$.  For the purpose of achieving decoherence suppression and arbitrary state preservation, the dynamics 
generated by $\tilde{H}(t)$ corresponds to unwanted (``error'') evolution, thus $\tilde{H}(t)\equiv H_e(t)$
in the framework of dynamical error suppression \cite{Kaveh2010,PazViola}.
In a multi-qubit setting, the noise acting on different (subsets of) qubits may exhibit different kinds of {\em temporal 
as well as spatial correlations}. In particular, two limiting situations may be envisioned for the coupling of the 
quantum bath to different qubits: 

\begin{itemize}
\item A {\it common} bath, in which case $[{B}_{\ell}(t),{B}_{\ell'}(t')] \neq 0$, 
$\forall \, \ell \ne \ell'$ for at least some $t,t'$; 

\item A {\em private} (or independent) bath, in which case $[{B}_{\ell}(t),{B}_{\ell'}(t')] = 0$,  
$\forall \, \ell\neq \ell'$, $\forall t,t'$. 
\end{itemize}
\noindent 
While intermediate situations are clearly possible, an additional distinction is relevant for the common-bath 
scenario depending on whether qubit-permutation symmetry is present.  Specifically, {\it collective} 
dephasing corresponds to ${B}_{\ell}(t) = B(t)$, $\forall \,\ell$, as extensively studied in the context 
of decoherence-free subspaces \cite{qec,dfs}.  Noise processes and their correlations can be further 
characterized by their spectral properties in the frequency domain.

\vspace*{1mm}

{\bf Spectral properties of classical phase noise.--} Classical noise processes may be compactly 
characterized by their statistical moments~\cite{Kubo}, obtained via the ensemble average 
over noise realizations, henceforth denoted by $\langle \cdot \rangle_c$.  Given the fluctuating 
Hamiltonian $H_S(t)$ in Eq. (\ref{gralHs}), we are most generally interested in cross-correlations 
of the form $\langle \zeta_{{\ell}_1}(t_{1})\cdots \zeta_{{\ell}_j}(t_{j}) \eta_{p_{j+1}}(t_{j+1}) \cdots 
\eta_{p_k}(t_k)\rangle_c$ for all $j$, $k$, where $p\equiv \{\ell,\ell' \}$ labels qubit pairs, and 
we include joint moments of a single noise source, 
$\zeta_{{\ell}}(t)$ or $\eta_{p}(t)$, as a particular case. 
A simpler description is obtained by using {\it cumulants}, which we denote 
$C^{(k)}( \zeta_{{\ell}_1}(t_{1})\cdots \zeta_{{\ell}_j}(t_{j}) \eta_{p_{j+1}}(t_{j+1}) \cdots \eta_{p_k}(t_k))$ and 
define via general moment-cumulant relations \cite{Kubo}. If $\{ A_i \}$ denotes a set of random 
variables, then 
\beq
\label{cumu}
\langle A_1 \rangle_c = C^{(1)}(A_1), \quad  \langle A_1\cdots 
A_n\rangle_c = \sum_{\pi \in \Pi(n)} \prod_i C^{(V_i)} (A_1\cdots A_n), 
\eneq	
where $\Pi(n)$ is the set of partitions of $n$ elements, $\pi = \{V_1,\ldots,V_r\}$, and each block $V_i$ contains 
the elements $\{ v_i(s)\}_{s=1}^{|V_i|}$ {\it ordered according to $A_1,\ldots,A_n$}, with 
\beq
C^{(V_i)} (A_1\cdots A_n) \equiv  C^{(| V_i |)}(A_{v_i (1)} \cdots A_{v_i (|V_i |)}).
\label{cumu2}
\eneq
For example:
\begin{eqnarray*}
\hspace*{-15mm} 
\langle A_3 A_1 A_2\rangle_c &=& C^{(3)}(A_3 A_1 A_2) + C^{(1)}(A_3)C^{(1)}(A_1)C^{(1)}(A_2) +\\ 
&& C^{(2)}(A_3 A_1) C^{(1)}(A_2) + C^{(2)}(A_3 A_2) C^{(1)}(A_1)+ C^{(2)}(A_1 A_2) C^{(1)}(A_3) ,
\end{eqnarray*}
and so on. The above definition of cumulant, which respects the order of the arguments, is equivalent to the traditional 
one when the random variables commute, but solves the ambiguity that arises when ordering becomes important,
as it does for non-commuting operator-variables in the quantum case. 
The expressions in Eq. (\ref{cumu}) can be inverted as usual in probability theory and statistics, allowing one to 
write the cumulants in terms of the moments~\cite{Kubo}.

A description of the noise process in frequency space may be given by considering the Fourier transform of cumulants. 
Specifically, the Fourier transform of a $k$-th order (cross-) cumulant defines the {\em $k$-th order polyspectrum} 
\cite{Brillinger,NGpaper}, 
$S^{\zeta,\eta}_{{\ell}_{1},\ldots,{\ell}_j,p_{j+1},\ldots,p_k} (\omega_{1},\ldots,\omega_{k})$, via  
\beqy
\nonumber 
S^{\zeta,\eta}_{{\ell}_{1},\ldots,{\ell}_j,p_{j+1},\ldots,p_k} (\vec{\omega}_{[k]}) & \equiv& 
C^{(k)}( \tilde{\zeta}_{{\ell}_1}(\omega_{1}) \cdots \tilde{\zeta}_{{\ell}_j}(\omega_{j})\tilde{\eta}_{p_{j+1}}
(\omega_{{j+1}}) \cdots \tilde{\eta}_{p_k}(\omega_{k})),
\eneqy 
where $\tilde{\zeta}_{{\ell}}(\omega)$, $\tilde{\eta}_{p}(\omega)$ are random variables obtained from  
${\zeta}_{{\ell}}(t)$, ${\eta}_{p}(t)$ via Fourier transform, and we have introduced the notation 
$\vec{\omega}_{[k]} \equiv (\omega_1,\ldots,\omega_k)$ to denote a vector of length 
$k$~\footnote{Noise cross-power spectra may be also defined in terms of the Fourier transform of the 
moments (and related high-order time-correlation functions)~\cite{Kubo}.  Following standard practice
in classical signal processing theory~\cite{Brillinger} and our previous analysis~\cite{PazViola}, 
we employ cumulants in what follows.  In addition, we will assume sufficient regularity for all the relevant 
Fourier transforms to be mathematically well-defined.}. 
The noise process is {\it stationary} if arbitrary cumulants, 
$C^{(k)}( \zeta_{{\ell}_1}(t_{1})\cdots \zeta_{{\ell}_j}(t_{j}) \eta_{p_{j+1}}(t_{{j+1}}) \cdots \eta_{p_k}(t_{{k}}))$, depend 
solely on time differences, say, $\tau_j \equiv t_{j+1} - t_1$ for $j=1, \ldots, k-1$, rather that absolute time values. 
Thus, the stationarity assumption translates into 
\beqy
\label{station} 
S^{\zeta,\eta}_{{\ell}_{1},\ldots,{\ell}_j,p_{j+1},...,p_k} (\vec{\omega}_{[k]}) = 2 \pi 
\delta(\omega_{1}+ \cdots + \omega_{k}) 
{S}^{\zeta,\eta}_{{\ell}_{1},\ldots,{\ell}_j,p_{j+1},\ldots,p_k} (\vec{\omega}_{[k-1]}),
\eneqy
where, by letting $d \vec{\tau}_{[k-1]} \equiv \prod_{i=1}^{k-1} {d \tau_i}$ 
and $\vec{\omega}\cdot\vec{\tau}\equiv \omega_1 \tau_1 +\ldots +\omega_{[k-1]} \tau_{[k-1]}$, 
we may write
\beq
\nonumber  {S}^{\zeta,\eta}_{{\ell}_{1},...,{\ell}_j,p_{j+1},...,p_k} (\vec{\omega}_{[k-1]}) = \int_{-\infty}^\infty 
\hspace*{-1mm} {d\vec{\tau}_{[k-1]}} 
\, e^{-i \vec{\omega}\cdot\vec{\tau}} \, 
C^{(k)}( \zeta_{{\ell}_1}(t) \zeta_{\ell+2}(t+\tau_1) \ldots \eta_{p_{k}} (t +\tau_{k-1}) ), \; \; \forall t.
\eneq
Of particular interest are {\it stationary zero-mean classical Gaussian noise processes}, whose statistical properties 
are completely characterized by the second-order cumulants.  That is, 
\beqy
\label{Gaussian} 
& C^{(k)}( \zeta_{{\ell}_1}(t) \zeta_{\ell+2}(t+\tau_1) \ldots \eta_{p_{k}} (t +\tau_{k-1}) ) =0, \quad 
\forall k \neq 2, \; \forall t, \tau_j .&
\eneqy
Accordingly, for Gaussian noise all the statistical properties are formally encapsulated by 
${S}^{\zeta,\eta}_{\ell_1,\cdots \ell_j,p_{j+1}, \cdots, p_k} (\omega)$ for $k =2$ and $j=0,1,2$; that is, physically, 
the noise power spectra associated to two-point correlations in the single-qubit and two-qubit energy fluctuations, 
${S}^{\zeta}_{\ell_1,\ell_2} (\omega)$ ($j=2$), ${S}^{\eta}_{p_1,p_2} (\omega)$ ($j=0$), and the 
cross-power spectrum ${S}^{\zeta,\eta}_{\ell_1,p_1} (\omega)$ ($j=1$).  If, additionally, 
the local and two-local classical noise sources, $\zeta_\ell(t)$ and $\eta_{p} (t)$, are 
statistically independent, one may further simplify to 
$S^{\zeta,\eta}_{\ell_1,\cdots \ell_j,p_{j+1}, \cdots, p_k} ({\omega}) = 0$ for $ 0 < j < k$. 

\vspace*{1mm}

{\bf Spectral properties of quantum phase noise.--} 
The spectral properties of a quantum dephasing environment may likewise be characterized 
in terms of statistical moments and cumulants.  The starting point is 
to define statistical averages with respect to the initial state of the bath, say, $\rho_B$, by 
using $\Tr_B (\cdot \rho_B)$ instead of $\langle \cdot \rangle_c$. While in the most general case 
initial system-bath correlations may be present (hence $\rho_B=\textrm{Tr}_S[\rho_{SB}(0)]$), 
throughout the present analysis we shall work under the standard assumption of an {\em initial 
factorized state} of the form $\rho_{SB}(0)\equiv \rho_S(0)\otimes \rho_B$. Thus, we are 
interested in the time-domain $k$-th order moments,
$$ \textrm{Tr}_B [ B_{{\ell}_1}(t_{1}) \cdots B_{{\ell}_k}(t_{k}) \rho_B]  \equiv \langle B_{{\ell}_1}(t_{1}) \cdots 
B_{{\ell}_k}(t_{k}) \rangle_q, $$
or in the frequency Fourier transform of the corresponding cumulant,
\beq 
S^{B}_{{\ell}_1,...,{\ell}_k} (\vec{\omega}_{[k]}) = C^{(k)}( \tilde{B}_{{\ell}_1}(\omega_1) \cdots \tilde{B}_{{\ell}_k}(\omega_k) ),
\label{qck}
\eneq 
where $\tilde{B}_{{\ell}}(\omega)$ is the Fourier transform of ${B}_{{\ell}}(t)$. Operator cumulants are defined recursively via the moment-cumulant relations given earlier, Eq.~\eqref{cumu}, upon replacing $\langle \cdot \rangle_c$ with $\langle \cdot \rangle_q$, e.g., 
\begin{eqnarray*}
\langle {{B}}_{{\ell_1}}(t_1) {{B}}_{{\ell_2}}(t_2) \rangle_q &=&  C^{(2)} ({{B}}_{{\ell_1}}(t_1) {{B}}_{{\ell_2}}(t_2)) + C^{(1)} ({{B}}_{{\ell_1}}(t_1)) C^{(1)} ( {{B}}_{{\ell_2}}(t_2)),\\
\langle {{B}}_{{\ell_2}}(t_2) {{B}}_{{\ell_1}}(t_1) \rangle_q &=&  C^{(2)} ({{B}}_{{\ell_2}}(t_2) {{B}}_{{\ell_1}}(t_1)) + C^{(1)} ({{B}}_{{\ell_2}}(t_2)) C^{(1)} ( {{B}}_{{\ell_1}}(t_1)). 
\end{eqnarray*}
In the generic case where $[B_\ell, H_B] \ne 0$, 
one can see that a necessary and sufficient condition for stationarity is that $[ \rho_B, H_B]=0$.  
In particular, a {\em stationary zero-mean quantum Gaussian noise process} may be defined by requiring that,  
for all $t$, $\langle B_\ell (t)\rangle_q =0,$ and, in analogy to Eq. (\ref{Gaussian}), that 
arbitrary  cumulants vanish for $k>2$.
The corresponding $k=2$ quantum power spectrum then obeys the relation
$S^B_{{{\ell}},{{\ell'}}} (\omega_1,\omega_2) = 2 \pi \delta(\omega_1+\omega_2) 
{S}^B_{{{\ell}},{{\ell'}}} (\omega_2)$, where
\beq
\nonumber  
{S}^B_{{{\ell}},{{\ell'}}} ({\omega_2}) = \int_{-\infty}^\infty  \hspace*{-2mm} d\tau \, e^{-i \omega_2 \tau} 
\langle {{B}}_{{\ell}}(t+\tau) {{B}}_{{\ell'}}(t ) \rangle_q , \quad \forall t   .
\eneq

In what follows, we focus on dephasing noise originating from bosonic sources, representing 
for instance phonon or photon modes.  In this context, a paradigmatic example is the  {\em linear 
spin-boson model}~\cite{Leggett1987,Palma1996,Viola1998,Duan1998,Reina2002,Hui,Irene2,Irene1}, defined by 
the following bath Hamiltonian and interaction operators in Eqs. (\ref{gralH}) and (\ref{gralHsb}) (in units 
where $\hbar=1$):
\beq
H_B = \sum_k \Omega_k a^\dagger_k a_k , \; \Omega_k \geq 0, \quad B_{\ell} = \sum_k 
(g^{\ell}_k a^\dagger_k +  g^{\ell\,*}_{k} a_k ), 
\label{spinboson}
\eneq
where $[a_k, a^\dagger_{k'}] = \delta_{k,k'}$, $[a_k, a_{k'}] = 0 = [a^\dagger_k, a^\dagger_{k'}]$, 
and $g^{\ell}_k$ describes the coupling strength between qubit $\ell$ and bath mode $k$.  
If the initial bath state $\rho_B$ is thermal, with inverse 
temperature $\beta  = 1/k_B T$, $\rho_B \propto e^{- \beta H_B}$, the family of operators 
$B_\ell (t) = \sum_k 
(g^{\ell}_k\, e^{i\Omega_k t} a^\dagger_k +  g^{\ell\,*}_{k} \,e^{-i\Omega_k t} a_k )$ describes 
a stationary noise process which also obeys free-field Gaussian statistics \cite{Hui}. 
The above quantum power spectrum is explicitly given by
\begin{equation}
{S}^B_{{{\ell}},{{\ell'}}} (\omega) 
\label{eq:spectralrelation}  
= 2 \pi \sum_k  \Big( g_k^\ell g_k^{\ell' *} \delta(\omega { + }\Omega_k)  \langle a^\dagger_k a_k 
\rangle_q + g_k^{\ell'} g_k^{\ell *} \delta(\omega { - }\Omega_k) \langle a_k a_k^\dagger \rangle_q \Big),
\end{equation}
where $\langle a^\dagger_k a_k \rangle_q = \left(\coth ({\beta \Omega_k}/{2 }) - 1\right)/2 =  
\langle a_k a^\dagger_k \rangle_q -1$ 
are equilibrium expectation values.  Assuming that the couplings are 
position-dependent as $g_k^{\ell} = |g_k| e^{i \vec{k} \cdot \vec{r}_\ell}$ \cite{Leggett1987}, 
where $\vec{k}$ is the momentum associated with the $k$-th mode and $\vec{r}_\ell$
the position of the $\ell$-th qubit, one gets 
\begin{equation}
g^\ell_k g_k^{\ell' *}  = |g_k|^2 e^{ i \vec{k} \cdot(\vec{r}_{\ell}-\vec{r}_{\ell'})} \equiv  |g_k|^2 e^{ i \Omega_k t_{\ell,\ell'}}, 
\label{ttime}
\end{equation}
where we have further expressed the exponent in terms of the mode frequency and the 
{\em transit time}~\cite{Reina2002} via the (linear) dispersion relation 
$\Omega_k t_{\ell,\ell'} = \vec{k} \cdot(\vec{r}_{\ell}-\vec{r}_{\ell'})$. 
Under these assumptions, 
\begin{eqnarray}
S^{B}_{\ell,\ell'} (\omega)  &=&2 \pi J(\omega) 
\begin{cases}  e^{{\bf -} i \omega t_{\ell,\ell'}} (\coth(\beta\omega/2) { + }1)/2, & \omega >0, \\
                        e^{{\bf -}  i \omega t_{\ell,\ell'}}(\coth(-\beta \omega/2) { - }1)/2, &  \omega <0, \end{cases}  
\label{Qspectrum}
\end{eqnarray}
where $J(\omega) = \sum_{k} |g_k|^2 [\delta (\omega-\Omega_k) + \delta (\omega+\Omega_k)] = J(-\omega)$ 
is the {\em spectral density function} of the oscillator bath in the continuum limit~\footnote{Note that, unlike classical noise, 
quantum noise is {\em spectrally asymmetric} in general, with the ratio $S^{B}_{\ell,\ell'} (\omega)/S^{B}_{\ell',\ell} (-\omega) = 
e^{  \beta \omega}$ enforcing detailed balance at thermal equilibrium.  }.

Since, by definition, $S^{B}_{\ell,\ell'} (\omega_1, \omega_2)$ is the second-order cumulant of the 
Fourier-transformed bath operators, we may write 
$S^{B}_{\ell,\ell'} (\omega) \equiv S^{B,+}_{\ell,\ell'} (\omega) + S^{B,-}_{\ell,\ell'} (\omega)$ 
upon separating operator products into anti-commutators and commutators. If the latter are, 
respectively, denoted by $[\: , \,]_\pm$, direct calculation yields 
\begin{align}
\Tr\left[ [\tilde{B}_{\ell}(\omega_1),\tilde{B}_{\ell'}(\omega_2)]_\pm \, \rho_B \right] & 
\equiv S^{B,\pm}_{\ell,\ell'} (\omega_1,\omega_2) 
= 2 \pi \delta(\omega_1+\omega_2) S^{B,\pm}_{\ell,\ell'} (\omega_2),  
\label{Sminus}
\end{align}
\begin{equation}
\hspace*{0mm}
S^{B,+}_{\ell,\ell'} (\omega) =  \pi  J(\omega) \hspace*{-1mm}
\begin{cases}   \;\;\: e^{-i \omega t_{\ell,\ell'}} \coth(\beta\omega/2),  \,\omega >0,\\
                            { - } e^{-i \omega t_{\ell,\ell'}}\coth(\beta\omega/2), \, \omega <0 ,
\end{cases} 
S^{B,-}_{\ell,\ell'} (\omega) =  \pi  J(\omega) \hspace*{-1mm}
\begin{cases}   \;\;\: e^{-i \omega t_{\ell,\ell'}}, \, \omega >0,\\
                            { - } e^{-i \omega t_{\ell,\ell'}}, \, \omega <0 .
\end{cases} 
\label{HQspect}
\end{equation}
Thus, $S^{B,+}_{\ell,\ell'} (\omega)$ ($S^{B,-}_{\ell,\ell'} (\omega)$) have 
symmetric (anti-symmetric) character in the sense that 
\begin{equation}
S^{B,+}_{\ell,\ell'} (- \omega) =  [ S^{B,+}_{\ell,\ell'} (\omega) ]^*,  \quad
S^{B,-}_{\ell,\ell'} (- \omega) = - [ S^{B,-}_{\ell,\ell'} (\omega) ]^*,  \quad \forall \ell, \ell', 
\label{HQspect1}
\end{equation}
a feature that will be useful in understanding their contribution to different dynamical aspects.

While the free dynamics under multi-qubit classical plus bosonic dephasing will be 
recovered  as a special case of the DD-controlled dynamics, two remarks are 
in order.  First, our approach may be applied to a wider class of quantum noise scenarios, for instance 
environments consisting of multiple interacting types of bosons, as long as
the relevant Hamiltonian can be mapped to Eq. (\ref{spinboson}) via an appropriate Bogoliubov
transformation~\cite{correbos}.  Second, for all the noise sources considered, 
a finite norm of the error dynamics to be filtered out will be assumed -- or, in more physical terms, a 
{\em finite correlation time} (``non-Markovian'' behavior in the sense of \cite{Viola1999Dec,Hui}) 
relative to which {\em fast control} regimes may be accessed. This is a crucial requirement for DD methods 
to be effective \cite{qec,Viola1999Dec,PazViola,limits,HuiLidar}, which will translate into assuming that all relevant 
noise  spectra decay sufficiently rapidly at high frequencies.

\subsection{Control protocols}
\label{control}

To facilitate presentation of the essential new results, we assume access to perfect instantaneous $\pi$ rotations, 
and postpone discussion of realistic effects to Sec. \ref{sec:realistic}. This ideal ``bang-bang'' 
DD setting has been extensively investigated in both single-qubit \cite{Viola1998,CDD,UDD,Irene1,Irene2} 
and multi-qubit systems~\cite{Viola1999Dec,Lea2008,Wang2011,WanPaz} (see \cite{qec} 
for a review). For the sake of completeness and consistency, we summarize 
the basic ingredients in this section, building in particular on Refs.~\cite{WanPaz,PazLidar}.  In the process, 
we introduce a new {\em composition rule} which unifies existing high-order sequence constructions and 
may be of independent interest within DD theory. 

For the dephasing noise models under consideration, control implemented via 
instantaneous rotations along a fixed axis (say, $x$), suffices.  Let a DD sequence, labeled by an 
integer index $s$, be specified in terms of a control operation $X (\equiv \sigma_x)$,
and a number $n(s)$ of control intervals $\{ \tau^{(s)}_{j} \equiv r_j^{(s)} T\}$, where $T$ is the total evolution 
time and $\{r_j^{(s)}\}_{j=0}^{n(s)-1}$ are positive numbers, that satisfy $\sum_j r_j^{(s)} =1$ and describe the 
relative pulse timings.  If $U_0(\tau^{(s)}_{j})$ denotes free evolution over $\tau^{(s)}_{j}$, 
the controlled propagator induced by such a sequence reads 
\beq 
\label{propaDD} 
U_s^{(X)} (T) = X^s  U_0(\tau^{(s)}_{n(s)-1})  X  \cdots X U_0(\tau^{(s)}_1) X U_0(\tau^{(s)}_0) ,
\eneq 
where operators are applied from right to left and 
a final pulse is included at time $t=T$ depending on the parity of $s$ in order to ensure a 
cyclic control propagator. 
Following \cite{PazViola,PazLidar}, we say that the DD sequence achieves {\em cancellation order} (CO) 
$\alpha$, with respect to the control operation $X$, if the norm of unwanted (non-commuting) error terms 
is suppressed up to order $\alpha$ in time, that is, 
$\vert \vert \,[X, U^{(X)}_{\alpha} (T)]\,  \vert \vert \sim \mathcal{O} (T^{\alpha+1})$. 
With this in mind, we shall henceforth identify $s\equiv \alpha$.
Consider two DD sequences, each acting on qubit $\ell$, $\ell'$ via control operations $X_\ell$, $X_{\ell'}$, 
and specified in terms of intervals $\{\tau^{(s)}_{j} \}$, $\{ \tau_{j'}^{(s')} \}$ respectively. 
We can build a new sequence over time $T$ by using the following {composition rule}:
\beq
\label{compo}
U_{s'}^{(X_{\ell'})}\circ  U_{s}^{(X_{\ell})} (T) \equiv X_{\ell'}^{s'} U^{(X_{\ell})}_{s}(\tau^{(s')}_{n(s')-1}) 
X_{\ell'}  \cdots X_{\ell'} U^{(X_{\ell})}_s(\tau^{(s')}_1) X_{\ell'}U^{(X_{\ell})}_s(\tau^{(s')}_{0}) .
\eneq
That is, each free evolution period in the original ``outer sequence'' $U_{s'}^{(X_{\ell'})} (T)$ is replaced 
by an ``inner sequence'' $U_{s}^{(X_{\ell})} $ of duration determined by the corresponding outer 
interval $\tau_{j'}^{(s')}$.  We are now ready to define the specific high-order sequences we will be using.

\vspace*{1mm}

{\bf Single-qubit DD sequences.-- } The two relevant sequences are single-axis CDD \cite{CDD} and Uhrig DD 
(UDD) \cite{UDD}.  The former may be defined recursively via 
$$ \text{CDD} ^{(X)}_{\alpha} (T) = \text{CDD} ^{(X)}_{1} \circ \text{CDD} ^{(X)}_{\alpha-1} (T) ,$$
\noindent 
with relative pulse timings $\{r_j^{(1)}\} = \{ {1}/{2},{1}/{2} \}$, so that, for example, 
\beqynn
\hspace*{-10mm}\text{CDD} ^{(X)}_{1} (T) &=& {X U_0(T/2)  X U_0(T/2)} , \\  
\hspace*{-10mm}\text{CDD} ^{(X)}_{2} (T) &=&  X \text{CDD} ^{(X)}_{1} (T/2) X \text{CDD} ^{(X)}_{1} (T/2) 
= U_0(T/4)  X U_0(T/2) X U_0(T/4), \quad 
\eneqynn
and so on. On the other hand, the $\alpha$-th order UDD protocol is defined by the pulse timings $t^{(\alpha)}_j = 
T \sin^2 [{\pi j}/{(2 \alpha+2)]}$, with $\tau_j^{(\alpha)} = t^{(\alpha)}_{j+1}-t^{(\alpha)}_j$ 
for $j=0,\ldots,\alpha$.

\vspace*{1mm}

{\bf Multi-qubit DD sequences.--} Both single-qubit \text{CDD}  and UDD sequences admit extensions to multiple qubits 
via the composition rule given in Eq.~\eqref{compo}. In particular, by letting $\vec{X} \equiv (X_1,\ldots,X_N)$, 
one can define the following $N$-qubit sequences: 
\beqy
\label{DD1} 
\mathring{U}_{(\alpha_1,\ldots,\alpha_N)}^{\vec{X}}(T)&\equiv& U^{(X_1)}_{\alpha_1} \circ \left( U^{(X_2)}_{\alpha_2} \circ 
\left( \cdots \left(U^{(X_{N-1})}_{\alpha_{N-1}} \circ U^{(X_{N})}_{\alpha_{N}} \right)\right) \cdots \right)(T),\\
\label{DD2}
{\mathring{U'}}_{(\alpha,\ldots,\alpha)}^{\vec{X}}(T)&\equiv&  \underbrace{\mathring{U}_{(1,\ldots,1)}^{\vec{X}} \circ 
\left( \mathring{U}_{(1,\ldots,1)}^{\vec{X}} \circ \left(  \cdots \circ \mathring{U}_{(1,\ldots,1)}^{\vec{X}} \right) \cdots 
\right)}_{\textrm{$\alpha$ times}}(T).
\eneqy
By construction, both achieve CO equal to $\alpha= \min\{\alpha_1,\ldots,\alpha_N\}$ with respect to $\{X_\ell\}$.
However, if the single-qubit sequence 
${U}^{(X_\ell)}_{\alpha_\ell}$ uses $n_P=n(\alpha_\ell)$ pulses, 
$\mathring{U}_{(\alpha_1, \ldots,\alpha_N)}^{\vec{X}}(T)$ in Eq. (\ref{DD1}) 
uses a total number of pulses 
$n_P^{\text{tot}} (N) = \prod_{\ell=1}^N n(\alpha_\ell) \geq [n(\alpha)]^N$, 
whereas ${\mathring{U'}}_{(\alpha, \ldots,\alpha)}^{\vec{X}}(T)$ in Eq. (\ref{DD2}) 
uses $n_P^{\text{tot}} (N) = [n(1)]^{\alpha N}$, 
with  $n^{\text{tot}}_P(N)$ growing exponentially with $N$ in either case. 
Depending on the sequences used as building blocks,  all 
known high-order DD sequences may be recovered. Specifically, we shall consider the following 
$\alpha$-th order $N$-qubit sequences:

\begin{itemize}
\item[(i)] Multi-qubit CDD~\cite{CDD}:  
$$\mathring{\text{CDD}}\hspace*{-.6mm}\mbox{  }_{(\alpha,\ldots,\alpha)}^{(\vec{X})}(T) = 
\underbrace{\text{CDD} _{(1,\ldots,1)}^{\vec{X}} \circ \left( \text{CDD} _{(1,\ldots,1)}^{\vec{X}} \circ \left(  
\cdots \circ \text{CDD} _{(1,\ldots,1)}^{\vec{X}} \right) \cdots \right)}_{\textrm{$\alpha$ times}}(T).$$

\vspace*{-5mm}

\item[(ii)] Nested UDD (NUDD)~\cite{Wang2011}:  
$$\text{NUDD}_{(\alpha_1,\ldots,\alpha_N)}^{(\vec{X})}(T)  = \text{UDD}^{(X_1)}_{\alpha_1} \circ \left( \text{UDD}^{(X_2)}_{\alpha_2} 
\circ \left( \cdots \left(\text{UDD}^{(X_{N-1})}_{\alpha_{N-1}} \circ \text{UDD}^{(X_{N})}_{\alpha_{N}} \right)\right) \cdots \right)(T).$$

\vspace*{1mm}

\item[(iii)] Nested CDD (NCDD)\footnote{Note that the sequences given in (ii) and (iii) are two extremes 
of the possible range of sequences that may be built using the composition rule of Eq. (\ref{compo}) and \text{CDD} 
as building blocks. The composition rule may be naturally extended to {\em multi-axis} DD sequences, 
recovering for instance concatenated UDD or quadratic DD \cite{qec}.}: 
$$\mathring{\text{NCDD} }\hspace*{-.6mm}\mbox{ }_{(\alpha_1,\ldots,\alpha_N)}^{(\vec{X})}(T) = 
\text{CDD} ^{(X_1)}_{\alpha_1} \circ \left( \text{CDD} ^{(X_2)}_{\alpha_2} \circ \left( \ldots 
\left(\text{CDD} ^{(X_{N-1})}_{\alpha_{N-1}} \circ \text{CDD} ^{(X_{N})}_{\alpha_{N}} \right)\right) \ldots \right)(T).$$
\end{itemize}
To date, NUDD is the most efficient known sequence, in terms of the required number of pulses 
capable of ensuring CO $\alpha$ for general multi-qubit dephasing noise as in Eqs. 
(\ref{gralHs})-(\ref{gralHsb}), at least for sufficiently hard noise spectral cutoffs~\cite{PasiniPowerlaw}.

\begin{figure}[t]
\hspace*{30mm}
\includegraphics[width=0.7\columnwidth]{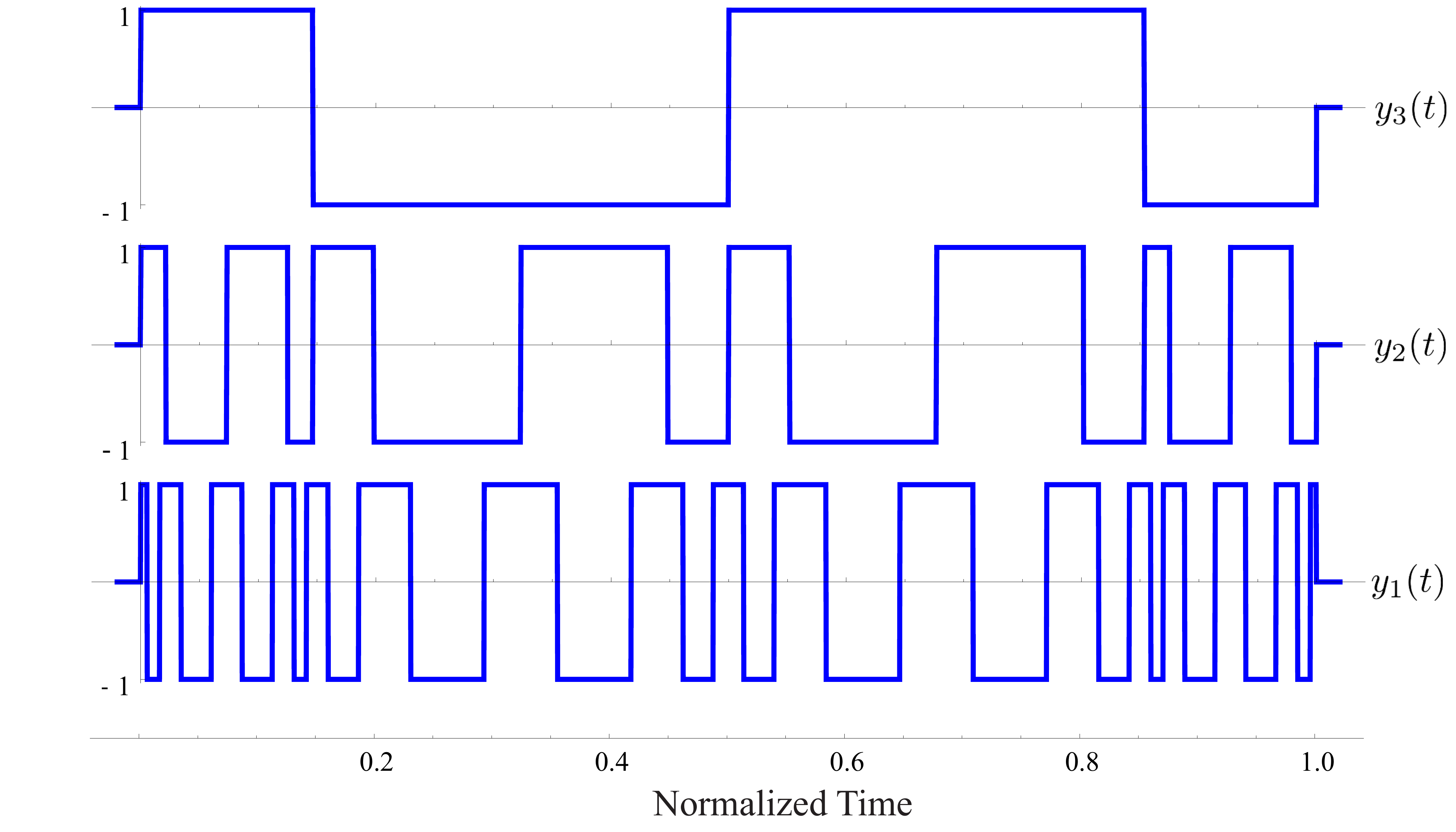}
\vspace*{-3mm}
\caption{
Control switching functions vs. normalized time for NUDD on $N=3$ qubits, NUDD$^{(X_1,X_2,X_3)}=$ 
UDD$^{(X_3)}_{\alpha_3} \circ ( \text{UDD}^{(X_2)}_{\alpha_2} \circ \text{UDD}^{(X_1)}_{\alpha_1})$, 
resulting in CO $\alpha =2$. Each crossing of the switching function with the time axis is associated with a 
control pulse on the corresponding qubit (for a total of $n_P^{\text{tot}}=4\times4\times2=32$ pulses). }
\label{fig:UDDvsCDD}
\end{figure}

Once a multi-qubit DD sequence is chosen, the effect of the control is most simply described 
in the interaction picture associated to the control propagator 
$U_c(t)$, which in the present setting simply leads to 
time-dependent ``error generators,'' namely,    
\begin{eqnarray*}
&&  \hspace*{-4mm}
Z_\ell   \mapsto Z_\ell(t) \equiv U_{c}^{\dagger}(t) Z_\ell U_{c}(t) = y_\ell(t) Z_\ell,\;\; \forall \ell, \\
&&  \hspace*{-4mm}
Z_\ell Z_{\ell'}  \mapsto Z_\ell Z_{\ell'} (t) \equiv 
U_{c}^{\dagger}(t) Z_\ell Z_{\ell'} U_{c}(t) = y_\ell(t) y_{\ell'} (t) Z_\ell Z_{\ell'},\;\; \forall \ell \ne \ell', 
\end{eqnarray*}
where the {\em control switching function} for the $\ell$-th qubit has the form
\begin{equation}
y_\ell (t)=\sum_{j=0}^{n_P^{(\ell)}} (-1)^j\Theta(t-t^{(\ell)}_{j}) \Theta(t^{(\ell)}_{j+1}-t). 
\label{yl}
\end{equation}
Here, $n_P^{(\ell)}$ denotes the total number of pulses applied on qubit $\ell$  and $\Theta$ is, as usual, 
the step function. We say that the control 
is (qubit) {\em non-selective} (or global) if $y_\ell(t) = y(t)$ $\forall \ell$, and that it is {\em selective} otherwise.  
Note that nested multi-qubit sequences necessarily require qubit-selective control.  A simple illustrative 
example is depicted in Fig. \ref{fig:UDDvsCDD}. 

\subsection{Exact representation of the controlled dynamics and fundamental filter functions}
\label{sub:exact}

In order to determine the reduced dynamics of the controlled multi-qubit system, the first step is to obtain an 
expression for the unitary ``error propagator'' resulting from Eq. (\ref{gralH}) under DD, evaluated in the combined 
interaction picture associated to the control and bath evolution.  Thanks to the fact that the bosonic algebra 
guarantees simple commutation relationships between noise operators, 
namely, $[B_\ell(t),B_{\ell'}(t')] \propto \mathbb{I}_B$ for all $t \ne t'$, the Magnus expansion \cite{qec} 
{\em truncates exactly at the second order}. Accordingly, the desired propagator, which we denote henceforth 
$\tilde{U}_e(T)$, may be compactly expressed as follows:
\begin{eqnarray}
\hspace*{-1cm}
\tilde{U}_e(T) & = & {\cal T_{+}}{\exp}\Big[-i \hspace*{-1mm}\int_{0}^{T} \hspace*{-1mm} U_c^\dagger (t) 
\left( \tilde{H}_{S}(t) + \tilde{H}_{SB}(t)  \right) U_c(t) \, dt\Big] \equiv \exp [-i H_{\text{eff}} T] \nonumber \\
& = & \textrm{exp} \bigg[  \sum_{ {\ell} } Z_{{\ell}} \Big(\overline{\zeta}'_{{\ell}}(T) + \overline{{B}}_{{\ell}} (T) \Big)
+  \sum_{{\ell} \ne {\ell'}} Z_{{\ell}} Z_{{\ell'}} 
\Big( {\overline{\eta}'_{{\ell},\ell'}(T)} + \overline{R}_{{\ell}, {\ell'}}(T)\Big) \bigg],
\label{Magnus}
\end{eqnarray}
where Eq. (\ref{Magnus}) explicitly defines the relevant effective Hamiltonian $H_{\text{eff}}$, and we have introduced  
new random variables as well as time-averaged quantities by letting
\begin{equation}
\label{newrv}
{\zeta}'_{{\ell}}(t) \equiv  d_{{\ell}}+ \zeta_{{\ell}}(t) , \quad 
{\eta'}_{{\ell,\ell'}}(t)   \equiv  d_{ \ell,\ell' }+ \eta_{ \ell,\ell' } (t) \vspace*{-4mm},
\end{equation}
\begin{equation}
\label{first} 
\hspace*{-3mm}
\overline{A}_\ell (T)   \equiv - i \int_0^T  \hspace*{-2mm} dt \, y_{{\ell}} (t) {A}_\ell(t) , \;
A_\ell \in \{ \zeta'_\ell, B_\ell \}, \;\; 
\overline{\eta}_{\ell , \ell'} (T)   \equiv - i \int_0^T \hspace*{-2mm} dt \, y_{{\ell}} (t) y_{\ell'}(t) {\eta}_{\ell(t), \ell'}(t)  \\
\end{equation}
\begin{equation}
\label{second}  
\overline{R}_{{\ell}, {\ell'}} (T) \equiv -\frac{1}{2} \int_0^T  \hspace*{-1mm} d t_1 \int_0^{t_1}  \hspace*{-1mm} d t_2 \,
y_{{\ell}} (t_1)y_{{\ell'}} (t_2) \, [{B}_{{\ell}}(t_1), {B}_{{\ell'}}(t_2)] .
\end{equation}
\noindent 
It is instructive to examine the propagator in the frequency domain, by exploiting the 
{\em fundamental FF} formalism introduced in Ref.~\cite{PazViola}.  Substituting the explicit forms of the Magnus terms 
in Eq. (\ref{first})-(\ref{second}) in the expression for $\tilde{U}_e(T)$ in Eq. (\ref{Magnus}) and taking the Fourier transform, 
one may rewrite (up to irrelevant global phase factors): 
\begin{align}
\nonumber 
\tilde{U}_e(T)  = \textrm{exp} & \Big[\hspace*{-1mm}
- i  \sum_{ {\ell} } Z_{{\ell}} \int_{-\infty}^\infty \frac{d \omega}{2\pi}  \, 
G^{(1)}_{Z_\ell} (\omega,T) (\tilde{\zeta'}_{{\ell}}(\omega)+ \tilde{B}_{{\ell}} (\omega)) \\ 
\nonumber & \; - i\sum_{ { {\ell}<{\ell'}  } }
Z_{{\ell}} Z_{{\ell'}} \Big( \int_{-\infty}^\infty \frac{d \omega}{2\pi} \, 
G^{(1)}_{Z_\ell Z_{\ell'}} (\omega, T) \tilde{\eta}'_{{\ell},\ell'}(\omega) \\ 
& \hspace{2.2cm} 
+ \int_{-\infty}^\infty \frac{d \omega_1}{2\pi} \int_{-\infty}^\infty \frac{d \omega_2}{2\pi}\, {G}^{(2)}_{Z_\ell,Z_{\ell'}} 
(\omega_1,\omega_2,T)  {[ \tilde{B}_{{\ell}} (\omega_1),  \tilde{B}_{{\ell'}} (\omega_2) ] \Big), }
\label{FFprop} 
\end{align}
where the relevant first- and second-order generalized FFs, $G^{(1)} (\omega, T)$ and $G^{(2)}(\omega_1, \omega_2, T)$,
may in turn be expressed in terms of fundamental FFs: 
\begin{align*}
-i G^{(1)}_{Z_\ell} (\omega,T) &=  F^{(1)}_{Z_\ell} (\omega,T), \quad \ell=1,\ldots, N, \\
-i G^{(1)}_{Z_\ell Z_{\ell'}} (\omega,T) &= F^{(1)}_{Z_\ell Z_{\ell'}} (\omega,T) + F^{(1)}_{ Z_{\ell'}Z_\ell} (\omega,T), \quad \ell\ne \ell', \\
-i G^{(2)}_{Z_\ell, Z_{\ell'}} (\omega_1,\omega_2,T) &=  
\frac{1}{2}\Big(F^{(2)}_{Z_\ell, Z_{\ell'}} (\omega_1,\omega_2,T) 
- F^{(2)}_{Z_{\ell'}, Z_{\ell}} (\omega_2,\omega_1,T)\Big), \quad \ell\ne \ell',  \\
\end{align*}
and the corresponding {\em first- and second-order fundamental FFs} are given by 
\begin{align}
F^{(1)}_{Z_\ell} (\omega,T)  &\equiv -i  \int_0^T \hspace*{-2mm}{dt} \, y_\ell (t) \, e^{i \omega t}  
= {-}
[F^{(1)}_{Z_\ell} (-\omega, T)]^\ast,  \label{FFF1} \\
 F^{(1)}_{Z_\ell Z_{\ell'}} (\omega,T)  &\equiv -i  \int_0^T \hspace*{-2mm}{dt} \, 
y_\ell (t) y_{\ell'} (t) \, e^{i \omega t}  
= {-}
[F^{(1)}_{Z_\ell Z_{\ell'}} (-\omega, T)]^\ast, \label{FFF1b}\\
F^{(2)}_{Z_\ell, Z_{\ell'}} (\omega_1,\omega_2,T) &\equiv -\int_0^T  \hspace*{-2mm}{dt_1} 
\int_0^{t_1} \hspace*{-2mm} {dt_2}   \,y_\ell (t_1) y_{\ell'} (t_2)  e^{i \vec{\omega}\cdot \vec{t}} = [F^{(2)}_{Z_\ell, Z_{\ell'}} (-\omega_1,-\omega_2,T)]^\ast. 
\label{FFF2}
\end{align}

We recall that this general filtering formalism allows one to determine the CO of a control protocol as well as 
its {\em filtering order} (FO) -- in particular, around $\vec{\omega}=0$, as relevant to DD.  If 
$$G^{(j)}_{O_{a_1},\ldots,O_{a_j}} (\omega_1,\ldots,\omega_j,T) = \mathcal{O} (m^{\Phi^{(j)}_{a_1,\ldots,a_j}} 
(\vec{\omega}) \, T^{\alpha^{(j)}_{a_1,\ldots,a_j}+1}),$$ 
\noindent
where $m^{\Phi^{(j)}_{a_1,\ldots,a_j}} (\vec{\omega})$ is a degree-${\Phi^{(j)}_{a_1,\ldots,a_j}}$ monomial in 
the components of $\vec{\omega}$, then the CO $\alpha = \min \{{\alpha^{(j)}_{a_1,\ldots,a_j}}\}$ and the 
(level-2-Magnus) FO $\Phi^{[2]} \equiv \Phi = \min \{{\Phi^{(j)}_{a_1,\ldots,a_j}}\}$, respectively. 
As discussed in \cite{PazViola},  it 
is crucial to analyze the suppression capabilities of a protocol in {\em both} the frequency and time domain to 
fully characterize the control  performance. From the filtering point of view, the high-order sequences described 
in the previous section have interesting properties: because only FFs of order $\alpha=1, 2$ are relevant to the analysis, 
one may verify that $\mathring{\textrm{CDD}}$ and NUDD sequences have CO equal to $\alpha$ and       
FO $\in \{ \alpha -1, \alpha \}$\footnote{We stress that this simple relationship between CO and FO is exclusive to 
dephasing scenarios. Under more general noise models, $\Phi$ and $\alpha$ are need {\em not} be tightly related 
and one can only guarantee that $\Phi^{[\infty]} \leq \alpha$ \cite{PazViola}. }.

The effective Hamiltonian $H_{\text{eff}}$ defined in Eq. (\ref{Magnus}) [or Eq. \eqref{FFprop} in Fourier space] 
comprises different  physical contributions. One-body terms, proportional to $Z_\ell$, are present for 
qubits coupled to arbitrary classical or private quantum baths. Two-body terms, proportional to $Z_\ell Z_{\ell'}$ 
and capable of inducing quantum correlations between different qubits, 
may have two distinct origins within our model:

\vspace*{-1mm}

\begin{itemize}
\item They may result from the ``direct'' Ising coupling in the free Hamiltonian, corresponding to the term 
$\overline{\zeta}'_\ell(T)$ in Eq. (\ref{Magnus}) and filtered by $G^{(1)}_{Z_\ell Z_{\ell'}} (\omega,T)$ in 
Eq. (\ref{FFprop}); 

\item They may result from the ``induced'' coupling mediated by the quantum bath via its non-commuting character,
corresponding to the term $\overline{R}_{\ell, \ell'}(T)$ in Eq. (\ref{Magnus})
and filtered by $G^{(2)}_{Z_\ell, Z_{\ell'}} (\omega_1,\omega_2,T)$ in Eq. (\ref{FFprop}).
\end{itemize}

\vspace*{-1mm}

\noindent
Even in the absence of direct coupling ($\eta'_{\ell, \ell'}(t)\equiv 0$), the bath-mediated interaction poses 
a main obstacle to designing less resource-intensive DD sequences, as we will see in the following section.  
Notwithstanding, it is important to iterate that, thanks to the bosonic algebra, the frequency-domain commutator 
$[\tilde{B}_{\ell}(\omega_1),\tilde{B}_{\ell'}(\omega_2)] \propto \mathbb{I}_B$ depends {\em only} on one frequency 
variable, and thus only $G^{(2)}_{Z_\ell, Z_{\ell'}} (\omega,-\omega,T)$ will be relevant to our analysis. 

Depending on the initial state of the system and/or bath, the cancellation or filtering order of a DD sequence may 
be {\em higher}~\cite{PazViola} than they are at the propagator level. 
In order to proceed to the exact solution for the reduced system dynamics under the initial factorization 
assumption $\rho_{SB}(0) = \rho(0) \otimes \rho_B$, let the initial $N$-qubit state  be 
\begin{equation}
\rho(0)=\sum_{{a},{b}}\ket{a_{1}\cdots a_{N}}\!\bra{b_{1}\cdots b_{N}}\rho_{a,b}(0) \equiv 
\sum_{{a},{b}}\ket{a}\!\bra{b}\rho_{a,b}(0),
\label{eq:initialstate}
\end{equation}
where the sum ranges over all possible binary strings ${a} \equiv a_1\cdots a_N$ and ${b}\equiv b_1\cdots b_N$ 
of length $N$.  The interaction-picture reduced density matrix at time $T$ may then be expressed as 
\beqy
\label{dyn}
\hspace*{-8mm}
\langle\rho(T)  \rangle_{c,q}  &=& \langle \textrm{Tr}_B \left[ \tilde{U}_e(T) \rho(0) \otimes \rho_B \tilde{U}_e(T)^\dagger   
\right] \rangle_c \equiv \sum_{{a},{b}} e^{- Z_{a, b}(T)} \rho_{a,b}(0) \ket{a}\!\bra{b} ,
\eneqy
where in general the complex factor 
$Z_{a, b}(T) \equiv \chi_{a, b}(T) + i \phi_{a,b}(T)$ allows for both non-trivial decay and phase evolution of each 
coherence element.  Specifically, direct calculation yields:
\beqy
e^{- Z_{a, b}(T)} &=& \langle e^{\sum_{ {\ell} < {\ell'}} \Delta[{a}_{{\ell}}+{a}_{{\ell'}},{b}_{{\ell}}+{b}_{{\ell'}}]  \bar{\eta}'_{{\ell},{\ell'}}(T)} \, 
e^{\sum_{{\ell}}  \Delta[{a}_{{\ell}},{b}_{{\ell}}] \bar{\zeta}'_{{\ell}}(T)  } \rangle_c 
\label{Z1} \\
&\; \cdot & \langle e^{\sum_{ {\ell}} \Delta[{a}_{{\ell}},{b}_{{\ell}}]  \bar{{B}}_{{\ell}}(T)} \rangle_q 
\langle e^{\sum_{ {\ell} < {\ell'}} \Delta[{a}_{{\ell}}+{a}_{{\ell'}},{b}_{{\ell}}+{b}_{{\ell'}}]  (\bar{R}_{{\ell},{\ell'}}(T) + 
\bar{R}_{{\ell'},{\ell}}(T))} \rangle_q 
\label{Z2}
\\ & \;\cdot &\langle e^{ \sum_{\ell,\ell'} \frac{(-1)^{|{a}_{{{\ell}}}|+|{b}_{{{\ell'}}}|}}{2} [\bar{{{B}}}_{{\ell}}(T), 
\bar{{B}}_{{\ell'}}(T)]  } \rangle_q\, ,     
\label{Z3}
\eneqy
where $\Delta[{u} , {v}] \equiv (-1)^{{u}} - (-1)^{{v}}$ and ${a}_{{\ell}}$ is the $\ell$-th entry of the string $a$. 
Formally, irrespective of whether we consider the classical ensemble average or 
the quantum statistical average with respect to $\rho_B$, the following cumulant expansion holds for a 
random variable or operator $Q$: 
\begin{equation}
\langle e^{Q}\rangle  = \bigg \langle \sum_{s=0}^\infty \frac{Q^s}{s!} \bigg \rangle = e^{\sum_{k=1}^\infty  
C^{(k)}(Q,\ldots,Q)/k!} ,
\label{Cexpansion}
\end{equation}
where $\langle \cdot \rangle$ stands for either $\langle\cdot \rangle_q$ or $\langle\cdot \rangle_c$. 
If $Q$ has {stationary zero-mean Gaussian} statistics, 
the only non-vanishing cumulant is $C^{(2)}(Q,Q) = \langle Q Q \rangle$, hence 
$\langle e^{Q}\rangle  = e^{\langle QQ\rangle/2}$. Note that all the quantities in the exponents that define  
$Z_{a,b}(T)$, which involve sums over single (classical or quantum) noise sources, inherit the properties 
of the corresponding noise processes: 
e.g., if $\zeta_\ell(t)$ is zero-mean Gaussian, the same holds for 
$Q =\sum_{{\ell}}  \Delta[{a}_{{\ell}},{b}_{{\ell}}] \bar{\zeta'}_{{\ell}}(T)$ 
in Eq. (\ref{Z1}). Furthermore, in view of the bosonic algebra, 
[Eq.~\eqref{Sminus}], 
the two contributions $\langle e^{\sum_{ { {\ell} < {\ell'}} } \Delta[{a}_{{\ell}}+{a}_{{\ell'}},{b}_{{\ell}}+{b}_{{\ell'}}]  
(\bar{R}_{{\ell},{\ell'}}(T) + \bar{R}_{{\ell'},{\ell}}(T))  }\rangle_q$ 
[Eq. (\ref{Z2})] and $\langle e^{ \sum_{\ell,\ell'} \frac{(-1)^{|{a}_{{{\ell}}}|+|{b}_{{{\ell'}}}|}}{2} 
[\bar	{{{B}}}_{{\ell}}(T), \bar{{B}}_{{\ell'}}(T)]  } \rangle_q$ [Eq. (\ref{Z3})] are {\em independent of $\rho_B$} 
and thus constants with respect to $\langle \cdot \rangle_q$. 
This allows us to obtain exact analytical expressions for the controlled 
dynamics when all noise sources are Gaussian. 

For non-Gaussian dephasing models in which the cumulant series does not truncate to the second order 
(including non-linear spin-boson models \cite{Chikako}), one needs to consider, in principle, all the infinite 
multiple-point correlations which, in the Fourier domain, implies dealing with an infinite hierarchy of FFs. 
Remarkably, however, for the family of classical plus linear bosonic noise models studied in this paper, 
arbitrary high-order cumulants can still be written in a compact way in terms of {\em only two generalized 
FFs, regardless of stationarity}:
\begin{align*}
\nonumber C^{(k)}(\bar{\zeta}_{{\ell_1}}(T) \cdots \ignore{\bar{\zeta}_{{\ell_j}}(T)} 
\bar{\eta}_{{p_{j+1}}}(T) \cdots \bar{\eta}_{{p_k}}(T)) &= \int_{-\infty}^\infty  \frac{d\vec{\omega}_{[k]}}{(2\pi)^k} \, 
S^{\zeta,\eta}_{\ell_1,\ldots,{{p_k}}} (\vec{\omega}_{[k]})  \,G^{(1)}_{Z_{\ell_1}} (\omega_1,T) \cdots 
G^{(1)}_{Z_{\ell_j}} (\omega_j,T)  \nonumber \\
& \hspace*{2.3cm} \cdot G^{(1)}_{Z_{\ell_{j+1}}Z_{\ell'_{j+1}}} (\omega_{j+1},T) \cdots 
G^{(1)}_{Z_{\ell_k}Z_{\ell'_k}} (\omega_k,T),\\
 C^{(k)}(  \bar{{B}}_{{\ell_1}}(T) \cdots \bar{{B}}_{{\ell_k}}(T)) &=\int_{-\infty}^\infty  \frac{d\vec{\omega}_{[k]}}{(2\pi)^k} \,S^B_{{{\ell}_1},
 \ldots,{{\ell}_k}} (\vec{\omega}_{[k]}) \, G^{(1)}_{Z_{\ell_1}} (\omega_1,T) \cdots  G^{(1)}_{Z_{\ell_k}} (\omega_k,T), 
\end{align*}
and
\beq
\nonumber  \bar{R}_{{\ell},{\ell'}}(T) + \bar{R}_{{\ell'},{\ell}}(T)  = \int_{-\infty}^\infty \frac{d\vec{\omega}_{[2]}}{(2\pi)^2} \,
 G^{(2)}_{Z_{\ell}, Z_{\ell'}} (\omega_1,\omega_2,T)  S^{B,-}_{\ell, \ell'} (\omega_1, \omega_2) , 
\eneq
where now both $G^{(2)}_{Z_{\ell}, Z_{\ell'}}$ and $S^{B, -}_{\ell,\ell'}$ are functions of two independent 
frequency variables. Similarly the $k$-th order polyspectra depend on the full set of $k$ frequency variables. 
Together with the exact expressions in Eqs. (\ref{Z1})--({\ref{Z3}), the above still provide a closed-form representation 
of the reduced multi-qubit dynamics, which may be used to infer general results or serve as 
a basis to build approximations via truncation of the cumulant expansion.

\section{Dynamical Decoupling versus Multi-qubit Dephasing Noise: Short-time memory}
\label{Decou}

Having introduced the necessary tools, we are now poised to address the first of the two control 
problems we set out to explore, namely, efficient short-term memory using DD.
Different DD protocols will be contrasted, and in each case we will characterize their  
decoherence suppression capabilities and extract their CO and FO directly from the 
relevant generalized FFs. While certain control strategies will only allow a fixed, small CO, 
we will see how, as sufficient structure is added to the control in terms of selectivity and symmetry, 
arbitrarily high CO is possible in principle. Beside re-establishing \text{CDD}  and NUDD sequences as 
capable of arbitrary CO against general dephasing noise~\cite{qec,UDD,CDD,WanPaz,Wang2011}, 
a main goal is to construct more efficient DD sequences tailored to dephasing noise with 
particular features. We will present a new family of multi-qubit sequences that, by satisfying a particular 
{\em displacement anti-symmetry} condition, can achieve arbitrary CO {\em using exponentially less resources}, 
so long as no direct Ising coupling is present. 
The symmetry that these new sequences possess will also prove fundamental in the 
context of building long-term multi-qubit memories [see Sec.~\ref{longterm}].

\subsection{Non-selective multi-qubit control sequences}
\label{sec:ns}

The most naive, yet often most readily available, strategy to decouple multiple qubits 
from their environment is to rely on non-selective control. In the formalism of Sec. 
\ref{control}, this means applying global control operations of the form 
$X_1 \otimes \ldots \otimes X_N \equiv X_1\cdots X_N$, 
with a corresponding controlled propagator $U_{\alpha}^{({X_1\cdots X_N})} (T)$. 
If $y_\ell(t)$ is the control switching function for qubit $\ell$, requiring that the 
CO $\alpha \geq 1$ implies that 
$G^{(1)}_{Z_\ell} (\omega,T) \sim \mathcal{O} (m^\alpha(\vec{ \omega}) T^{\alpha+1})$~\cite{PazViola}.
Since, however, the same pulse is applied synchronously to each qubit, $y_\ell (t) = y (t)$ for all $\ell$, 
hence $y_{\ell} (t) y_{\ell'}(t) = 1$ for all $t$. This affects the ability of non-selective DD to 
suppress time-independent noise: while $G^{(1)}_{Z_\ell} (0,T) = 0$,
it follows from Eq. (\ref{FFF1b}) that $G^{(1)}_{Z_\ell Z_{\ell'}} (0,T) = {2 T} \neq 0, $
i.e., the noise induced by any constant two-qubit direct coupling [$d_{\ell, \ell'} \ne 0$ in Eq. (\ref{gralHs})] 
{\em cannot} be suppressed by non-selective DD. This is expected, as global pulses 
commute with the direct (system-only) coupling term and thus cannot affect it. 

Interestingly, the bath-induced two-qubit coupling and the phase evolution ensuing from the second-order 
Magnus term {\em can} nevertheless be suppressed, 
albeit not according to an arbitrarily high power-law behavior.  If either a CDD$_\alpha$ 
or UDD$_\alpha$ non-selective protocol is applied, the FF suppressing the induced coupling term, 
$G^{(2)}_{Z_\ell, Z_{\ell'}} (\omega,-\omega,T)$, is found to scale as $\mathcal{O} (m^1(\vec{ \omega}) T^3)$, 
similar to free evolution (note that, using Eq. (\ref{FFF2}),
$G^{(2)}_{Z_\ell, Z_{\ell'}} (\omega,-\omega,T)$ is a real function). 
Thus, non-selective DD cannot suppress the induced coupling term any better than free evolution, namely, 
only up to a fixed order independent of $\alpha$. While at first it would then seem that increasing $\alpha$ would 
not improve error suppression, as claimed in some two-qubit analysis~\cite{Pan2012,Kim2000}, numerical exploration
clearly demonstrates that as $\alpha$ increases, the 
absolute value $|G^{(2)}_{Z_\ell, Z_{\ell'}} (\omega_1,\omega_2,T)|$ {\em does} decrease, i.e., better suppression is 
achieved (see Fig.~\ref{preZeno}) . 

\begin{figure*}[tbp]
\centering
\hspace*{18mm}\includegraphics[width=0.85\columnwidth]{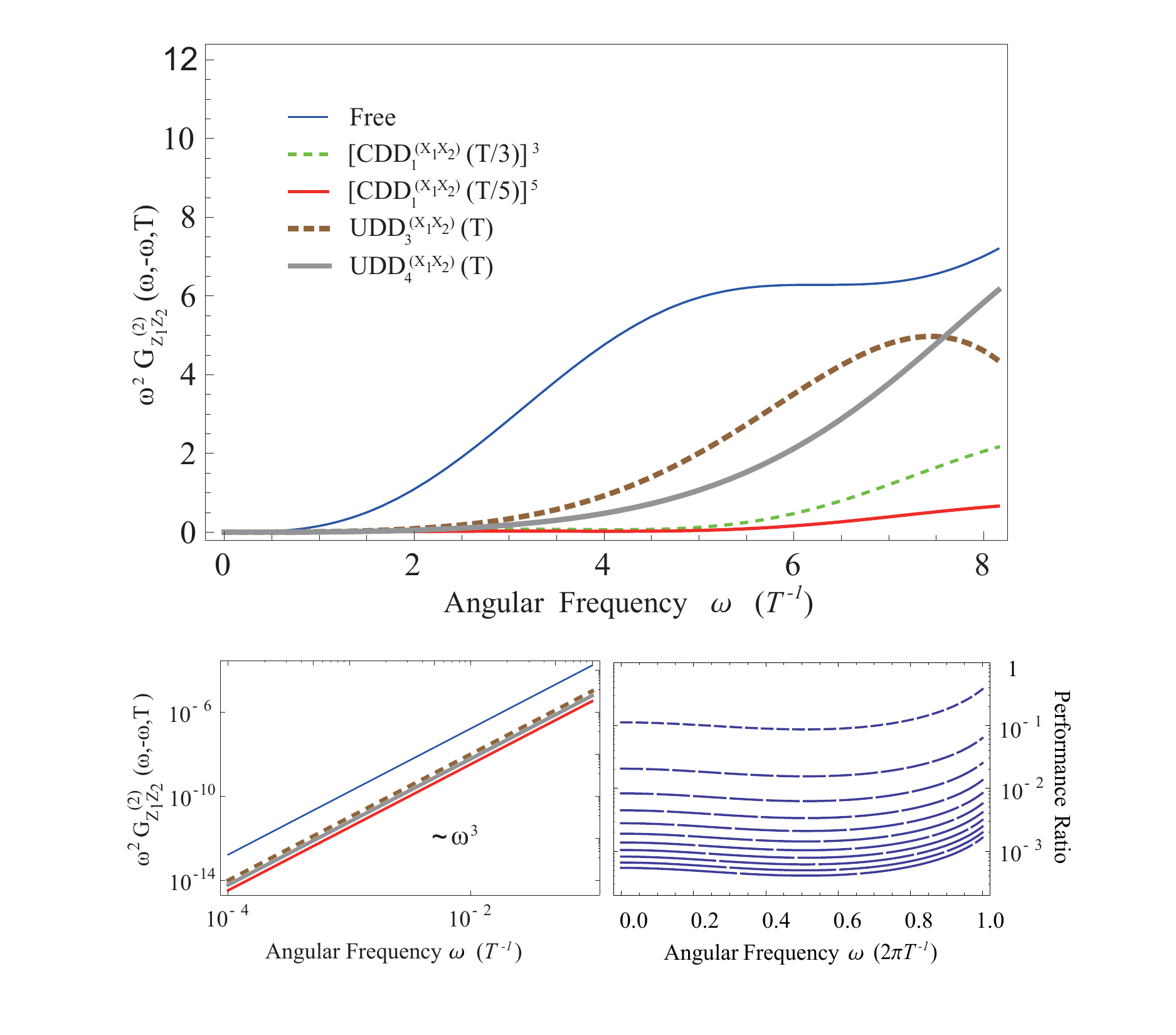}
\vspace*{-8mm}
\caption{Effect of non-selective DD sequences on two-qubit bath-induced phase evolution.
Top: FFs for sequences with comparable minimum interpulse separation, 
$\tau$: $\tau \sim  0.14 T$ for UDD$_3^{(X_1X_2)} (T)$ and [CDD$_1^{(X_1X_2)} (T/3)]^3$, 
while $\tau \sim 0.09 T$ for UDD$_4^{(X_1X_2)} (T)$ and [CDD$_1^{(X_1X_2)} (T/5)]^5$. 
In all cases the CO and FO are the same as for free evolution (lower left panel), but the value of the 
FF decreases as the number of pulses within a fixed time $T$ grows.
This is  verified by studying $M$ repetitions of CDD$_1$ 
($2 \tau_M =2 T/M)$ within $T$. Lower right panel: Log-plot of the ratio 
$|G^{(2)}_{Z_\ell, Z_{\ell'}} (\omega, -\omega,T = 2 M \tau_M )|/ |G^{(2)}_{Z_\ell, Z_{\ell'}} 
(\omega, - \omega,T = 2 \tau_1)|$ for $M = 3 + 4 s'$ and $s' \in[0,10]$. Plots with longer 
dashes correspond to larger values of $M$ and exhibit smaller ratios, hence better suppression, of the 
induced coupling.
}
\label{preZeno}
\end{figure*}

This behavior may be understood in terms of a formal analogy between DD and a Zeno effect resulting from the 
increasingly more frequent pulses within $[0,T]$, as suggested in~\cite{Viola1998} for a single qubit. 
Consider implementing $M$ repetitions of a given non-selective DD sequence $U_1^{{X_1\ldots X_N}} 
(\tau_M =T/M)$, that is, an overall sequence $[U_1^{{X_1\ldots X_N}} (T/M)]^M$ over a fixed time $T$, 
in the continuous limit where $M\rightarrow \infty$. While the CO is fixed, 
the number of pulses grows as $2M$. In the simplest case of $U_1^{X_1\ldots X_N}=
\text{CDD}_1$, direct calculation yields
$$ \hspace*{-2mm}
|G^{(2)}_{Z_\ell, Z_{\ell'}} (\omega,-\omega,T)|  =   \frac{ 
2 \left\vert \omega T - \tan \left(\frac{\omega T}{4 M}\right) 
\left[ \sin (\omega T) \tan \left (\frac{\omega T}{4 M}\right) + 4 M \right] \right\vert }{\omega^2} 
\sim \mathcal{O} \Big(\frac{\omega T^3 }{M^2} \Big).$$
Thus, while increasing $M$ does not change the power law of $\omega$ or $T$, 
a suppression effect still takes place, stemming from the increasing number of pulses within the storage time.  

While, in practice, if no direct Ising couplings are present, the quality of the achievable 
short-time suppression under non-selective control will depend upon specific system and noise details, 
selective control sequences become increasingly important and ultimately necessary 
as technological limitations prevent access to sufficiently short minimum switching times.

\subsection{Selective two-qubit control sequences}
\label{sub:selective}

When qubit-selective control is available, multi-qubit DD protocols with arbitrarily high CO may be devised, 
using the composition rule in Eq. (\ref{compo}).  Both for clarity and relevance to 
near-term implementations, we first consider two-qubit systems, though by using notation that may be  
easily adapted to the multi-qubit case (Sec. \ref{sub:multisel}).  
Let the DD sequences being composed have CO $\alpha_1$, $\alpha_2$, and require $n(\alpha_1)$, 
$n(\alpha_2)$ pulses.
The composition guarantees a high CO, $\alpha =\min \{\alpha_1,\alpha_2\}$, hence 
 all the generalized FFs are $\mathcal{O} (T^{\alpha+1})$. 
We seek strategies that achieve the same CO,  but that also ensure 
high FO and require {\em less} than $n_P^{ \text{tot} }= n(\alpha_1) n(\alpha_2)$ pulses.

\subsubsection{Vanishing direct qubit-qubit coupling.}

Assume that no direct inter-qubit coupling is present, i.e., $d_{\ell,\ell'} = 0 = \eta_{\ell,\ell'}(t)$ for all $t$ 
(hence, no contribution to decay due to $\bar{\eta}'_{\ell, \ell}(T)$ in Eq. (\ref{Z1})).
Similar to non-selective control protocols $U_{\alpha}^{(X_\ell X_{\ell'})}$ considered above,
one can see that simply executing the single-qubit sequences 
$U_{\alpha_{\ell}}^{(X_{\ell})} (T)$ and $U_{\alpha_{\ell'}}^{(X_{\ell'})}(T)$ {\em independently} on the two qubits 
achieves the desired power-law dependence for the local (one-body) terms but {\em not}  for the induced (two-body) 
coupling terms: that is, 
$G^{(1)}_{Z_\ell} (\omega,T) \sim \mathcal{O} ({\omega}^\alpha T^{\alpha+1})$, 
whereas $G^{(2)}_{Z_\ell, Z_{\ell'}} (\omega_1,\omega_2,T) 
\neq \mathcal{O} (m^\alpha(\vec{\omega}) T^{\alpha+1})$ in general. 
However, adding more structure to the control design can overcome the issue, yielding arbitrarily high FO and CO 
for all the generalized FFs relevant to the problem, as we show next.

$\bullet$ {\bf Mirror anti-symmetry.--} While, as remarked, the commutator spectrum 
$S^{B,-}_{\ell, \ell'}(\omega_1, \omega_2)$ that enters the 
error integrals for two-body terms, Eq.~\eqref{FFprop}, is in general complex albeit anti-symmetric in the sense of 
Eq. (\ref{HQspect1}), a simplified solution is possible if the noise is known to obey additional symmetry properties.
Suppose that 
\begin{equation}
{S}^{B,-}_{{\ell},{\ell'}} (\omega) = - S^{B,-}_{{\ell},{\ell'}} (-\omega),\quad \forall \omega,
\label{symmnoise}
\end{equation}
and consider any DD protocol that satisfies the following {\it mirror anti-symmetry} condition~(see also top panel in Fig. 
\ref{fig:Symmetry} for a pictorial illustration):
\beq
\label{mirasym}
y_{{\ell}} (T/2 + t_1) y_{{\ell'}}(T/2 +t_2) = - y_{{\ell}}  (T/2 - t_1) y_{{\ell'}}(T/2 - t_2),\quad \forall t_1, t_2 \in [0, T/2], 
\eneq
with {\em mirror symmetry}, corresponding to a reflection about $t=T/2$, 
being correspondingly described by a $+$ sign in the right hand-side of Eq. (\ref{mirasym}). 
Then it is easy to verify that  
\begin{align}
\nonumber &   { - 2 i }\int_{-\infty}^\infty \hspace*{-1.5mm} d \omega_1 \int_{-\infty}^\infty \hspace*{-1.5mm}
d \omega_2  \, G^{(2)}_{Z_{\ell}, Z_{\ell'}} (\omega_1,\omega_2,T) 
S^{B,-}_{{\ell},{\ell'}} (\omega_1, \omega_2)  \\
\nonumber =& \int_{-\infty}^{\infty} \hspace*{-1.5mm} d \omega \left[ \int_0^{T} \hspace*{-1.5mm}
dt_1 \int_0^{t_1} \hspace*{-1.5mm} dt_2 \, y_\ell(t_1) y_{\ell'}(t_2)  
- \int_{T}^{0} dt_2 \int_T^{t_2} dt_1 \, y_\ell(t_1) y_{\ell'} (t_2) \right] \hspace*{-.8mm}
e^{i \omega(t_1 -t_2)} {S}^{B,-}_{{\ell},{\ell'}} (\omega)
= 0. \nonumber 
\end{align}
Thus, mirror anti-symmetry guarantees that no contribution to the propagator in Eq.~\eqref{FFprop} 
arises from the above integral. A simple recipe to generate a control protocol with the required 
anti-symmetry is to execute independent \text{CDD} or UDD sequences with CO $\alpha_\ell$ and $\alpha_{\ell'}$ obeying 
the ``odd-parity condition'' that $\alpha_\ell + \alpha_{\ell'}$ is odd. If so, $y_\ell(t)$ has mirror symmetry, 
$y_\ell(T/2+t) = y_{\ell}(T/2-t)$, whereas $y_{\ell'}(t)$, has mirror anti-symmetry, $y_{\ell'}(T/2+t) = - y_{\ell'}(T/2-t)$, 
their product then obeying Eq.~\eqref{mirasym}. Let us denote this type of independent sequences by $U^{X_\ell}_{\alpha_\ell} \times U^{X_{\ell'}}_{\alpha_{\ell'}} (T)$. The only contributing generalized FF, $G^{(1)}_{Z_\ell} (\omega,T) \sim 
\mathcal{O} (  \omega^\alpha T^{\alpha+1})$, 
implies then that the sequence achieves CO $=$ FO $= \alpha = \min \{\alpha_\ell, \alpha_{\ell'}\}$ by using 
{\em only $n(\alpha_\ell) + n(\alpha_{\ell'})$ pulses}, as opposed to the $n(\alpha_\ell) \, n(\alpha_{\ell'})$ 
for a sequence built via composition.  For example, using the UDD sequences as building blocks, NUDD would require 
$(\alpha_1+1)(\alpha_2+1)$ pulses, in contrast to $(\alpha_1+\alpha_2+2)$ pulses for mirror anti-symmetric DD. 

\begin{figure}
\centering
\hspace*{24mm}\includegraphics[width=0.65\textwidth]{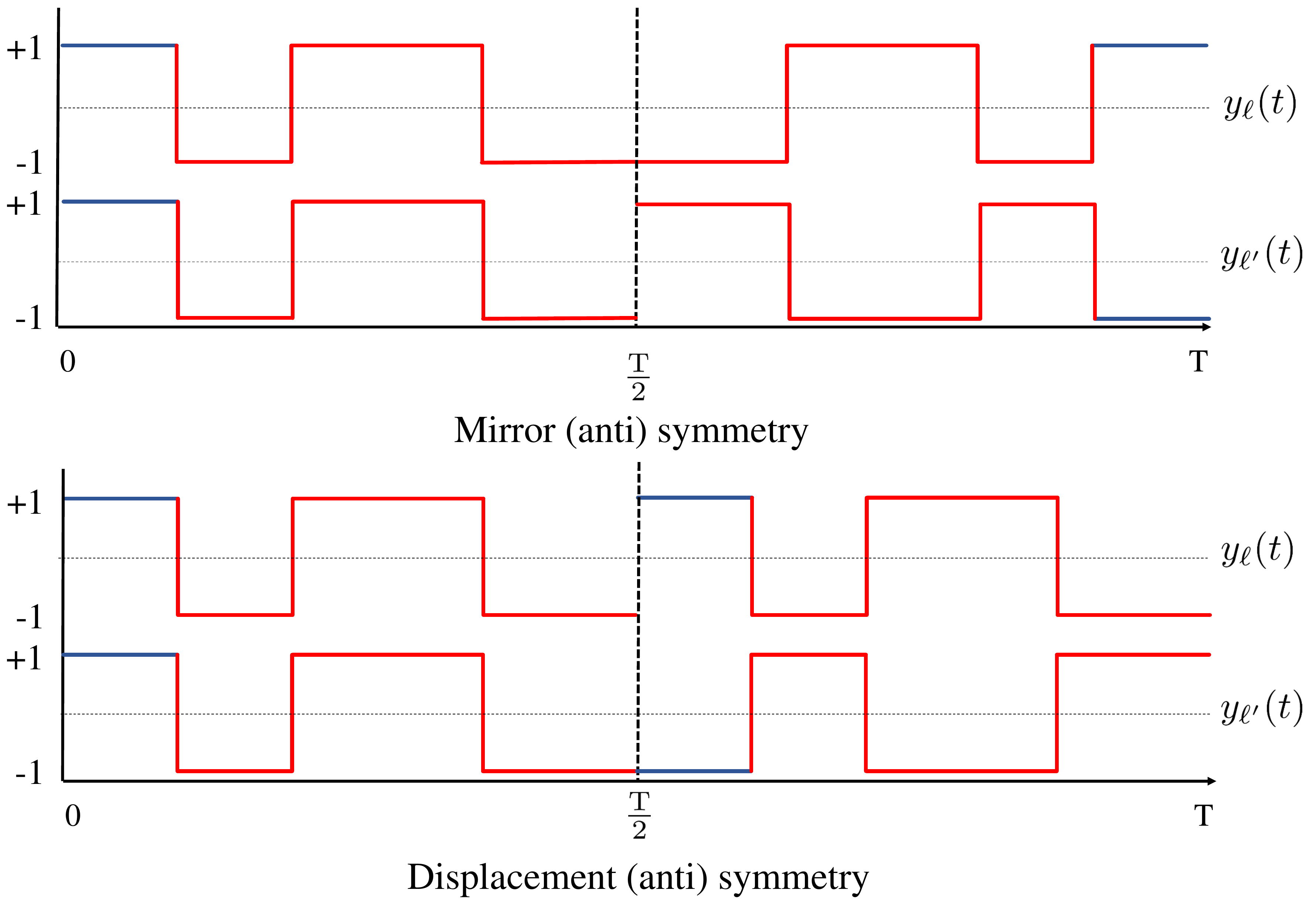}
\vspace*{-3mm}
\caption{Mirror- (or displacement-) {\em symmetric} patterns (top row in each 
diagram) vs. mirror- (or displacement-) {\em anti-symmetric} pattern (bottom rows). 
Mirror symmetry [Eq. (\ref{mirasym})] entails a reflection of the pattern with respect to $T/2$, whereas 
displacement symmetry [Eq. (\ref{displ})] is achieved via a translation in $[0,T/2]$. 
Anti-symmetry requires an extra inversion operation -- effectively multiplying the pattern by $-1$. 
The segment highlighted in blue is a visual aid. 
}
\label{fig:Symmetry}
\end{figure}

The symmetry condition imposed on the noise spectra by Eq. (\ref{symmnoise})
corresponds to requiring that $g_k^\ell g_k^{\ell' *} \in {\mathbb R}$. In view of Eq. (\ref{ttime}), this in turn 
implies {\em real coupling strengths}, $\textrm{Im} [g_k^\ell] = 0$ for all $\ell$, with collective dephasing as a 
special case. Interestingly, it had already been shown \cite{Pan2012} that 
enhanced DD protocols obeying the above-mentioned odd-parity condition guarantee high CO for two qubits 
exposed to collective spin-boson dephasing from a Gaussian (thermal) bath state.
Beside shedding light on the underlying symmetry enabling this result, our analysis at the 
propagator level shows something stronger:  as long as Eq. (\ref{symmnoise}) is obeyed, 
mirror anti-symmetric sequences achieve arbitrary CO 
irrespective of the initial state $\rho_{B}$, hence including 
arbitrary {\em non-Gaussian} bath states.

\vspace*{1mm}

$\bullet$ {\bf Displacement anti-symmetry.--} Since the environmental couplings may not be precisely characterized 
in realistic scenarios, or need not comply with Eq. (\ref{symmnoise}), assuming such a symmetry is too 
stringent in general.  We thus proceed to design {\em model-robust} efficient high-order sequences without 
any further symmetry assumptions beyond those stemming from the bosonic algebra.  
Demanding a suitable {symmetry on the control switching functions} will still be crucial to achieve this goal. 
Building on the explicit form of the second-order fundamental FF relevant to the problem, 
$F^{(2)}_{Z_\ell, Z_{\ell'}} (\omega_1,\omega_2,T)$ in Eq. (\ref{FFF2}), we propose the use of a 
{\it displacement anti-symmetry}, defined as follows~(see also bottom panels in Fig.~\ref{fig:Symmetry}): 
\begin{equation}
y_{{\ell}} (T/2 + t_1) y_{{\ell'}}(T/2 +t_2) = - y_\ell(t_1) y_{{\ell'}}(t_2), \quad \forall t_1, t_2 \in [0, T/2], 
\label{displ}
\end{equation} 
with displacement symmetry being instead associated to the $+$ sign in the right hand-side of the 
above equation. Direct calculation shows that the second-order fundamental FF obeys 
\begin{align}
\nonumber
F^{(2)}_{Z_\ell, Z_{\ell'}} (\omega_1,\omega_2,T) 
\nonumber & \hspace*{-1mm}=  - \bigg[ \int_0^{T/2}  \hspace*{-2mm}{dt_1} 
\int_0^{t_1} \hspace*{-2mm} {dt_2}+ \int_{T/2}^T  \hspace*{-2mm}{dt_1} 
\int_0^{T/2} \hspace*{-2mm} {dt_2} + \int_{T/2}^T  \hspace*{-2mm}{dt_1} 
\int_{T/2}^{t_1} \hspace*{-2mm} {dt_2} \bigg] y_\ell (t_1) y_{\ell'} (t_2)  e^{i \vec{\omega}\cdot \vec{t}}
\nonumber \\ 
\nonumber &= F^{(2)}_{Z_\ell, Z_{\ell'}} (\omega_1,\omega_2,T/2) -\int_{0}^{T/2}  \hspace*{-2mm}{dt'_1} 
\int_0^{T/2} \hspace*{-2mm} {dt_2}   \,y_\ell (T/2+t'_1) y_{\ell'} (t_2)  e^{i \vec{\omega}\cdot \vec{t}} 
e^{i \frac{\omega_1 T}{2}}\\
\nonumber & \hspace*{24mm} -\int_0^{T/2}  \hspace*{-2mm}{dt'_1} 
\int_0^{t'_1} \hspace*{-2mm} {dt'_2}   \,y_\ell (T/2+t'_1) y_{\ell'} (T/2+t'_2)  e^{i \vec{\omega}\cdot \vec{t}} 
e^{i \frac{(\omega_1+\omega_2) T}{2}},
\end{align}
where $t'_j=t_j-T/2$ for $j=1,2$. Thus, the  anti-symmetry requirement implies that
\begin{align*}
F^{(2)}_{Z_\ell, Z_{\ell'}} (\omega_1,\omega_2,T) & \hspace*{-0.6mm}= \hspace*{-0.6mm} 
(1-e^{i \frac{(\omega_1 + \omega_2) T}{2}}) 
F^{(2)}_{Z_\ell, Z_{\ell'}} (\omega_1,\omega_2,{T}/{2}) {-} e^{i  \frac{\omega_1 T}{2}} F^{(1)}_{Z_\ell} 
(\omega_1,{T}/{2}) F^{(1)}_{Z_{\ell'}} (\omega_2,{T}/{2}),\\
 F^{(2)}_{Z_{\ell'}, Z_{\ell}} (\omega_2,\omega_1,T) &\hspace*{-0.6mm}= \hspace*{-0.6mm} 
 (1-e^{i \frac{(\omega_1 + \omega_2) T}{2}}) 
 F^{(2)}_{Z_{\ell'}, Z_{\ell}} (\omega_2,\omega_1,{T}/{2}) { +} e^{i \frac{\omega_2 T}{2}} F^{(1)}_{Z_{\ell'}} 
 (\omega_2,{T}/{2}) F^{(1)}_{Z_{\ell}} (\omega_1,{T}/{2}),
\label{sb} 
\end{align*}
with each fundamental FF appearing in an integral over $(\omega_1$, $\omega_2)$ and being multiplied 
by $S^{B,-}_{\ell,\ell'} (\omega_1,\omega_2) \propto \delta(\omega_1 + \omega_2)$. 
Accordingly, $1-e^{i \frac{(\omega_1 - \omega_1) T}{2}} =0$ 
and the first term in each of the above expressions never contributes. It follows that the relevant 
second-order generalized FF {\em factorizes into a product of first-order fundamental FFs}:
\beq 
\label{sb} 
{- i} G^{(2)}_{Z_\ell, Z_{\ell'}} (\omega, - \omega,T) = - \cos ( \omega T/2) \,
F^{(1)}_{Z_\ell} (\omega,T/2) F^{(1)}_{Z_{\ell'}} (-\omega,T/2).
\eneq
Let the two basic single-qubit DD sequences correspond to 
CO (FO) equal to $\alpha_\ell$ ($\Phi_\ell$) and $\alpha_{\ell'}$ ($\Phi_{\ell'}$) in the intervals $[0,T/2]$ 
and $[T/2,T]$, respectively.  Eq. (\ref{sb}) then implies that  
$$G^{(2)}_{Z_\ell, Z_{\ell'}} (\omega,-\omega,T) \sim \mathcal{O} (\omega^{ \Phi_\ell + \Phi_{\ell'}} \,
T^{ \alpha_\ell + \alpha_{\ell'} {+2}}),$$
\noindent 
and thus the resulting two-qubit sequence achieves 
CO $=\min \{ \alpha_1, \alpha_2\} $ and FO $= \min \{ \Phi_1, \Phi_2\}$, 
using {\em only $2[n(\alpha_1) + n(\alpha_{2})]$ pulses}. 
Notice that it is straightforward to build a control sequence possessing 
the required displacement anti-symmetry: given a DD sequence over an interval of duration $T/2$, say 
$U_{\alpha_\ell}^{(X_\ell)} {\times} U_{\alpha_{\ell'}}^{(X_{\ell'})}(T/2)$ or $U_{\alpha}^{(X_\ell X_{\ell'})}(T/2)$, 
define  
\begin{align}
\label{2disl} 
U_{\alpha_\ell,\alpha_{\ell'}}^{d, (X_\ell, X_{\ell'})} (T) &\equiv X_{\ell} \Big( U_{\alpha_\ell}^{(X_\ell)} {\times } 
U_{\alpha_{\ell'}}^{(X_{\ell'})}(T/2) \Big) \, X_{\ell}^\dagger \left ( U_{\alpha_{\ell}}^{(X_{\ell})} {\times} 
U_{\alpha_{\ell'}}^{(X_{\ell'})}(T/2) \right)  , \\
\label{2diss} 
U_{\alpha}^{d, (X_\ell X_{\ell'})}(T) &\equiv X_{\ell} \Big( U_{\alpha}^{(X_1 X_2)} (T/2) \Big) \, X_{\ell}^\dagger 
\Big( ( U_{\alpha}^{(X_1 X_2)}(T/2) \Big) ,
\end{align}
where the conjugation of the second half of the sequence, over time $[T/2, T]$, by $X_{\ell'}$ guarantees that 
$y_{{\ell}}(T/2 +t_1) = - y_{{\ell}}(t_1)$ while $y_{{\ell'}}(T/2 +t_2) =  y_{{\ell'}}(t_2)$, 
altogether ensuring that the product $y_{{\ell}} (T/2 + t_1) y_{{\ell'}}(T/2 +t_2) = - y_\ell(t_1) y_{{\ell'}}(t_2)$ 
in agreement with Eq. (\ref{displ}).

While the relevant first- and second-order generalized FFs give us information on the worst-case filtering and 
cancellation capabilities that a DD protocol can ensure under minimal knowledge about the noise 
process, higher {\it effective} CO and FO can be achieved if additional assumptions hold \cite{PazViola}. Recall, 
in particular, the pure spin-boson stationary Gaussian noise model on two qubits discussed in the Appendix.
As remarked there, $G^{(1)}_{Z_\ell} (\omega,T)$ never appears alone in the reduced dynamics and thus, 
assuming e.g. that qubit $1$ corresponds to an anti-symmetric switching function, 
we have the {\it effective} generalized FFs 
\begin{eqnarray*}
& G^{(1)}_{Z_1} (\omega, T) G^{(1)}_{Z_{1}} (-\omega, T) \sim \omega^{2 (\alpha_{1} +1)} T^{2(\alpha_1+1)+2}, \label{d1} \\
& G^{(1)}_{Z_2} (\omega, T) G^{(1)}_{Z_{2}} (-\omega, T) \sim \omega^{2 \alpha_{2}} T^{2\alpha_2 + 2}, \label{d2} \\
& G^{(1)}_{Z_1} (\omega, T) G^{(1)}_{Z_{2}} (-\omega, T) \sim \omega^{(\alpha_{1} +1) + \alpha_2} 
T^{(\alpha_1+1)+\alpha_2 +2} , \label{d3} \\
& G^{(2)}_{Z_1,Z_2} (\omega,-\omega,T) \sim \omega^{\alpha_{1} + \alpha_2} T^{\alpha_1 + \alpha_2 +2} .
\end{eqnarray*}
Thus, the extra symmetry associated with the zero-mean Gaussian nature of the initial bath state has 
effectively increased  the CO and FO of the sequence. The above predicted power-law behaviors are 
demonstrated in Fig. \ref{FFsb}.

\vspace*{1mm}

\begin{figure*}[h]
\centering
\hspace*{20mm}\includegraphics[width=13.6cm]{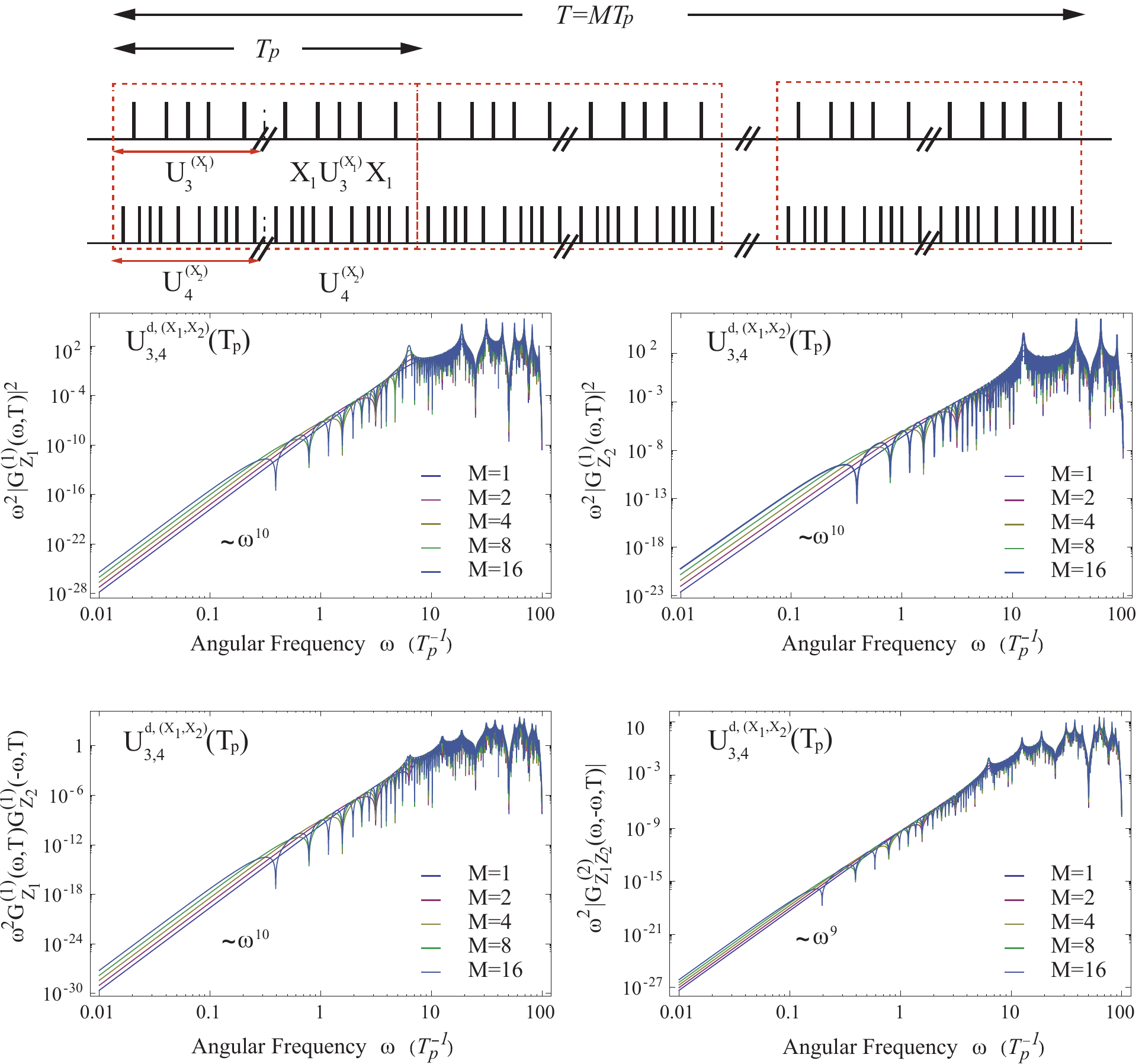}
\caption{
Top: Pulse structure for $M$ repetitions of a $U_{3,4}^{d,(X_1,X_2)}$ sequence with displacement anti-symmetry,
Eq. (\ref{2disl}), and \text{CDD}$_{\alpha_\ell}$, $\alpha_1=3, \alpha_2=4$, as building blocks. 
Bottom: Effective FFs entering the two-qubit reduced dynamics 
vs. frequency, for a different $M$. 
$G^{(2)}_{Z_1,Z_2} (\omega, -\omega,T)$ is purely imaginary whenever displacement anti-symmetry is imposed,
whereas $G^{(1)}_{Z_1} (\omega,T)G^{(1)}_{Z_2} (-\omega,T)$ is purely real, due to 
the fact that $\alpha^{(1)} + \alpha^{(2)} = 7$ is odd.  
}
\label{FFsb}
\end{figure*}

\vspace*{1mm}

$\bullet$ {\bf Sequence comparison.--} In order to gain quantitative insight, 
it is useful to compare the more efficient displacement-anti-symmetric DD sequences we have just built with 
other DD sequences applicable to our noise model.
The term which is most relevant to such a comparison is the induced coupling term, 
since guaranteeing a high CO for this term is the whole point of 
introducing anti-symmetry or using the composition rule. Specifically, let us focus on 
the noise-induced phase evolution $i\phi^0(T)$ over time $T$, namely, from Eq. (\ref{P0}), 
\beq
\label{interest}
2\pi i \phi^0(T) = -i \int^{\infty}_{-\infty} d \omega \,G^{(2)}_{Z_1,Z_2} (\omega,-\omega,T) S^{B,-}_{1,2} (\omega) 
\equiv - {i}  \int^{\infty}_{-\infty} d \omega I (\omega,T) 
\eneq 
evaluated for a sub-Ohmic noise spectrum $S^{B,-}_{1,2} (\omega)$, inspired by phenomenological treatments of 
nuclear-spin-induced dephasing in semiconductor quantum-dot qubits \cite{MikeBluhm,Memory}.
For fixed total time $T$ and minimum switching time $\tau$, we construct sequences with the highest possible 
CO $\alpha$ within these constraints, using \text{CDD} as building blocks and incorporating mirror anti-symmetry, 
displacement anti-symmetry, or using nesting -- see Table \ref{table:sequences} for a summary.

\begin{table}[t]
\centering
\begin{tabular}{l | ll}\hline 
\vspace*{-4mm} \\
{Free evolution} & CO$\,=2$, FO$\,=1$ & {\small No DD applied}\\
CDD$_{\alpha_1}^{(X_1)} \times$ CDD$_{\alpha_2}^{(X_2)} (T)$ & CO$\,=1$, FO$\,=0$ & 
{\small Mirror anti-symmetric DD, $\alpha_1 = 3,5$, $\alpha_2 = 2,4$} \\
CDD$_{\alpha}^{(X_1 X_2)} (T)$ & CO$\,=2$, FO$\,=1$ & {\small Non-selective CDD, $\alpha = 3,5$} \\
CDD$^{d,(X_1,X_2)}_{2,1} (T)$  & CO$\,=5$, FO$\,=4$ & {\small Displacement anti-symmetric DD, $T=T_1$} \\
CDD$^{d,(X_1,X_2)}_{4,4} (T)$  & CO$\,=9$, FO$\,=8$ &  {\small Displacement anti-symmetric DD, $T=T_2$} \\ 
$\mathring{\textrm{NCDD}}^{(X_1,X_2)}_{2,1} (T)$ & CO$\,=1$, FO$\,=0$ & {\small Nested CDD, $T=T_1$} \\ 
$\mathring{\textrm{NCDD}}^{(X_1,X_2)}_{3,2} (T)$ & CO$\,=5$, FO$\,=4$ & {\small Nested CDD, $T=T_2$ }\\ 
$\mathring{\textrm{CDD}}^{(X_1,X_2)}_{1,1} (T)$ & CO$\,=2$, FO$\,=1$ & {\small Multi-qubit CDD, $T=T_1$}\\
$\mathring{\textrm{CDD}}^{(X_1,X_2)}_{2,2} (T)$ & CO$\,=3$, FO$\,=2$ &  {\small Multi-qubit CDD, $T=T_2$} 
\vspace*{.5mm} \\
\hline
\end{tabular}
\caption{DD sequences applied in Fig. \ref{fig:filtering}, along with the corresponding CO and FO for the 
second-order FF $G^{(2)}_{Z_1,Z_2} (\omega,-\omega,T)$ relevant for noise-induced phase evolution. }
\label{table:sequences}
\end{table} 

\begin{figure}[t]
\centering
\hspace*{15mm}
\includegraphics[width=0.78\textwidth]{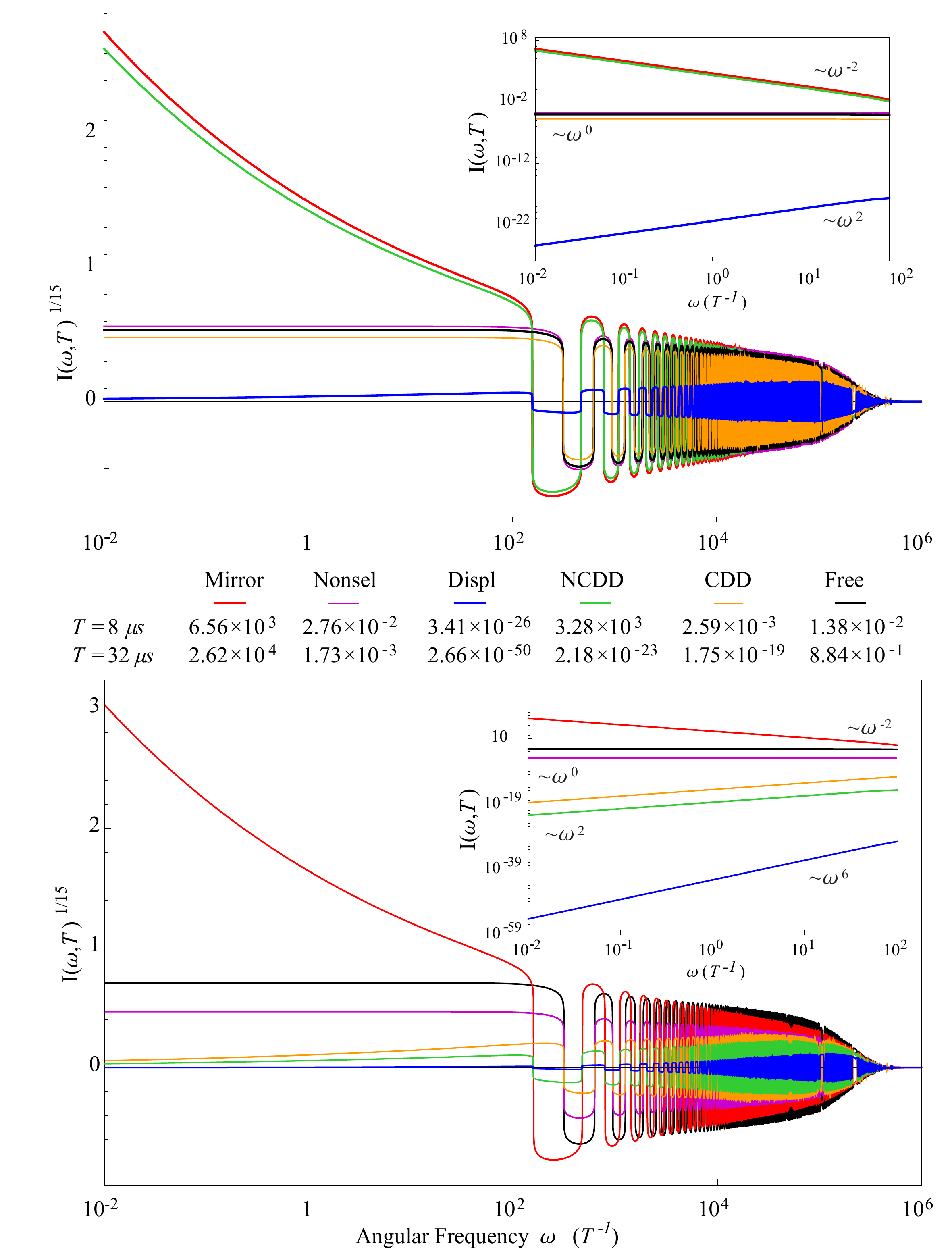}
\vspace*{-2mm}
\caption{
Comparison between different DD sequences capable of suppressing the induced coupling term 
vs. free evolution, for times $T\equiv T_1= 8 \mu$s (top) and $T\equiv 
T_2= 32 \mu$s (bottom), at same minimum switching time $\tau=1 \mu$s. 
A sub-Ohmic spectrum with Gaussian cutoff is used, $S^{B,-}_{1,2} (\omega) 
\simeq e^{i \omega t_{1,2}} g (\omega/\omega_c)^{-2} e^{-\omega^2/\omega_c^2}$, 
with $g/\omega_c  =  0. 2078$,  $\omega_c= 2\pi 10^4 \,\textrm{Hz}$, and $t_{1,2} =10^{-2}$ s.  
The main panels show $I(\omega,T)^{1/15}$ [Eq. (\ref{interest})]; 
insets are in log scale to facilitate the visualization of power-law behavior. The in-line table lists 
{the value of $|\int d \omega I(\omega, T)| =2\pi \phi^0(T)$ for 
the corresponding sequence. }
Specifically, we compare (see also Table \ref{table:sequences}):  
CDD$_{3}^{(X_1)} \times$ CDD$_{2}^{(X_2)} (T)$, 
CDD$_{3}^{(X_1)} \times$ CDD$_{3}^{(X_2)} (T)$,
CDD$^{d,(X_1,X_2)}_{2,1} (T)$, 
$\mathring{\textrm{NCDD}}^{(X_1,X_2)}_{2,1} (T)$,
and $\mathring{\textrm{CDD}}^{(X_1,X_2)}_{1,1} (T)$ 
for $T= T_1$; 
and CDD$_{5}^{(X_1)} \times$ CDD$_{4}^{(X_2)} (T)$, 
CDD$_{5}^{(X_1)} \times$ CDD$_{5}^{(X_2)} (T)$, 
CDD$^{d,(X_1,X_2)}_{4,4} (T)$, 
$\mathring{\textrm{NCDD}}^{(X_1,X_2)}_{3,2} (T)$, 
and $\mathring{\textrm{CDD}}^{(X_1,X_2)}_{2,2} (T)$ 
for $T\equiv T_2$.  
}
\label{fig:filtering}
\end{figure}

Representative results are shown in Fig. \ref{fig:filtering}.
While absolute values of the unwanted contribution $I(\omega,T)$ are quoted in the in-line table, 
we plot a small power of the integrand, { $I(\omega,T)^{1/p}$}, $p =15$, 
rather than using a logarithmic scale as 
in Fig.~\ref{FFsb}, in order to better visualize the full range of values while avoiding issues associated to the 
negative values of $I(\omega,T)$ at large frequencies. 
As one may see, guaranteeing a high FO (hence a high CO) for $G^{(2)}_{Z_1,Z_2} (\omega,-\omega,T)$ 
is key to effectively suppressing phase evolution, with similar conclusions 
holding for any spectrum $S^{B,-}_{1,2}(\omega)$ that is heavily weighted around $\omega=0$. Because 
displacement-anti-symmetric DD achieves the highest CO and FO for the induced coupling term, 
it outperforms all other strategies. Interestingly, two suppression mechanisms are evident: 
the first, stemming from the FO and CO, is manifest 
in the relative amplitudes of the oscillating functions, especially at low frequency; 
the second is the oscillatory character of the integrand itself, 
with positive and negative contributions partially canceling each other.

The absolute values of $I(\omega, T)$ quoted in Fig. \ref{fig:filtering} further show the importance of the FO: 
amongst all sequences, displacement anti-symmetry-enhanced DD yields substantially better suppression. 
While one might naively expect such a difference in performance to originate solely from the difference 
in CO, it has already been shown in \cite{PazViola} that sequences with the same CO can have very different 
performance because their FO differs. Here, we observe a related behavior. 
Observe that $G^{(2)}(\omega,-\omega, T)\sim  \mathcal{O} (\omega^{\phi^{(2)}} T^{\alpha^{(2)}+1})$ must have 
dimensions of $[T^2]$. If a sequence has overall CO $= \alpha$, one may guarantee that $\alpha^{(2)} \geq \alpha$, 
while dimensional analysis implies that  $-\phi^{(2)} + \alpha^{(2)} =1$. The sequence with the highest value of $\phi^{(2)}$ 
will thus also have the largest $\alpha^{(2)}$. Accordingly, a higher FO does not only imply better protection 
around $\omega=0$, but indeed a higher protection in the sense of the power law in $T$. 
Also notice that when the FO/CO of a sequence are below those of free evolution, DD may become a liability,
effectively increasing the unwanted noise effect; e.g., this is the case for both the nested CDD sequence and for 
the mirror anti-symmetric sequences for $T= 8 \mu$s.

\subsubsection{Non-vanishing direct coupling.}

In the presence of a direct Ising coupling in the system Hamiltonian, i.e., $d_{\ell,\ell'} \neq 0 \neq \eta_{\ell\ell'}(t)$, 
DD sequences applied independently to each qubit will not suppress such terms to arbitrary order.  
Using the displacement anti-symmetry-enhanced sequences described earlier and recalling Eq. (\ref{FFprop}), 
one finds that for the static direct coupling, proportional to $d_{\ell,\ell'}$, the relevant generalized FF 
$G^{(1)}_{Z_\ell Z_{\ell'}} (\omega=0,T) = 2 \int_0^{T} {dt} \, y_\ell(t) y_{\ell'}(t) = 0 , $
as desired.  However, for the FF corresponding to the {time-dependent} noise component, 
proportional to $\eta_{\ell,\ell'}(t)$, we find that, in general, 
$$G^{(1)}_{Z_\ell Z_{\ell'}} (\omega,T) = 2 \int_0^{T} {dt} \, y_\ell(t) y_{\ell'}(t) e^{i \omega t} \ne \mathcal{O} 
((\omega T)^{\alpha_\ell + \alpha_\ell'+1}).$$ 
Thus, displacement anti-symmetry alone guarantees a high FO and CO {\em only in the presence of 
a static Ising coupling}. In such a case, $U^{d, (X_\ell,X_{\ell'})}_{\alpha_\ell,\alpha_{\ell'}}$ achieves 
CO $=$ FO $= \alpha = \min \{\alpha_\ell,\alpha_{\ell'}\}$.  

If a time-dependent noisy coupling is present, it is necessary to resort to 
nested two-qubit DD sequences built via composition, i.e., 
$\mathring{U}_{(\alpha_\ell,\alpha_{\ell'})}^{\vec{X}}(T)$ or ${\mathring{U}}_{(\alpha,\alpha)}^{'\,\vec{X}}(T)$. 
These sequences, which include NUDD and multi-qubit CDD, are the only known protocols capable of 
achieving arbitrary CO $\alpha$. However, their FO need not be maximum: direct calculation of 
$G^{(2)}_{Z_\ell,Z_{\ell'}} (\omega,-\omega,T)$, combined with dimensional analysis, shows 
that indeed $\Phi = \alpha$ for $\mathring{\text{CDD}}_{(\alpha, \alpha)}^{(\vec{X})} (T)$, but 
$\mathring{\text{NUDD}}_{(\alpha_\ell, \alpha_{\ell'})}^{(\vec{X})} (T)$ in general only achieves 
$\Phi =\alpha-1$. 
As before, by further imposing a displacement anti-symmetry on NUDD, the FO 
can be maximized to $\Phi = \alpha$.

\subsection{Selective multi-qubit control sequences}
\label{sub:multisel}

\subsubsection{General construction.}

The above results may be extended beyond the $N=2$ qubit scenario. Doing so requires generalizing 
displacement anti-symmetry to multiple qubits. Since satisfying the requirement of pairwise displacement 
anti-symmetry, Eq. (\ref{displ}), for all pairs simultaneously is clearly impossible, we seek a control structure 
capable of satisfying an analogue to displacement anti-symmetry {\em at different timescales}. 
We start by subdividing the total evolution time $T$ in $2^{N-1}$ segments of length $\tau_0$, i.e., $T= 2^{N-1} \tau_0$.
One can then demand that for every $\ell$ and $\tau_s = 2^{s} \tau_0$, $s=0,\ldots,N-2$, and $t \in [0,\tau_s)$,
$$ y_\ell(T/2 - m \tau_s + t ) = (-1)^{P^{(N)}(\ell,s)} y_\ell  (T/2 + (m-1) \tau_s  + t),  \quad 
1 \leq m \leq 2^{N-s-2}, $$
\noindent 
where $P^{(N)}(\ell,s)$ are the entries of a $N \times (N-1)$ binary matrix.  In this notation, the $N=2$ 
qubit displacement anti-symmetry is simply represented by $P^{(2)}= {0 \choose 1}$, which yields 
\begin{align*}
y_1( t ) =  (+1) y_1  (T/2 + t), \quad 
y_2( t ) = (-1) y_2  (T/2 + t).  
\end{align*}
We can generalize to multiple qubits via an appropriate matrix
\begin{equation}
P^{(N)} (\ell, s)= \left( \begin{array} {c c c c c } 
0 & 0 & 0 & \cdots  & 0\\
1 & 0 & 0 & \cdots  & 0\\
0 & 1 & 0 & \cdots  & 0\\
&& \cdots&& \\
0 & 0 & \cdots & 0  & 1\\
\end{array}\right).
\label{Pmatrix}
\end{equation}
In this way, for every pair $\ell,\ell'$ there exists an $s$ such that, when $t_1$ and $t_2$ are in 
an interval $[T/2 - m \tau_s, T/2 -(m-1) \tau_s)$,
\beqy
\nonumber 
\hspace*{-5mm}
y_\ell(t_1) y_{{\ell'}}(t_2) &=& (-1)^{P^{(N)}(\ell,s) + P^{(N)}(\ell',s)} \, y_{{\ell}} ( (2m-1) 
\tau_s + t_1) y_{{\ell'}}( (2m-1) \tau_s +t_2) \\
\label{par} 
\hspace*{-5mm}&=& (-) \, y_{{\ell}} ( (2m-1) \tau_s + t_1) y_{{\ell'}}( (2m-1) \tau_s +t_2),
\eneqy
for all $1 \leq m \leq 2^{N-s-2}$, resulting in what we call {\em generalized displacement anti-symmetry}.

The next step is to give a systematic procedure to build $N$-qubit DD sequences that incorporate the above 
symmetry constraint. We do so by introducing an auxiliary $N \times 2^{N-1}$ matrix $Q^{(N)}$ associated to $P^{(N)}$, with 
the $\ell$-th row, $Q_\ell$, defined as follows. 
Let the $2^{N-2}$ entry of the row $Q_\ell$, $[Q_\ell]_{2^{N-2}}$, be set to $0$. First, $[Q_\ell]_{2^{N-2}+1}$ 
is chosen such that  $[Q_\ell]_{2^{N-2}} = [Q_\ell]_{2^{N-2}+1} \oplus P(\ell,1)$, where $\oplus$ denotes addition modulo two. Next, 
$[Q_\ell]_{2^{N-2}-1}$ and $[Q_\ell]_{2^{N-2}+2}$ are chosen 
such that  $\{[Q_\ell]_{2^{N-2}-1},[Q_\ell]_{2^{N-2}}\}=  \{[Q_\ell]_{2^{N-2}-1} \oplus 
P(\ell,2), [Q_\ell]_{2^{N-2}} \oplus P(\ell,2)\}$. 
We then proceed recursively: given   
$\{ [Q_\ell]_{2^{N-2}-(2^{y-1}-1)}, \ldots \ignore{[Q_\ell]_{2^{N-2}}, [Q_\ell]_{2^{N-2}+1}, \ldots}, 
[Q_\ell]_{2^{N-2}+2^{y-1}}\}$, 
we may build a {\em unique} string $\{ [Q_\ell]_{2^{N-2}-(2^{y}-1)} \ldots [Q_\ell]_{2^{N-2}}, 
[Q_\ell]_{2^{N-2}+1}, \ldots, [Q_\ell]_{2^{N-2}+2^{y}}\}$ such that $\{[Q_\ell]_{2^{N-2}-(2^{y}-1)}, 
\ldots,[Q_\ell]_{2^{N-2}}\} =  \{[Q_\ell]_{2^{N-2}}\oplus P(\ell,y), \ldots ,[Q_\ell]_{2^{N-2}+2^{y} } \oplus P(\ell,y)\}$. 
The resulting matrix $Q^{(N)}$ gives us a way to build-in the symmetries and anti-symmetries 
of the switching functions $y_\ell(t)$ at different timescales; e.g., in the two qubit 
case described earlier, 
$$ Q^{(2)} = \left( \begin{array} {c|c}   0 & 0  \\ 0 & 1   \end{array} \right).$$ 
\noindent 
Given a base DD sequence on an interval $T_0\equiv T/2$, say  $U_{\alpha_\ell}^{(X_\ell)} {\times} U_{\alpha_{\ell'}}^{(X_{\ell'})}(T/2)$, 
the displacement anti-symmetry-enhanced version is then given by
\begin{align*}
\Big(X_1^{Q_{1,1}} X_2^{Q_{2,1}} U_{\alpha_\ell}^{(X_\ell)} {\times } U_{\alpha_{\ell'}}^{(X_{\ell'})}(T/2) X_1^{Q_{1,1}} X_2^{Q_{2,1}} \Big) \hspace*{-1mm}
\left(X_1^{Q_{1,2}} X_2^{Q_{2,2}} U_{\alpha_\ell}^{(X_\ell)} {\times } U_{\alpha_{\ell'}}^{(X_{\ell'})}(T/2) X_1^{Q_{1,2}} X_2^{Q_{2,2}} \right), \nonumber
\end{align*}
in agreement with our construction in Eq.~\eqref{2disl}.  For $N$ qubits, the generalized-displacement 
anti-symmetry-enhanced sequence 
${U}^{d,(\vec{X})}_{\vec{\alpha}} (T) \equiv {U}^{d,(X_1,\cdots,X_N)}_{\alpha_1,\cdots,\alpha_N} (T = 2^{N-1} T_0)$ 
of a base DD sequence ${U}^{(\vec{X})}_{\vec{\alpha}} (T_0)$ is similarly given by
\begin{align} 
\label{dispanti}
{U}^{d,(\vec{X})}_{\vec{\alpha}} (T= 2^{N-1} T_0) &= \prod_{s=1}^{2^{N-1}}  \left(\otimes_{\ell=1}^N 
X_\ell^{Q(\ell ,s)}\right)  {U'}^{(\vec{X})}_{\vec{\alpha}} (T_0) \left(\otimes_{\ell=1}^N X_\ell^{Q(\ell ,s)}\right)^\dagger.
\end{align} 

It remains to show that the above construction achieves in general the desired arbitrary-order noise suppression, similar 
to the two-qubit case. Our strategy is to first show the result for a {\em fixed} pair of qubits for which generalized 
displacement anti-symmetry at scale $\tau_{s}$ holds, in the sense of Eq. (\ref{sb}), and then generalize the argument 
by noticing that, by construction, Eq.~$\eqref{par}$ is satisfied for at least one $s$ for every $\ell ,\ell'$ pair. 
Given pair $(\ell, \ell')$, one then has
\begin{align*}
{\color{blue}-}F^{(2)}_{Z_\ell, Z_{\ell'}} (\omega_1,\omega_2,T) 
& = \sum_{q=0}^{2^{N-s-1}-1} \int_{q \tau_s}^{(q+1) \tau_s} dt_1 \int_{q \tau_s}^{t_1} dt_2 \, y_\ell(t_1) y_{\ell'}(t_2) 
e^{ i \vec{\omega}\cdot \vec{t}} \\
&\,\,\,\,+\sum_{q>r=0}^{2^{N-s-1}-1} \int_{q \tau_s}^{(q+1) \tau_s} dt_1 \int_{r \tau_s}^{(r+1) \tau_s} dt_2 \, y_\ell(t_1) y_{\ell'}(t_2) 
e^{ i \vec{\omega}\cdot \vec{t}}\\
& = \sum_{q=0}^{2^{N-s-2}-1}  \int_{0}^{\tau_s} dt_1 \int_{0}^{t_1} dt_2 \,\left( e^{i q (\omega_1 +\omega_2) \tau_s}   
y_\ell(q \tau_s + t_1) y_{\ell'}(q \tau_s+ t_2) e^{ i \vec{\omega}\cdot \vec{t}} \right. \\
&\,\,\,\, \left.+ \, e^{i (2q-1) (\omega_1 +\omega_2) \tau_s}  y_\ell((2q-1) \tau_s + t_1) y_{\ell'}((2q-1) \tau_s+ t_2) 
e^{ i \vec{\omega}\cdot \vec{t}}\right)  \\
&\,\,\,\,+\sum_{q>r=0}^{2^{N-s-1}-1} \int_{q \tau_s}^{(q+1) \tau_s} dt_1 \int_{r \tau_s}^{(r+1) \tau_s} dt_2 \, y_\ell(t_1) y_{\ell'}(t_2) 
e^{ i \vec{\omega}\cdot \vec{t}}.
\end{align*}
Each integral in the second summation is just a time-translated version of 
$F^{(1)}(\omega_1,\tau_s)F^{(1)}(\omega_2,\tau_s)$, 
while the first summation can be simplified by invoking displacement anti-symmetry:
\begin{align*}
&\sum_{q=0}^{2^{N-s-2}-1}  \left( e^{i q (\omega_1 +\omega_2) \tau_s} \int_{0}^{\tau_s} dt_1 \int_{0}^{t_1} dt_2 \, 
y_\ell(q \tau_s + t_1) y_{\ell'}(q \tau_s+ t_2) e^{ i \vec{\omega}\cdot \vec{t}} \right. \\
&\left.+ \, e^{i (2q-1) (\omega_1 +\omega_2) \tau_s} \int_{0}^{\tau_s} dt_1 \int_{0}^{t_1} dt_2 \, y_\ell((2q-1) 
\tau_s + t_1) y_{\ell'}((2q-1) \tau_s+ t_2) e^{ i \vec{\omega}\cdot \vec{t}}\right) = \\
&\sum_{q=0}^{2^{N-s-2}-1}  \left( e^{i q (\omega_1 +\omega_2) \tau_s} -e^{i (2q-1) (\omega_1 +\omega_2) 
\tau_s} \right) \int_{0}^{\tau_s} dt_1 \int_{0}^{t_1} dt_2 \, y_\ell(q \tau_s + t_1) y_{\ell'}(q \tau_s+ t_2) e^{ i \vec{\omega}\cdot \vec{t}},  
\end{align*}
which vanishes when $\omega_1 + \omega_2 =0$, as relevant to stationary noise. 
Accordingly, in the absence of a time-dependent direct Ising coupling, generalized 
displacement-antisymmetry-enhanced DD achieves CO and FO equal to $\alpha = \min\{\alpha_\ell\}$, as desired.

\subsubsection{Example and resource scaling.} 

We illustrate the above general construction in the simplest non-trivial multi-qubit setting, $N=3$, in which case 
we have 
\begin{eqnarray*}
P^{(3)}(\ell,s) = \left( \begin{array} {cc}  0 & 0  \\ 1 & 0  \\ 0 & 1  \end{array} \right),  & \quad & 
Q^{(3)} = \left( \begin{array} {cc|cc}  0 & 0 & 0 & 0 \\ 1& 0 & 1 & 0  \\ 1& 0 & 0 & 1  \end{array} \right).  
\end{eqnarray*}
Explicitly, the following set of symmetries are enforced by $P^{(3)}$: 
at time scales $\tau_0$, for $t \in [T/2 - m \tau_0, T/2 -(m-1) \tau_0)$, we have 
\begin{align*}
y_1( t ) &= (+1) \,y_1  ((2 m -1) \tau_0 + t), \\ 
y_2( t ) &= (-1) \, y_2  ((2 m -1) \tau_0 + t),\\
y_3( t ) &= (+1)\, y_3  ((2 m -1) \tau_0 + t),
\end{align*} 
for $ m= 1,2$, while at time scale $\tau_1 = 2 \tau_0$, $t \in [T/2 - m \tau_1, T/2 -(m-1) \tau_1)$,  we have 
\begin{align*}
y_1( t ) &= (+1) \, y_1  ( \tau_1 + t), \\ 
y_2( t ) &= (+1) \, y_2  (\tau_1 + t),  \\
y_3( t ) &= (-1) \, y_3  ( \tau_1 + t),  
\end{align*} 
for $m=1$. Starting from $y_3(t)$, for instance, and with $\tau_0=T/4$, these conditions require 
\begin{align*}
y_3( t ) &= (+1) \, y_3  ( T/4 + t), \quad t\in[T/4, T/2), \\ 
y_3( t ) &= (+1) \, y_3  (3T/4 + t), \quad t\in [0, T/4) \\
y_3( t ) &= (-1) \, y_3 ( T/2 + t), \quad t \in [0, T/2),   
\end{align*}
which may be simultaneously obeyed by the pattern
$$ y_3(t) = -y_3(T/4 +t) = -y_3(T/2 +t) = y_3(3T/4 +t), \quad t \in  [ 0, T/4), $$ 
\noindent 
see Fig.~\ref{fig:disp}. With $T_0=T/4$, we may then build an enhanced DD sequence 
using Eq. \eqref{dispanti} and the (independent) base sequence 
$U^{(\vec{X})}_{\vec{\alpha}} (T/4) = U^{(X_1)}_{\alpha_1} \times U^{(X_2)}_{\alpha_2} 
\times U^{(X_3)}_{\alpha_3} (T/4)$.  Explicitly:
$${U}^{d,(\vec{X})}_{\vec{\alpha}} (T) =  X_3X_2 U^{(\vec{X})}_{\vec{\alpha}} (T/4) X_3X_2 
U^{(\vec{X})}_{\vec{\alpha}}(T/4) X_2  U^{(\vec{X})}_{\vec{\alpha}}(T/4)  X_2 X_3 
U^{(\vec{X})}_{\vec{\alpha}}(T/4) X_3. $$

\begin{figure}[t]
\centering
\begin{subfigure}
\centering
\hspace*{24mm}\includegraphics[width=.35\textwidth]{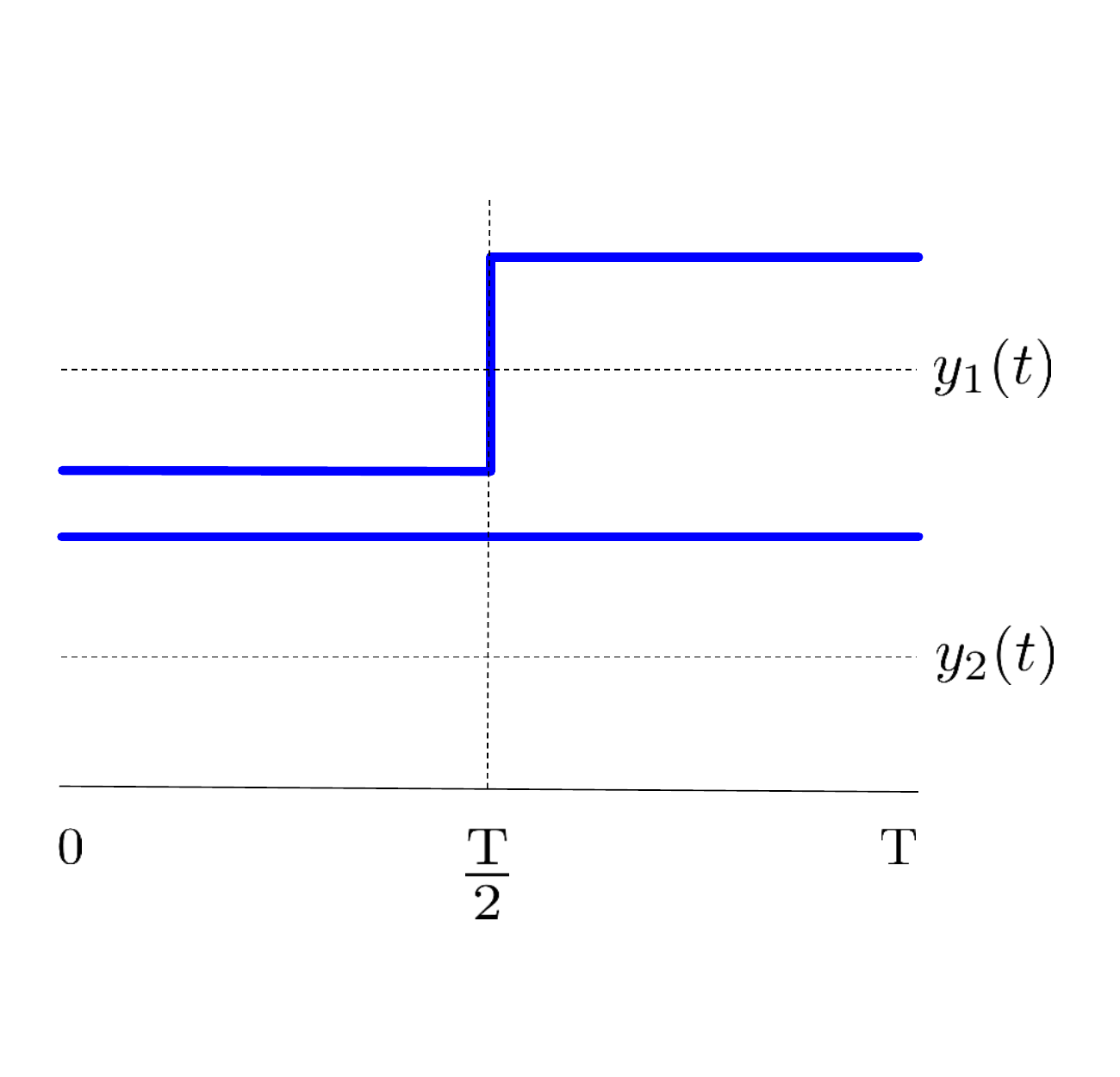}
\end{subfigure}
\begin{subfigure}
\centering
\includegraphics[width=.36\textwidth]{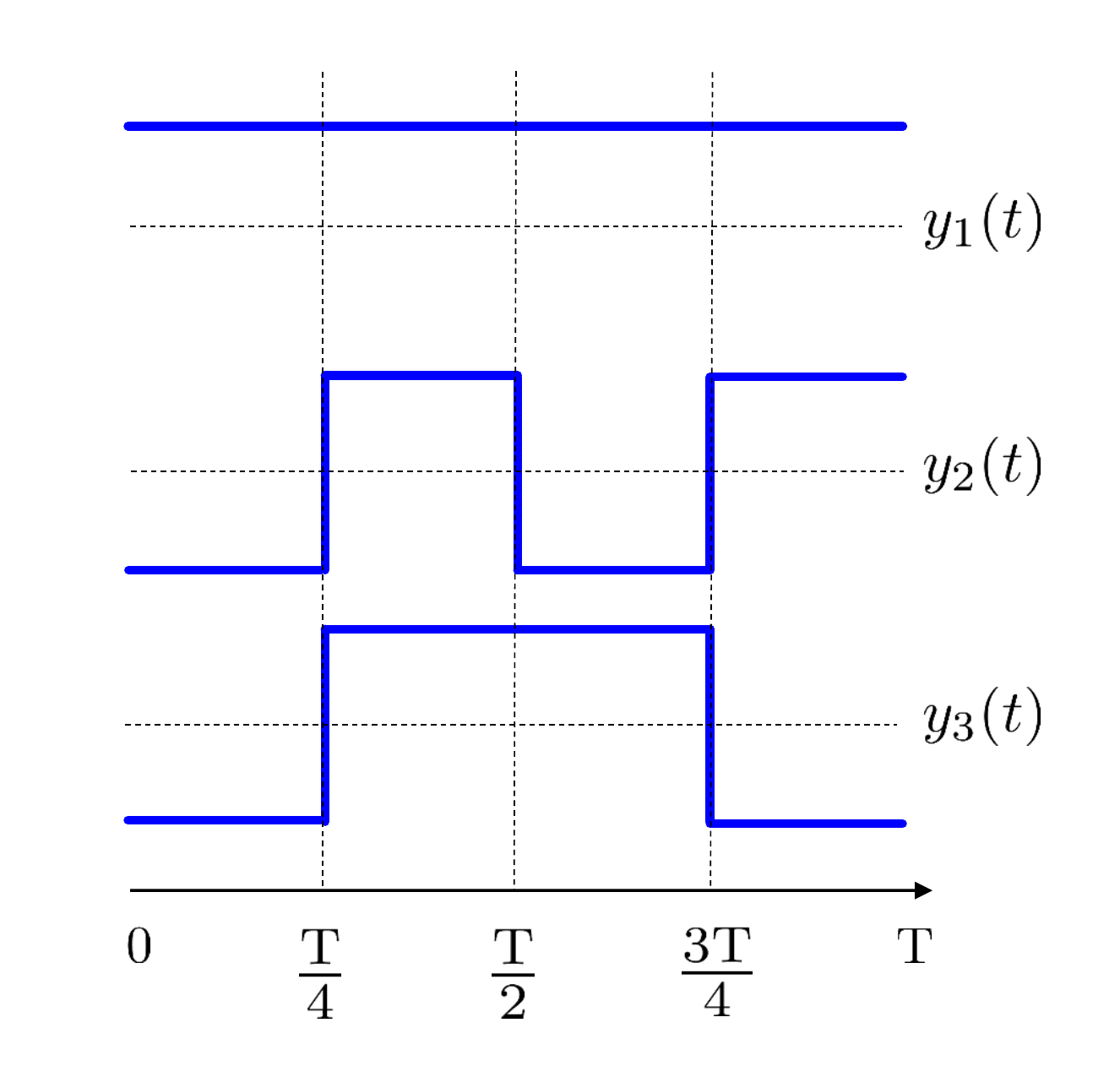}
\end{subfigure}
\vspace*{-5mm}
\caption{Schematic representation of the sign pattern induced in each of the $y_\ell(t)$ by 
the displacement anti-symmetry for two (left panel) vs three (right panel) qubits. For example, for 
$N=3$: $y_3(t) = - y_3(T/4+t ) = - y_3(T/2 +t) = y_3(3 T/4+t)$, $t \in [0, T/4)$. }
\label{fig:disp}
\end{figure}

As in Sec. \ref{control}, let $n(\alpha_\ell) \geq n(\alpha)$ denote the number of pulses in a single-qubit 
DD sequence achieving CO $\alpha_\ell$, with $n_P^{\text{tot}} (N) =  \prod_{\ell=1}^N n(\alpha_\ell) 
\geq [n(\alpha)]^N$  for multi-qubit $\mathring{\text{CDD}}$ or NUDD. 
For a sequence that incorporates generalized 
displacement anti-symmetry, the required number of pulses grows instead as 
$n_P^{d,\text{tot}} (N) = 2^{N-1} \sum_{\ell=1}^N n(\alpha_\ell) \geq 2^{N-1} N\, n(\alpha)$.
While this is still exponential in $N$, it may translate into a significant resource reduction:
$$ \frac{n_P^{d, \text{tot}(N)} }{ n_P^{\text{tot}} (N)} \geq  \frac{N \,2^{N-1} 
n(\alpha)}{ [n(\alpha)]^N } = \frac{N}{ [n(\alpha)/2]^{N-1}} , $$
\noindent 
which represents an {\em exponential reduction} in the required pulse number 
whenever  $\alpha > 1$. 

Beside affording a smaller total pulse number, the presence of displacement anti-symmetry also 
implies less stringent requirements on the minimum switching time of the corresponding protocols.  For instance, 
the minimum switching time for a multi-qubit NUDD sequence achieving minimum CO $\alpha$ scales as 
$\tau_\alpha^N \sim ({1}/{\alpha^2})^N$, whereas an enhanced UDD$^{d, (X_1\cdots X_N)}_{\alpha}$ sequence 
built out of Eq. (\ref{dispanti}) has minimum switching time scaling as $\tau_\alpha^N={\tau_\alpha}/{2^{N-1}} 
\sim ({1}/{\alpha^2})^N ({\alpha^2}/{2})^{N-1}$; that is, our sequences also have an {\em exponentially larger 
minimum switching time}, making them potentially more amenable to experimental implementation in multi-qubit systems. 

As in the two-qubit case, the above advantages in terms of pulse number and timing constraints are meaningful 
only so long as an arbitrarily high CO/FO is achievable, which in our noise model is possible 
only when no fluctuating direct coupling is present.   Even if the latter is non-zero, however, it is beneficial 
to incorporate displacement anti-symmetry in  NUDD or $\mathring{U'}_{\alpha,\ldots\alpha}^{(X_1,\ldots,X_N)}$ 
sequences which can achieve arbitrary CO: 
by doing so, the enhanced sequences $\mathring{U}_{\alpha_1,\ldots,\alpha_N}^{d, (X_1,\ldots,X_N)}$ and 
$\mathring{U'}\hspace*{0.2mm} \mbox{}_{\alpha,\ldots,\alpha}^{d, (X_1,\ldots,X_N)}$ achieve {\em both} CO and FO 
equal to $\alpha$ for the most general version of the noise model under consideration.
It is also worth noting that, out of the multi-qubit sequences described in Sec.~\ref{control}, 
the multi-qubit $\mathring{\text{CDD}'} \mbox{}_{\alpha,\cdots, \alpha}^{(X_1,\cdots,X_2)} (T_\alpha)$
already has a form of displacement anti-symmetry built in. 
This is easily seen by noticing that $\mathring{\text{CDD}'} \mbox{}_{\alpha,\ldots, \alpha}^{(X_1,\ldots,X_2)} (T_\alpha)$ 
may be thought of as the result of imposing anti-symmetry on a sequence  
$\mathring{\text{CDD}'} \mbox{}_{\alpha-1,\ldots, \alpha-1}^{(X_1,\ldots,X_2)} 
(T_{\alpha-1})$ using an ``enlarged'' $P^{(N)}$ matrix, with $N \times N$ entries, given by 
$$\tilde{P}^{(N)} = \left( \begin{array} {c c c c c } 
1 &  \cdots  & 1 & 1 & 1\\
1 &  \cdots  & 1 & 1 & 0\\
1 &  \cdots  & 1 & 0 & 0\\
&& \cdots&& \\
1 &  \cdots  & 0 & 0 & 0\\
\end{array}\right). $$
The concatenated structure results in the ``enlarged'' displacement anti-symmetry and guarantees 
FO $\Phi= \alpha$ using $n_P^{\text{tot}} =2^{2 N (\alpha+1)}$ pulses.  While the latter 
is exponentially larger than than for 
$U^{d,(X_1\cdots X_2)}_{\alpha}$ sequences (which use $2^{N-1} N n(\alpha)$ pulses), multi-qubit CDD is   
capable of suppressing an arbitrary (direct and induced) coupling term. 
As we shall see next, 
displacement anti-symmetry is also the key to establishing our results for long-term storage 
of multiple qubits, hence our results will apply in particular to 
$\mathring{\text{CDD}'} \mbox{}_{\alpha,\cdots, \alpha}^{(X_1,\cdots,X_2)} (T)$ sequences.

\section{Dynamical Decoupling versus Multi-qubit Dephasing Noise: Long-time Storage}
\label{longterm}

We now turn attention to the problem of high-fidelity long-time storage of information, under  
realistic timing and access-latency constraints. That is, we are not solely interested in protecting an arbitrary 
multi-qubit state for a fixed (potentially arbitrarily long) storage time $T$, but also in being able to retrieve it 
on demand, with guaranteed high fidelity and sufficiently small latency. This problem was studied 
for single-qubit Gaussian noise in \cite{Memory}, where a systematic approach to meet these requirements 
was proposed, based on periodic repetition of a high-order DD sequence $U_{\alpha}^{(X_1)} (T_p)$, with 
$T=M T_p$ and access latency capped at $T_p \ll T$.  
Provided that the relevant noise spectrum decays to zero sufficiently rapidly at high 
frequency (ideally, with a ``hard cutoff'' of the form $S(\omega) \propto \omega^s \,\Theta (\omega - \omega_c)$), 
the key observation is that a coherence plateau may be engineered by appropriately choosing the base DD sequence, 
so that the following conditions are obeyed:
\beq
\label{1qplateau}
\omega_c T_p < 2\pi,  \quad  s + 2 \alpha  > 1. 
\eneq
Here, we show how this result extends beyond the single-qubit Gaussian scenario, by identifying 
conditions that ensure, in principle, a {\em fidelity plateau} for stationary multi-qubit 
dephasing from classical and/or spin-boson (Gaussian and non-Gaussian) noise sources. 
The starting point is to recall that the expression for the system's reduced dynamics, 
Eq.~\eqref{dyn}, does not make any special assumption on the statistical properties of the noise beyond those 
imposed by stationarity and the spin-boson algebra, along with the standard initial system-bath factorization.
Thus, we may separately analyze each of the factors in Eq.~\eqref{dyn} in order to derive plateau conditions 
for classical and quantum noise sources independently.

\subsection{Fidelity plateau conditions for multi-qubit classical stationary dephasing}
\label{p1}

Let us first analyze the classical noise contributions, described by the fluctuations of 
single-qubit and two-qubit energies, $\zeta'_\ell(t)$ and $\eta'_{\ell, \ell'}(t)$. 
We would like to show that when the control protocol consists of $M$ repetitions of a base DD sequence 
with duration $T_p$, all classical contributions to the decay  $e^{-Z_{a,b}(T)}$, 
$ \langle e^{\sum_{ {\ell} , {\ell'}} \Delta[{a}_{{\ell}}+{a}_{{\ell'}},{b}_{{\ell}}+{b}_{{\ell'}}]  \bar{\eta'}_{{\ell},{\ell'}}(T)} 
\times  e^{\sum_{{\ell}}  \Delta[{a}_{{\ell}},{b}_{{\ell}}] \bar{\zeta'}_{{\ell}}(T)  } \rangle_c , $
may be made to approach a {\em constant} value in the large-$M$ limit (formally, as $M \rightarrow \infty$).  
Recalling the cumulant expansion in Eq. (\ref{Cexpansion}), one sees that it suffices to show 
that each $k$-th order cumulant 
$C^{(k)}(\bar{\zeta'}_{{\ell_1}}(T) \cdots \bar{\zeta'}_{{\ell_j}}(T),\bar{\eta'}_{{p_{j+1}}}(T) \cdots \bar{\eta'}_{{p_k}}(T))_c$ 
approaches a constant value in such a limit, for every $j=0, \ldots, k$.
We use the fact that for any periodic control strategy, first-order FFs obey a simple composition rule under repetition \cite{Memory}:
$$G^{(1)}_{Z_{\ell}} (\omega_\ell, T = M T_p) = \frac{1- e^{ i M  \omega_\ell T_p }}{{1- e^{ i \omega_\ell T_p }}} 
\, G^{(1)}_{Z_{\ell}} (\omega_\ell, T_p),$$
\noindent 
and similarly for $G^{(1)}_{Z_{\ell} 	Z_{\ell'}} (\omega_{\ell,\ell'},T = M T_p)$. Defining the index vector 
$\vec{L}_{[k]} = \{ L_r \} \equiv  (\ell_1,...,\ell_j,{p_{j+1}},...,{{p_k}})$ as a 
$k$-component vector with entries $L_r$, direct calculation yields:
\begin{align}
& C^{(k)}(\bar{\zeta}_{{\ell_1}}(T) \cdots \bar{\zeta}_{{\ell_j}}(T),\bar{\eta}_{{p_{j+1}}}(T) \cdots \bar{\eta}_{{p_k}}(T))_c 
\nonumber \\
&= \int_{-\infty}^\infty  \frac{d\vec{\omega}_{[k-1]}}{(2\pi)^{k-1}} \left(\prod_{r=1}^{k-1}  \frac{\sin 
{ (M\omega_{L_r}} T_p/2)}{\sin { ( \omega_{L_r} T_p/2)}}\right)  
G^{(1)}_{Z_{\ell_1}} (\omega_{L_1},T_p) \cdots 
G^{(1)}_{Z_{\ell_{k-1}}Z_{\ell'_{k-1}}} \hspace*{-2mm}(\omega_{{L_{k-1}}},T_p) 
\label{comb1}\\
&\,\,\,\, \times G^{(1)}_{Z_{\ell_k}Z_{\ell'_k}}\hspace*{-1mm} \Big(\hspace*{-1mm}-\sum_{r=1}^{k-1} \omega_{L_r},T_p \Big)
\, \frac{\sin { \Big( M(\sum\limits_{r=1}^{k-1} \omega_{L_r} )\frac{T_p}{2}} \Big)}{\sin {  \Big( (\sum\limits_{r=1}^{k-1} 
\omega_{L_r})\frac{T_p}{2}} \Big)}   
\,S^{\zeta,\eta}_{\ell_1,...,\ell_j,{p_{j+1}},...,{{p_k}}} (\omega_{L_1},...,\omega_{{L_{k-1}}} ).  
\label{comb2}
\end{align} 

Upon making the change of variables $M\omega_{L_r} T_p/2 \mapsto \omega_{L_r}$ and taking the large-$M$ limit, 
the product of ratios of $\sin$ functions in Eq. (\ref{comb1}) may be replaced by a product of 
$\textrm{sinc} $ functions, which approaches a multi-dimensional frequency comb with peaks 
of height $\mathcal{O} (M)$ at $\omega_{L_r} = q_{L_r} 2 \pi /T$, $q_{L_r} \in \mathbb{Z}$. 
In order for the above $k$th-order cumulant to have a finite value, one must then avoid the 
$\mathcal{O} (M)$ divergences at the peaks located at every 
\begin{equation}
\vec{\omega}_{[k]} (\vec{q}) = \frac{2\pi}{T_p} \Big(q_{L_1},...,q_{L_{k-1}},-\sum_{ r=1}^{k-1}q_{L_r}\Big) .
\label{peaks}
\end{equation} 
In practice, meeting this requirement will be impossible, since no DD sequence is known, such that its  
FFs have a high-order zero for {\em all} $\{ q_{L_r}\}$, as required.  However, 
building on the single-qubit case \cite{Memory}, it is unrealistic to expect that a plateau regime may 
emerge at all, unless the noise is sufficiently well-behaved at high frequencies. Let us work in  
the simplest (idealized) scenario, where a hard frequency cutoff exists for each frequency variable, 
such that the power spectra may be Taylor-expanded around the $\vec{\omega}=0=\vec{q}$ as follows:
\beqy
S^{\zeta,\eta}_{L_1,\cdots,{{L_k}}} (\omega_{L_1},\ldots,\omega_{L_{k-1}}) = 
\prod_{r=1}^{k-1} \omega_r^{s_{L_r}} \, \Theta (|\omega_{{L_r}} -\omega_{c,{L_r}}|) ,
\eneqy
where $\omega_{c,{L_r}}$ is a high-frequency cutoff parameter and
$s_{L_r}$ characterizes the power-law behavior as $\omega_{L_r}\rightarrow 0$. Similarly, 
we assume that the low-frequency behavior of the relevant FFs reads
\begin{eqnarray}
G^{(1)}_{Z_\ell} (\omega_{L_r},T) \sim \mathcal{O} ( \omega_{L_r}^{\alpha_{L_r}} ), 
\quad  G^{(1)}_{Z_\ell, Z_{\ell'}} (\omega_{L_r, L_{r'}},T) \sim \mathcal{O} ( \omega_{L_r}^{\alpha_{L_r}} ), 
\label{G1ZZ} 
\end{eqnarray}
in terms of the CO $\alpha_{L_r}$ corresponding to each frequency variable. 
By demanding that, in analogy to Eq. \eqref{1qplateau},  
$\omega_{c,{L_r}} T_p < 2 \pi$, for all $L_r$, we may then ensure that {\em only} the divergence at $\{ q_{L_r}\} = 0$ 
contributes to the $k$-th order cumulant. The strategy is then to decompose the integral in 
Eqs. (\ref{comb1})-(\ref{comb2}) into a sum of integrals over hypercubes of side $2 \pi/T_p$ centered around 
$\vec{q}=0$ and, as in \cite{Memory}, to make sure that the dependence upon each frequency variable is 
sufficiently well-behaved close to zero for the corresponding integration to be convergent. That is, we require 
that each frequency integration admits a power law $\omega_{L_r}^x$, with $x> -1$, close to zero.  Lengthy but 
straightforward calculation yields the following conditions:
\begin{align}
\label{gplatclass2}
\omega_{c,L_r} T_p < 2 \pi,\quad 
\sum_{r=1}^{k-1} s_{L_r} + \sum_{r=1}^{k} \alpha_{L_r} >  1 , \quad    \forall \{ L_r\}, \forall k.
\end{align}

Accordingly, a plateau behavior may be engineered by choosing a base DD sequence such 
that all first-order FFs $G^{(1)}_{L_r} (\omega_{L_r},T)$ and $G^{(1)}_{Z_\ell, Z_{\ell'}} (\omega_{L_r},T)$ 
have zeroes of sufficiently high order at each $\omega_{L_r}=0$.  
Suitable choices were described in Sec.~\ref{Decou}. We note, 
in particular, that collective as well as independent single-qubit 
sequences suffice as long as no direct Ising coupling is present [$\eta'_{\ell, \ell'}(t) \equiv 0$], otherwise  
sequences built out of composition (such as NUDD or multi-qubit CDD) will be required to ensure high 
CO $\alpha_{p_j}=\alpha_{\ell, \ell'}$.  Still, since only first-order FFs are involved, displacement 
anti-symmetry does not play an essential role. This contrasts with the case where bosonic noise sources 
are present, as we shall see in Sec. \ref{p2}.  

The conditions in Eq. \eqref{gplatclass2} contain, as a special case, the ones derived in \cite{Memory} for 
a single qubit exposed to zero-mean stationary Gaussian noise, due to either a classical noise source or a 
quantum bosonic environment.  In such simple cases, with $k=2$ and $\vec{L}_{[2]} = (1,1)$, 
there is only one power spectrum -- namely, $S^{\zeta}_{1,1} ( \omega,-\omega) 
\equiv S^{\zeta} ( \omega)$ or, respectively, $S^{B,+}_{1,1}(\omega) = \pi 
J(\omega) \coth(\beta \omega/2)$, $\omega \geq 0$, hence our conditions translate into 
$2 \alpha_1 + s_1  > 1$ and $\omega_{c,1} T_p < 2 \pi$, in agreement with Eq. \eqref{1qplateau}. 
Since, for a single qubit, dephasing evolution is fully characterized by first-order FFs 
$G^{(1)}_{L_r} (\omega_{L_r},T)$, the plateau conditions in  Eq. \eqref{gplatclass2} apply, in fact, to the 
more general scenario where noise is {\em non-Gaussian}.  

Paradigmatic examples of classical non-Gaussian dephasing 
arise when a qubit is exposed to random telegraph noise \cite{1overf} or is non-linearly coupled to a 
Gaussian noise source, e.g., $\zeta(t)= [\xi(t)]^2$, with $\xi(t)$ Gaussian as for a qubit operated at an 
optimal point \cite{CywinskiOpt}.  Consider, for illustration, the latter case. 
If the power spectra $S^\xi_{1,1}(\omega)$ of $\xi(t)$ has a cutoff $\omega_c\equiv \Lambda$, it follows that  
$S^\zeta_{\vec{1}}(\omega_1,\cdots,\omega_{k-1})$ has a cutoff at $2\Lambda$ for all $k$. To see this, observe that 
$\langle \zeta(t_1) \cdots \zeta(t_k) \rangle = \langle \xi(t_1)\xi(t_1) \cdots \xi(t_k) \xi(t_k) \rangle$ can be written, by virtue 
of the Gaussianity of $\xi(t)$, as a sum of terms of the form $\langle \xi(t_{i_1}) \xi(t_{i_2})\rangle \cdots \langle 
\xi(t_{i_{k-1}}) \xi(t_{i_k})\rangle$, where $i_s \in [1,\cdots,k]$. After Fourier transforming, direct calculation shows 
that each of these terms is of the form  
\begin{align*}
& \int_{-\infty}^\infty d\omega_s  S^\xi(\omega_{s})S^\xi(\omega_{s} - \omega_{i_1})\cdots S^\xi(\omega_{s} -
\omega_{i_1}- \omega_{i_2} - \cdots  -\omega_{i_{s-1}})  \delta(\omega_{i_1} + \cdots +\omega_{i_s}) \\
&\,\,\,\,\, \times S^\xi(\omega_{i_s+1}) \cdots S^\xi(\omega_{i_{k}})\delta_{\omega_{i_{s+1}}} \cdots \delta_{\omega_{i_{k}}}, 
\quad   1\leq s \leq k.
\end{align*}   
Because of the assumed cutoff  in $S^\xi (\omega)$, it then follows that 
the integral is non-vanishing for $\omega_s \in [-\Lambda,\Lambda]$, and as long as 
$-\Lambda \leq  \omega_s - \omega_{i_1}  \leq \Lambda,$ 
$-\Lambda \leq  \omega_s - \omega_{i_1} - \omega_{i_2} \leq \Lambda,$ 
$\ldots ,$
$-\Lambda \leq  \omega_s - \omega_{i_1} - \cdots - \omega_{i_s} \leq \Lambda.$
Therefore, $\omega_{i_1}$ can at most be $2 \Lambda$, $\omega_{i_2} \in [-2 \Lambda,2 \Lambda]$, and so on. 
That is, any high-order spectra of $\zeta(t)$ has a cutoff of at most $2 \Lambda$ in all its frequency variables and, 
furthermore, is stationary, that is, $\omega_1 + \cdots + \omega_k =0$, at order $k$. Also notice that from the form 
of the above expressions we may infer that the functional dependence with respect to the frequencies is such 
that {$\sum_r {s_{L_r}}$} grows at most linearly with $k$, making it possible for the 
plateau constraints to be satisfied.

For bosonic noise, stationarity demands that $[\rho_B, H_B]=0$, hence $\rho_B$ is diagonal in the 
multi-mode Fock basis.  A simple example of non-Gaussian dephasing 
arises when $\rho_B$ is a mixture of thermal (Gaussian) components at different temperatures, e.g., 
$\rho_B =w_1 \rho_{\beta_1} + w_2 \rho_{\beta_2}$, with $\sum_i w_i=1$ \cite{NGpaper}.  
While detailed analysis is beyond the scope of this work, and may be most meaningfully carried out 
for a concrete qubit device, our approach provides in each case sufficient conditions for a coherence plateau 
to be engineered in principle.

\subsection{Fidelity plateau conditions for multi-qubit classical plus spin-boson dephasing} 
\label{p2}

Multi-qubit dephasing arising from combined 
classical and quantum bosonic noise sources may be analyzed by generalizing the strategy of Sec. \ref{p1}. 
Again, given the convenient form of Eq.~\eqref{dyn}, it is possible to analyze the additional effect of the 
spin-boson interaction by analyzing the contribution of the factors resulting from a quantum average.  
In order to guarantee the existence of a fidelity plateau in the large $M$ limit we additionally require that, 
at order $k$ in the cumulant expansion (recall Eq. \eqref{qck} and the stationarity assumption), 
\begin{align}
\label{e1} 
\int_{-\infty}^\infty  \frac{d\vec{\omega}_{[k-1]} }{(2\pi)^{k-1}} S^B_{{{\ell}_1},\ldots,{{\ell}_k}} 
(\omega_1,\ldots,\omega_{k-1}) G^{(1)}_{Z_{\ell_1}} (\omega_1,t) \cdots  
G^{(1)}_{Z_{\ell_k}} (-\hspace*{-1mm}\sum_{r=1}^{k-1} \omega_r,t)  &
\longrightarrow \textrm{\: constant}, \\
\label{e2}  
\int_{-\infty}^\infty  \frac{d{\omega}}{2\pi} S^{B,-}_{{{\ell}},{{\ell'}}} (\omega) 
G^{(1)}_{Z_{\ell}} (\omega, T) G^{(1)}_{Z_{\ell'}} (-\omega, T)&
\longrightarrow \textrm{\: constant}, \\
\label{e3} 
\int_{-\infty}^\infty \frac{d{\omega}}{2\pi}  S^{B,-}_{{{\ell}},{{\ell'}}} (\omega) G^{(2)}_{Z_{\ell},Z_{\ell'}} (\omega, -\omega, T)  
&\longrightarrow \textrm{\: constant}.
\end{align}
It is straightforward to determine a set of plateau conditions for Eqs.~\eqref{e1}-\eqref{e2} by following a similar 
analysis to the classical noise, as the equations are basically the same. 
In particular, we also demand a hard cut-off for the quantum noise spectra, thus a structure of the form
\begin{eqnarray*}
S^{B}_{\ell_1,\ldots,\ell_k} (\omega_{\ell_1},\ldots,\omega_{\ell_{k-1}}) \sim  
\prod_{r=1}^{k-1} \mathcal{O} (\omega_{{\ell_{r}}}^{\tilde{s}_{\ell_{r}}}) \, \Theta (|\omega_{{\ell_r}} -\omega_{c,{\ell_r}}|) , \\
S^{B,-}_{{{\ell}},{{\ell'}}} (\omega) \sim \mathcal{O}(\omega^{\tilde{s}^{-}_{\ell,\ell'} }) \,\Theta (|\omega -\omega_{c,-}|) ,
\end{eqnarray*}
where $\tilde{s}_{\ell_{r}}$ and $\tilde{s}^{-}_{\ell,\ell'}$ characterize the relevant low-frequency power-law behaviors 
and $\omega_{c,{\ell_r}}, \omega_{c,-}$ are high-frequency cut-offs.  
A similar analysis to the classical case then leads to 
\begin{eqnarray}
\label{gplatquant0}
&& \hspace*{2.5cm}\omega_{c,r} T_p < 2 \pi, \, \forall r, \quad \omega_{c,-} T_p < 2\pi,  \\
\label{gplatquant1}
&&\sum_{r=1}^{k-1} \tilde{s}_{{\ell_{r}}} + \sum_{r=1}^{k} \alpha_{{\ell_{r}}} > 1 
\ignore{\forall \,\, \ell_1,\cdots, \ell_k}, \, \forall k, 
\quad \tilde{s}^{-}_{{\ell_r},{{\ell_{r'}}} }  + \alpha_{{\ell_{r}}} + \alpha_{{\ell_{r'}}} >1, 
\, \forall \, r \neq r' .
\end{eqnarray} 

Obtaining plateau conditions for Eq.~\eqref{e3} requires more work. 
Using the periodicity of $y_\ell(t)$ over $T=MT_P$, we may rewrite  
\begin{align}
\nonumber 
{ 2i } G^{(2)}_{Z_{\ell},Z_{\ell'}} (\omega,-\omega, T) &=  G^{(1)}_{Z_{\ell}} (\omega, T_p) 
G^{(1)}_{Z_{\ell'}} (-\omega, T_p) \,\frac{e^{i M \omega T_p} - e^{-i M \omega T_p} - M  (e^{i \omega T_p} - 
e^{-i \omega T_p})}{(e^{i \omega T_p/2}-e^{-i  \omega T_p/2})^2}\\
& \label{G2} 
+ M { 2i } G^{(2)}_{Z_{\ell},Z_{\ell'}} (\omega,-\omega, T_p),
\end{align}
which, however, includes contributions with an \em explicit linear $M$-dependence}. 
Accordingly, $G^{(2)}_{Z_{\ell},Z_{\ell'}} (\omega,-\omega, T)$ grows with $M$ and, in the $M \rightarrow \infty$ limit we 
are interested in, it diverges. One would like to show that under appropriate symmetry of $y_\ell(t)$, 
and consequently of the FFs in each $T_p$ interval, the terms linear in $M$ vanish. 
As it turns out, the displacement anti-symmetry comes to our aid here as well, leading to the desired cancellations, 
as we show below.  We stress that the linear $M$-dependence is a generic feature, and unless 
additional symmetry is built into the applied control sequence, it necessarily leads to a divergence, forbidding the 
existence of a plateau regime. For example, repeating a high-order multi-qubit NUDD sequence without demanding 
further structure will {\em not} lead to a fidelity plateau in the presence of quantum noise, whereas a displacement 
anti-symmetry-enhanced NUDD sequence will achieve it in principle, provided that the appropriate 
conditions are obeyed.

\subsubsection{Two-qubit setting.}

\begin{figure}[t]
\centering
\hspace*{15mm}\includegraphics[width=0.75\columnwidth]{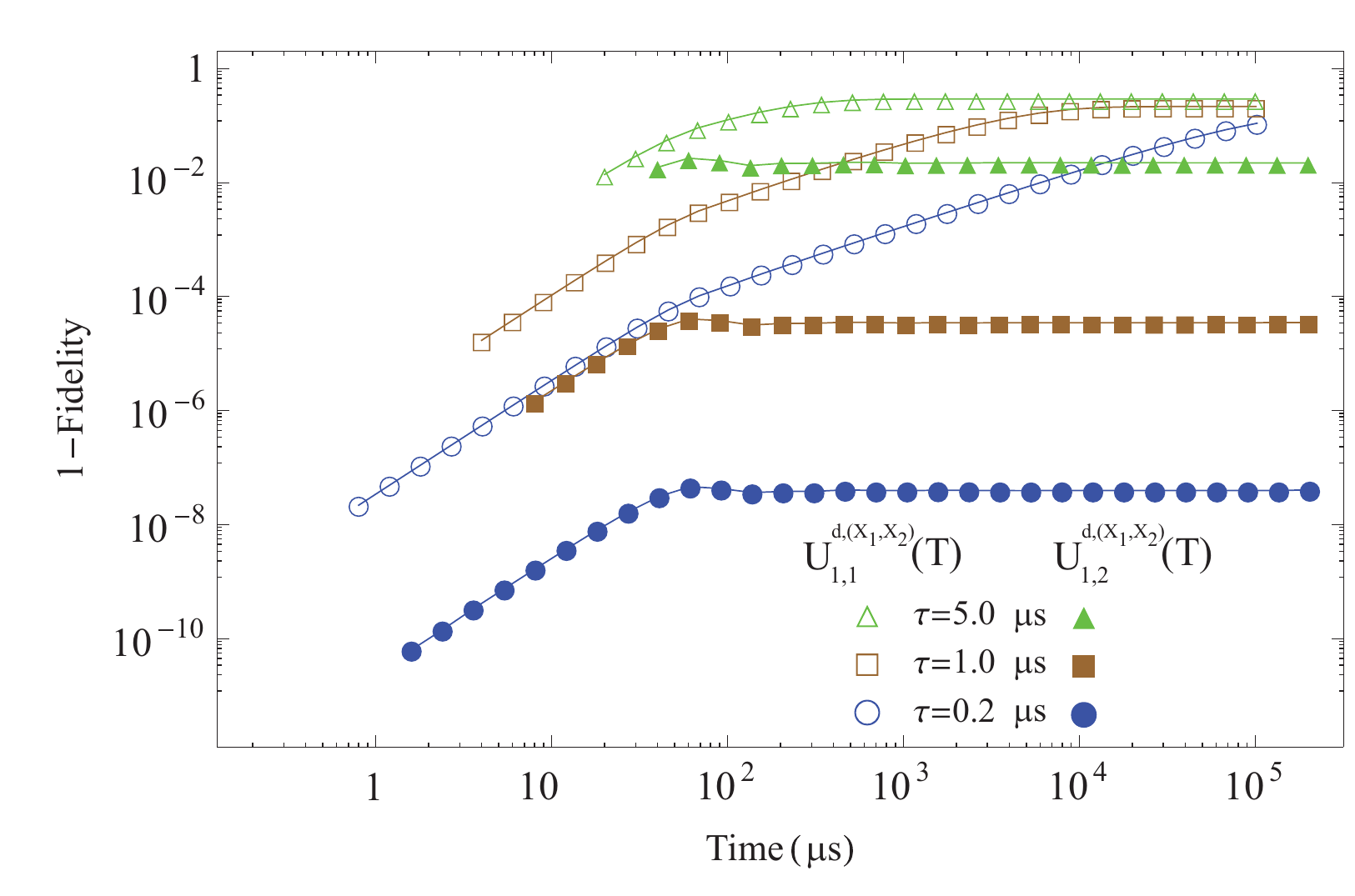}
\vspace*{-4mm}
\caption{ 
Emergence of a fidelity plateau under sequence repetition in the long-time limit. 
Two DD sequences incorporating displacement anti-symmetry are considered,
${U}_{1,1}^{d,(X_1,X_2)}(T)$ and ${U}_{1,2}^{d,(X_1,X_2)}(T)$ (empty vs. filled symbols), 
operating at different minimum pulse intervals, $\tau=0.2, 1.0, 5.0 \,\mu$s.  
A purely Gaussian bosonic spectrum in the low-temperature limit is assumed [$\coth({ \beta \omega }/2) \simeq 1$], 
of the form $S^{B}_{Z_1,Z_2} (\omega) = 2 \pi e^{i \omega t_{1,2}} \omega_c ({\omega}/{\omega_c})^{-2} 
\Theta (\abs{\omega- \omega_c})$, with $\omega_c =  2\pi 10^{4}\,$Hz
and $t_{1,2} = 10^{-2} s$. The fidelity is calculated using Eq. (\ref{fidelity}),  
by averaging over $10^3$ random pure initial states of the form 
$|\Psi\rangle = \sum_{a=0}^3 c_{a}\ket{a_1a_2}$,  
where for simplicity we have assumed real coefficients 
$c_1\equiv \cos{\theta}$, $c_2\equiv \sin{\theta}\cos{\theta'}$, $c_3\equiv\sin{\theta}\sin{\theta'}\cos{\theta''}$, 
$c_4\equiv\sin{\theta}\sin{\theta'}\sin{\theta''}$, with $\theta,\theta'\in [0,\pi]$ and $\theta''\in [0,2\pi]$ 
uniformly random.  }
\label{practplateau}
\end{figure}

To illustrate how the displacement anti-symmetry comes into play, let us analyze the two-qubit case first. 
Recall that displacement anti-symmetry can be built into any control sequence by the construction 
detailed in Eq.~\eqref{dispanti}. 
The specific choice of base sequence depends on the noise model: as noted, in the presence of 
time-dependent direct coupling multi-qubit high-order DD sequences, such as \text{CDD} or NUDD, 
are needed, while if the latter vanishes we may employ the  more efficient multi-qubit sequences 
described in Eqs.~(\ref{2disl})-(\ref{2diss}). Regardless, imposing displacement anti-symmetry 
over $T_p$, one has that 
\begin{align*}
G_{Z_\ell}^{(1)} (\omega,T_p)& =(1  - e^{i \omega T_p/2}) \,G_{Z_\ell}^{(1)} (\omega,T_p/2),\\
G_{Z_{\ell'}}^{(1)} (\omega,T_p)& = (1 + e^{i \omega T_p/2}) \,G_{Z_{\ell'}}^{(1)} (\omega,T_p/2), \\
2i  \,G_{Z_\ell, Z_{\ell'}}^{(2)} (\omega, -\omega,T_p)& =- 2 \cos(\omega T_p/2) \,
G_{Z_{\ell}}^{(1)} (\omega,T_p/2) G_{Z_{\ell'}}^{(1)} (-\omega,T_p/2).
\end{align*}
Collecting all the terms proportional to $M$ in Eq.~\eqref{G2} leads then to 
\begin{align}
 \label{2qzero} 
G_{Z_{\ell}}^{(1)} \Big(\hspace*{-0.5mm}\omega,\frac{T_p}{2}\Big)G_{Z_{\ell'}}^{(1)} 
\Big(\hspace*{-1.5mm}-\omega,\frac{T_p}{2}\Big)\hspace*{-1.5mm}
\left [  \cos(\omega T_p/2) \hspace*{-1mm} - \hspace*{-1mm}\frac{ e^{i \omega T_p} - 
e^{-i \omega T_p} }{2 (e^{i \omega T_p/2} - e^{-i \omega T_p /2})^2 }  2i \sin(\omega T_p/2)
\right] = 0, 
\end{align}
which leaves us with 
\begin{align*}
\nonumber 
 2i \,G^{(2)}_{Z_{\ell},Z_{\ell'}} (\omega,-\omega, T) &=  G^{(1)}_{Z_{\ell}} (\omega, T_p) 
G^{(1)}_{Z_{\ell'}} (-\omega, T_p) \,
\frac{e^{i M \omega T_p} - e^{-i M \omega T_p} }{(e^{i  \omega T_p/2}-e^{-i  \omega T_p/2})^2} \\
&= -G^{(1)}_{Z_{\ell}} (\omega, T_p/2) G^{(1)}_{Z_{\ell'}} (-\omega, T_p/2) \,
\frac{\sin (M \omega T_p)}{\sin({\omega T_p/2})}  .
\end{align*}
By exploiting this structure, the relevant integral in Eq. \eqref{e3} becomes: 
\begin{align*}
\hspace*{-1.5mm}& 
\int_{-\infty}^\infty \frac{d \omega}{2\pi}  S^{B,-}_{{{\ell}},{{\ell'}}} (\omega) \,
2i  G^{(2)}_{Z_{\ell},Z_{\ell'}} (\omega, -\omega, T)  = 
2i \hspace*{-1mm} \sum_{r=-\infty}^\infty \int_{(2r-1) \pi /T_p }^{(2r+1) \pi /T_p} \frac{d \omega}{2\pi}  
S^{B,-}_{{{\ell}},{{\ell'}}} (\omega) \,
G^{(2)}_{Z_{\ell},Z_{\ell'}} (\omega, -\omega, T) \\
\hspace*{-1.5mm}& 
=\hspace*{-0.5mm} -\hspace*{-2mm}
\sum_{r=-\infty}^\infty \int_{(2r-1) \pi /T_p }^{(2r+1) \pi/T_p} \frac{d \omega}{2\pi}  
S^{B,-}_{{{\ell}},{{\ell'}}} (\omega) \,
G^{(1)}_{Z_{\ell}} (\omega, T_p/2) \,G^{(1)}_{Z_{\ell'}} (-\omega, T_p/2) \frac{\sin (M \omega T_p)}{\sin({\omega T_p/2)}}\\
\hspace*{-1.5mm}& 
\underrightarrow{{ \textrm{ large { \it M} }}} \,\,- \frac{2}{T_p} \sum_{r=-\infty}^\infty \frac{(-1)^ r}{2\pi} 
S^{B,-}_{{{\ell}},{{\ell'}}} \hspace*{-1mm}
\Big(\frac{2r \pi}{ T_p} \Big) G^{(1)}_{Z_{\ell}} \Big(\frac{2r \pi}{ T_p}, \frac{T_p}{2}\Big) 
G^{(1)}_{Z_{\ell'}} \Big(\hspace*{-1mm}- \frac{2r \pi}{ T_p}, \frac{T_p}{2}\Big).
\end{align*}
Assuming that Eqs.~\eqref{gplatquant0}-\eqref{gplatquant1} are satisfied, 
the two-qubit displacement anti-symmetry guarantees that the contribution due to the 
bath-induced coupling term is finite, without imposing additional constraints on the 
COs $\{ \alpha_{L_r}\} $.  
To better appreciate the role of the plateau conditions in the presence of the quantum bath, consider two 
DD sequences on which the displacement anti-symmetry is imposed: 
\begin{align*}
{\textrm{CDD}}_{1,1}^{d,(X_1,X_2)}(T) &= \textrm{CDD}^{(X_1)}_1 \times \textrm{CDD}^{(X_2)}_1  (T/2) \, X_2 \, 
\left( \textrm{CDD}^{(X_1)}_1 \times \textrm{CDD}^{(X_2)}_1 (T/2) \right) X_2 , \\ 
{\textrm{CDD}}_{1,2}^{d,(X_1,X_2)} (T) &= \textrm{CDD}^{(X_1)}_1 \times \textrm{CDD}^{(X_2)}_2 (T/2) \, X_2 \, 
\left( \textrm{CDD}^{(X_1)}_1 \times \textrm{CDD}^{(X_2)}_2 (T/2)  \right) X_2 , 
\end{align*}
Representative results for the fidelity behavior when these are applied to two qubits subject to quantum 
Gaussian noise with a sub-Ohmic spectrum are shown in Fig.~\ref{practplateau}, for a value of the transit 
time $t_{1,2}$ corresponding to generic (neither collective nor independent) coupling strengths.  For any 
two-qubit initial pure state $\rho(0)\equiv |\Psi\rangle\langle \Psi|$, the fidelity is computed as 
\begin{equation}
\label{fidelity}
{F}(T=M T_p)=\langle \Psi | \rho(M T_p) | \Psi\rangle = 
\sum_{a,b}\vert\rho_{a,b}(0)\vert^2e^{-\chi_{a,b}(M T_p)}e^{i\phi_{a,b}(M T_p)} ,
\end{equation}
in terms of the appropriate decay and phase terms (see Appendix for explicit expressions). 
Only one of the sequences satisfies the plateau conditions, and indeed the plateau is seen to 
appear only for such a sequence. Also, notice that the value at which the fidelity saturates, i.e., the quality 
of the plateau, deteriorates as $\tau$ (hence $T_p$) increases 
and approaches the upper bound imposed by Eq.~\eqref{gplatquant0}.

\begin{figure*}[t]
\centering
\hspace*{10mm}\includegraphics[width=0.95 \textwidth]{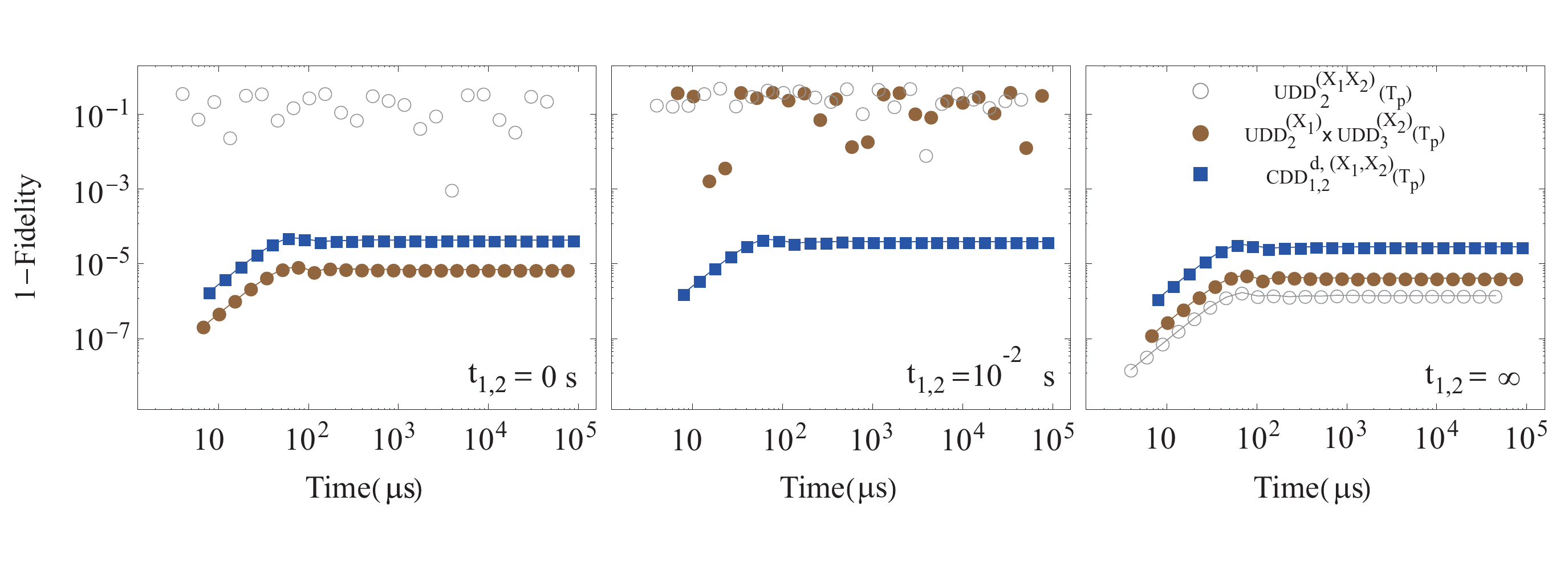}
\vspace*{-12mm}
\caption{Comparison between different protocols in achieving a fidelity plateau: non-selective control 
(UDD$_2^{(X_1 X_2)} (T_p)$, empty symbols), sequences with mirror anti-symmetry (UDD$^{(X_1)}_{2} \times $UDD$^{(X_{2})}_{3} (T_p)$, filled brown symbols), and with displacement anti-symmetry ({CDD}$_{1,2}^{d,(X_1,X_2)} (T_p)$, filled blue symbols). Only displacement anti-symmetry-enhanced DD achieves a plateau in the general case (middle panel). The fidelity loss, calculated by averaging over $10^3$ random pure initial states as in Fig. \ref{practplateau}, is plotted as a function of total time for the same Gaussian spin-boson noise spectrum 
$S^{B}_{Z_1,Z_{2}}(\omega)$ considered therein, for representative values of the transit time $t_{1,2}$. 
In each case, $T_p$ is fixed by the choice of minimum 
switching time, assumed here to be $\tau = 1 \mu$s, so that the appropriate plateau condition is met.
}
\label{comparison}
\end{figure*}

Fig.~\ref{comparison} further illustrates how the control protocol incorporating displacement anti-symmetry is also the 
only one exhibiting {\em model robustness}, namely, the only one guaranteeing that the plateau may be achieved for 
arbitrary spin-boson couplings.  Specifically, a protocol with displacement anti-symmetry (built out of 
${U}_{1,2}^{d,(X_1,X_2)} (T)$), one with mirror anti-symmetry (built out of $U^{X_1}_{2} \times U^{X_{2}}_{3} (T)$), and a 
non-selective control protocol ($U_2^{X_1X_2} (T)$) are tested against the same Gaussian noise spectrum of 
Fig. \ref{practplateau} for different values of the transit time -- equivalently, different spatial separation between the two qubits. 
In the $t_{1,2} \rightarrow \infty$ limit, which corresponds to private baths, all strategies work equally well, as expected. 
In the opposite limit of a collective coupling, $t_{1,2} \rightarrow 0$, the non-selective control protocol fails to achieve 
a plateau since it cannot make $G^{(2)}_{Z_1,Z_2} (\omega,-\omega,T)$ finite as  
$M\rightarrow \infty$. In the general case ($0< t_{1,2}<\infty$), only the displacement anti-symmetry 
enhanced protocol achieves a high-fidelity plateau. 

\subsubsection{Multi-qubit setting.} 
 
In order to extend the approach to the $N$-qubit setting, the key step is to show that, once the generalized 
displacement anti-symmetry introduced in Sec. \ref{sub:multisel} is imposed, 
a cancellation analog to Eq.~\eqref{2qzero} holds. 
Consider the terms proportional to $M$ arising in $G^{(2)}_{Z_{\ell},Z_{\ell'}} (\omega, -\omega, T)$ for each qubit pair 
[see Eq.~\eqref{G2}], and let $\max\{\ell,\ell'\} = \ell$ without loss of generality. Since, for the matrix defined in 
Eq. \eqref{Pmatrix}, $P^{(N)}(\ell, s) = \delta_{s,\ell-1}$, to every qubit $\ell>1$ let us associate $N_{\ell-1} = 2^{N-(\ell-1)}$ 
and intervals of length $\tau_{\ell-1}$, such that $\tau_{\ell-1} N_{\ell-1} = T_p$.
By dividing all integrals into sub-integrals over length-$\tau_{\ell-1}$ intervals and using the properties of 
$P^{(N)}(\ell, s)$, direct calculation shows that for every $\ell > \ell'$ the terms linear in $M$
reduce to 
\begin{align}
\nonumber &  
G^{(1)}_{Z_{\ell}} (\omega,\tau_{\ell-1}) G^{(1)}_{Z_{\ell'}} (-\omega, \tau_{\ell-1}) 
\bigg[ \sum_{r=1}^{N_{\ell-1}-1} \sum_{s=0}^{r-1} \Big( (-1)^r e^{i (r-s) \omega \tau_{\ell-1}}  -  (-1)^s e^{i (s-r) \omega \tau_{\ell-1}} \Big) \\
\nonumber &  - \bigg(\sum_{r=0}^{N_{\ell-1}-1} \sum_{s=0}^{N_{\ell-1}-1} (-1)^r e^{i (r-s) \omega \tau_{\ell-1}} \frac{e^{i \omega T_p} - e^{-i \omega T_p}}{(e^{i \omega T_p/2}-e^{-i  \omega T_p/2})^2} \bigg) \bigg] = 0.
\end{align}
Consequently, for any $\ell \neq \ell'$, we may re-express 
\begin{align}
\nonumber 
2i  G^{(2)}_{Z_{\ell},Z_{\ell'}} (\omega, -\omega, T) &=  G^{(1)}_{Z_\ell} (\omega, T_p) G^{(1)}_{Z_{\ell'}} (-\omega, T_p)  \, 
\frac{e^{i M \omega T_p} - e^{-i M \omega T_p}}{(e^{i\omega T_p/2}-e^{-i \omega T_p/2})^2}  \\
\nonumber 
&=  G^{(1)}_{Z_{\ell}} (\omega, \tau_{\ell-1}) G^{(1)}_{Z_{\ell'}} (-\omega, \tau_{\ell-1}) 
\bigg[ \frac{ - 2 i \sin^2(\frac{\omega T_p}{2})}{ \sin (\omega \tau_{\ell-1})} \times 
\frac{e^{i M \omega T_p} - e^{-i M \omega T_p}}{(e^{i {\omega T_p}/{2}}-e^{-i  
{\omega T_p}/{2}})^2}   \bigg]\\
\label{mgral} 
&= -  G^{(1)}_{Z_{\ell}} (\omega,\tau_{\ell-1}) G^{(1) }_{Z_{\ell'}} (-\omega, \tau_{\ell-1})  
\bigg[ \frac{\sin (M \omega T_p)}{ \sin (\omega \tau_{\ell-1})}  \bigg].
\end{align}
It then follows that the generalized displacement 
anti-symmetry guarantees that, in the large-$M$ limit, any contribution due to bath-induced qubit coupling, 
\begin{align*}
& 2i  \int_{-\infty}^\infty \frac{d \omega}{2\pi} \, S^{B,-}_{{{\ell}},{{\ell'}}} (\omega) 
G^{(2)}_{Z_{\ell},Z_{\ell'}} (\omega, -\omega, T) = \\
& - \frac{N_{\ell-1}}{T_p} \sum_{r=-\infty}^\infty \frac{(-1)^ r}{2\pi} S^{B,-}_{{{\ell}},{{\ell'}}} 
\Big(\frac{r \pi}{\tau_{\ell-1}} \Big) G^{(1)}_{Z_{\ell}} \Big(\frac{r \pi}{\tau_{\ell-1}},\tau_{\ell-1}\Big) 
G^{(1)}_{Z_{\ell'}} \Big(- \frac{r \pi}{ \tau_{\ell-1}}, \tau_{\ell-1}\Big), 
\end{align*}
is finite.  Thus, a plateau regime may still exists, provided that the conditions identified for the 
two-qubit case, Eqs. (\ref{gplatquant0})-(\ref{gplatquant1}) remain valid.  Remarkably, this result does not 
depend on the specific choice of base sequence, but only on the displacement anti-symmetry, further highlighting 
its fundamental role in multi-qubit control protocols for dephasing noise.

\section{Further Considerations}
\label{further}

\subsection{Controlled entanglement generation and storage}
\label{sec:ent}

Entanglement is a crucial resource across QIP, 
thus devising ways to reliably generating and storing it is an important task. 
Entanglement can be generated directly, via tunable or always-on couplings between qubits or, 
if the latter are not readily available, {\em indirectly}, with the aid of a common quantum environment. 
Various schemes for indirect generation of bipartite entanglement, as well as weaker quantum correlations 
quantified by discord, have been proposed \cite{Braun2002,Oh2006,Roszak2}.
Once created, such quantum correlations have to be stored (e.g., to be later used 
for quantum tasks of interest), and protected from unwanted decoherence. 
Several ways to do this have been invoked, e.g., employing the quantum Zeno effect \cite{PhysRevLett.100.090503}, 
quantum feedback \cite{Feedback1,Yamamoto2007981}, and DD control \cite{Franco2014}. 

While emphasis of existing work is on two-qubit settings and bipartite entanglement, 
the formalism we developed allows in principle to generate and store {\it multi-partite entanglement} with 
high-fidelity for a long time.  Let us assume a multi-qubit Hamiltonian that includes classical and 
bosonic dephasing, as in Eq.~\eqref{gralHs}-\eqref{gralHsb}, with vanishing direct coupling, 
$d_{\ell, \ell'}=0=\eta_{\ell,\ell'}$ (or else one could just use that to create entanglement), and no 
assumption on the initial bath state (in particular, no thermal equilibrium).  Our proposed strategy 
consists of two steps:

{\it Stage 1: Entanglement generation.} 
This can be achieved by using $M_g$ repetitions, each of duration $T_p$, of a high-order multi-qubit DD sequence 
such as \text{CDD} or NUDD {\it without displacement anti-symmetry}. In this way, the contribution of all dephasing 
terms is suppressed to high order, except for the one stemming from the induced coupling 
term -- which grows linearly with $M_g$ [recall Eq.~\eqref{G2}]. 
As noted, the induced coupling is ruled by  $S_{\ell,\ell'}^{B,-} (\omega)$ and, as such, it does not 
depend on $\rho_B$ but only on the actual dephasing Hamiltonian, in particular the spectral density function 
$J(\omega)$. The independence of the protocol upon $\rho_B$ is an important prerequisite for generating entanglement 
{\em on demand}, which may be achieved provided that accurate knowledge of the relevant power spectra 
$S_{\ell,\ell'}^{B,-} (\omega)$ is available.  In principle, this may be obtained by extending noise spectroscopy 
protocols for single-qubit~\cite{PhysRevLett.107.230501,NGpaper} and two-qubit classical dephasing \cite{cywinski2015} 
to general multi-qubit dephasing \cite{multiNSpaper}.

{\it Stage 2: Entanglement storage.} 
Once the state of the multi-qubit system after time $T_g=M_g T_p$ is sufficiently close to an 
entangled state of interest, we may switch to a DD sequence {\em with displacement anti-symmetry}, 
in order to achieve protection for long storage times, say, $T_s= M_s T'_p$, 
provided that the plateau conditions of Eqs. (\ref{gplatquant0})-(\ref{gplatquant0}) 
are satisfied, and where we allow for the duration of the storage base sequence to differ in general. 
Basically, the displacement anti-symmetry acts like an on/off switch for entanglement generation. 
There is, however, an important subtlety in the analysis that must be pointed out. 
In our derivation of the the plateau conditions, we relied on the assumption that the initial joint state 
was of the form $\rho_S \otimes \rho_B $; this need not be the case at the end of the 
entanglement generation stage, with $\rho_{SB} (T_g)$ involving entanglement between 
$S$ and $B$ in general (hence making $\rho_S(T_g)$ mixed). To move forward, it is necessary to 
re-examine the derivation when the total evolution time $T$ is divided in two consecutive stages, 
without and with displacement anti-symmetry, respectively. One can see that:
\begin{align*}
G^{(1)}_{O_a} (\omega,T) &=  G^{(1)}_{O_a} (\omega,M_g T_p) + e^{i \omega M_g T_p}
G^{(1)}_{O_a} (\omega, M_s T'_p), \quad 
O_a \in \{ Z_\ell, Z_{\ell}Z_{\ell'}\}, \\
2i \,G^{(2)}_{Z_\ell,Z_{\ell'}} (\omega_1, \omega_2, T) &= 2i \, G^{(2)}_{Z_\ell,Z_{\ell'}} (\omega_1, \omega_2, M_g T_p) 
+  2i \, G^{(2)}_{Z_\ell,Z_{\ell'}} (\omega_1, \omega_2, M_s T'_p) \\
&\,\,\,\,\, + e^{i \omega_1 M_g T_p} \, G^{(1)}_{Z_\ell} (\omega_1, M_s  T'_p) G^{(1)}_{Z_{\ell'}} 
( \omega_2, M_g  T_p ).
\end{align*}
By using these expressions in Eqs. (\ref{e1})-(\ref{e3}), each product of FFs, say with $k$ factors, 
becomes a sum of products which are {\em at most} $\mathcal{O} (M_s^k)$. Since we showed that 
$\mathcal{O} (M_s^k)$ terms are finite if the plateau conditions hold, it follows that terms 
with lower order in $M_s$ are also finite. 

By employing the above two-step control strategy, we can thus conclude that for any finite number 
$M_g$ of entanglement-generation cycles, it is possible to store the resulting multi-qubit entanglement 
with high fidelity for an arbitrary number $M_s$ of storage cycles in principle.  A number of important aspects 
require additional, more detailed analysis -- notably, the degree of purity and nature of the multipartite 
entangled states reachable over time under the assumed dephasing Hamiltonian, along with consideration of 
relevant time scales and resource scaling.  While we leave this to a separate investigation, indirect 
``environment-assisted'' generation and storage of multi-qubit states close to paradigmatic multi-partite 
entangled states of interest (such as W and GHZ states) would be especially interesting, and complementing  
ongoing effort on steady-state entanglement generation using 
engineered dissipation, see e.g. \cite{FT,Hakan,Reiter}.

\subsection{Realistic considerations}
\label{sec:realistic}

The noise model assumed in Eqs. (\ref{gralHs})-(\ref{gralHsb}) covers important sources of decoherence in a 
variety of quantum systems of relevance to QIP. Classical fluctuations, like the one- and two-body stochastic processes 
$\zeta_\ell(t)$ and $\eta_{\ell, \ell'}(t)$, provide an effective description of dephasing when the effects of back-action 
from the system on the environment are negligible. In the simplest case, the noise takes the form of fluctuations in 
an externally applied field, as it does for magnetic-field fluctuations in trapped-ions \cite{Mike2009,Mike2009b,Szwer},
or in NMR qubits \cite{ChuangNMR}. 
More generally, so long as appropriate conditions are satisfied (e.g., effective high temperature),
classical processes may be used to approximate the dephasing effects of more complex interactions -- 
ranging from background charge, magnetic flux, and critical current fluctuations in superconducting qubits 
\cite{PhysRevB.77.174509, Nakamura2002,Bylander2002}, to interactions between nuclear spins and 
their surrounding spin environment in NMR \cite{Suter2012}, and interactions between singlet-triplet 
electron spin states in quantum dots and, again,  
their surrounding nuclear-spin bath \cite{Bluhm2011,Yacoby2011,Marcus2012,Yacoby2005,MikeBluhm}. 
We note also that control imperfections, such as fluctuations in the frequency of a master oscillator, often 
result in effective dephasing that can be accurately modeled as classical noise \cite{Mike2009,SoarePRA,MikeLO}. 

The quantum contributions to our noise Hamiltonian describe a dephasing interaction between the multi-qubit system and a 
bosonic environment of quantum oscillators. Interactions with quantum oscillators, 
in the form of lattice vibrational modes, arise naturally in solid-state systems. A prominent example is the coupling 
between excitonic qubits in quantum dots and acoustic vibrations, with electron-phonon being is the dominant source 
of dephasing in this type of system \cite{Krumm,Vagov,Roszak1,Lastra,Lovett}.
Even beyond oscillator environments, the linear spin-boson model has been widely applied to describe 
open-system dynamics in situations where the system-environment coupling is weak enough for a linear approximation to be valid \cite{Stamp}. For example, the dephasing effects of current and voltage fluctuations in Josephson junction qubits are accurately described in terms of spin-boson interactions \cite{Schon}.

It is often the case that the noise is generated by a large number of independent, weakly coupled sources,
and can therefore be assumed to be Gaussian. Generic noise, however, is non-Gaussian; in particular,  
as mentioned in Sec. \ref{p1}, the assumption of Gaussianity breaks down for qubits operated at an optimal point 
\cite{CywinskiOpt}, as well as for strongly coupled bistable fluctuators of the sort that generate $1/f$ in 
Josephson nano-circuits \cite{1overf,PhysRevB.77.174509}. For quantum environments, both the initial state 
and the nature of the interaction with the system will determine whether the noise is Gaussian or not \cite{NGpaper}. 
In these cases, the interplay between generalized FFs and higher-order noise spectra needs to be carefully considered.

Despite the broad applicability of the dephasing noise model on which we have based our analysis, it necessarily neglects important aspects 
that will affect the evolution of the system in reality. In particular, it is generally true that interactions between a quantum system and its environment will generate both dephasing and relaxation effects, with associates characteristic time scales $T_2$ and $T_1$, respectively. 
So long as an appreciable separation between them exists, however -- that is, $T_{1}\gg T_{2}$, 
as it is the case for most of the above-mentioned QIP settings -- relaxation will not be a significant factor limiting the validity of our results,   including those pertaining to long-time quantum storage \cite{Memory}. By contrast, if the dephasing noise is strongly Markovian, the lack of temporal correlations will limit the usefulness of the proposed protocols.  However, as already remarked in Sec. \ref{sub:models}, this holds for DD techniques generally, and it is well understood that significant Markovian error contributions must be countered using closed-loop 
quantum error correction strategies \cite{qec}.  

Even if the above conditions are obeyed, the simplifying assumption of the existence of a hard noise cutoff $\omega_{c,{L_r}}$ 
for each noise spectrum frequencies $\omega_{{L_r}}$, made in Sec. \ref{p1}, will hardly be obeyed in realistic scenarios.
The purpose of this assumption was to avoid the singular behavior associated with the resonant frequencies defined by 
Eq. \eqref{peaks}. In the absence of these cutoffs, the contribution of any noise at these frequencies will be amplified in a way 
that grows linearly with the number $M$ of base sequence repetitions, ultimately making a long-time plateau unsustainable. 
However, building on the quantitative analysis carried out for the single-qubit case \cite{Memory}, 
if the the noise beyond the resonant frequencies decays sufficiently rapidly, one can expect the errors associated 
with high frequency noise to grow slowly with $M$.  This may still allow for the maintenance of a fidelity plateau for storage times 
that are still sufficiently long to be practically useful.

Lastly, we briefly turn to considerations of realistic control limitations.  While the precise nature and extent of 
the discrepancies resulting the simplifying assumption of perfect control resources we have made will depend 
on the particular physical system being investigated, 
it is worthwhile highlighting those featuress that have the greatest potential to limit the efficacy of the 
proposed quantum storage protocols across all potential QIP platforms.
First, it is clear that any real control pulse has a non-zero duration. As a result, the application of each pulse in a DD 
sequence will introduce depolarization errors, in addition to those that have a purely dephasing effect. It will not then be 
possible to describe the modulating influence of a pulse sequence in terms of simple, uniaxial switching functions of the 
form exemplified in Eq. \eqref{yl}, and, consequently, 
the Magnus expansion for the controlled propagator $\tilde{U}_e(T)$ will no longer truncate. Strictly speaking, although it is possible to design sequences of finite-width pulses that can achieve a non-zero CO \cite{EDD}, the FO of any DD protocols derived from a two-axis decoherence model, as needed to include such realistic pulse effects, will be zero \cite{PazViola}. This problem will generally become more significant as the number of applied pulses increases.

For short-term state preservation and relatively small numbers of qubits $N$, additional errors introduced by finite-duration 
pulses need not be a major concern, as the low-FO contributions may be negligible, so that the ``effective FO" remains high. 
For long-term memory and/or large $N$, however, the accumulation of pulse-induced errors may significantly 
reduce the attainable fidelity and the duration of any plateau. There are two readily apparent approaches to addressing this 
problem, which will likely have to be used in tandem in practical dephasing settings.
The first, and a motivation for this work, is to use DD protocols that utilize the minimum number of pulses to achieve a desired level 
of error suppression for a given dephasing environment and system size $N$. The second is to replace ``primitive'' DD pulses 
with dynamically corrected gates or composite pulses \cite{Kaveh2009,Kaveh2010,Suter2011b,Chingiz,Mike2014}, 
so that error cancellation is maintained {\em during} the duration of all pulses, up to a sufficiently high order.  
For single-qubit storage, this approach has been shown to successfully counter pulse-width effects which would otherwise 
prevent or degrade a coherence plateau \cite{Memory}.  In multi-qubit DD sequences with selective control, different COs
may be required for different qubits, to account for qubit-dependent modulation that an overall control sign pattern requires. 

Another important source of control error is timing ``jitter''. This may take the form imprecision in the timing of the individual pulses that comprise a DD sequence. As the analysis in \cite{MikeFF} implies, even when pulses are assumed to have vanishing duration, and imprecision is restricted to small rounding errors in the timing of each pulse, the actual CO and FO will go to zero. This is a most serious for optimized sequences, such as those based on UDD, which are particularly sensitive to pulse placement. A second form of timing jitter, that affects long-time storage in particular, is premature or belated memory access time. The long-time memory protocols discussed in Sec. \ref{longterm} are designed to allow for high-fidelity access to stored information only at integer multiples of the base sequence time, i.e., when $T=MT_{p}$, with access latency capped at $T_{p}\ll T$. If there is some small timing offset $\delta t$, so that $T=MT_{p}+\delta t$, the sub-optimal access time can have a similar effect on fidelity as pulse timing imprecision \cite{Memory}. It is therefore essential that DD protocols, particularly those that aim for long-time storage, be clocked by high-resolution timing systems with minimum jitter.

\section{Conclusion and Outlook}
\label{conclusion}

We have provided criteria for the design of effective and efficient control protocols for the preservation of arbitrary 
multi-qubit states in a relevant class of dephasing models, that combines the effects of classical noise and a linear interaction 
with a bosonic bath, not necessarily in thermal equilibrium.
Under the assumption that (nearly) instantaneous DD pulses may be selectively applied to arbitrary subsets of qubits, 
we showed that the reduced dynamics of a multi-qubit system can be expressed in terms of a hierarchy of noise spectra --
that capture the statistical properties of quantum and classical dephasing sources in the frequency domain -- and of a small 
set of first- and  second-order FFs -- that the describe the modulating effect of the applied control. 
These results allow for a relatively straightforward, exact 
characterization of the performance of an arbitrary DD protocol in both the time and frequency domains, and serve as the starting 
point for the derivation of conditions for the construction of resource-efficient high-order DD protocols.

Specifically, we showed that multi-qubit DD sequences may be constructed that are able to achieve high-order error 
suppression using exponentially fewer pulses than the most efficient existing protocols, so long as any direct qubit-qubit 
coupling is constant. This reduction in pulse number offers tremendous practical advantages in terms of the required 
minimum switching time and the cumulative effect of pulse errors. The improvement in efficiency derives from the property of \emph{displacement anti-symmetry} with which the new protocols are endowed. Importantly, through the imposition of this 
form of temporal symmetry, it is also possible to ensure that a sequence achieves maximal filtering order in the frequency 
domain. This contrasts with previously proposed multi-qubit protocols based on nesting and concatenation, for which maximum filtering order cannot always be guaranteed.

We also demonstrated that the method of producing long-time quantum memory via DD sequence repetition, described previously for 
single-qubit systems under Gaussian noise, can be generalized to multi-qubit systems possibly subject to general non-Gaussian noise. 
For strictly classical, Gaussian noise on $N$ qubits, the conditions for engineering the required fidelity plateau are a natural extension of 
those derived for the single-qubit case \cite{Memory}. However, when a quantum spin-boson interaction is included, the fidelity plateau 
cannot be maintained unless additional structure is imposed on the base sequence from the outset. We found that those sequences possessing displacement anti-symmetry have the necessary structure and, therefore, can be used to generate a fidelity plateau in a combined classical and quantum noise environment. On the basis of these observations, we have outlined a simple switched control protocol for the the generation and storage of entangled multi-qubit states.

The central role of that the displacement anti-symmetry plays in suppressing the effects of noise associated with the 
genuinely quantum (non-commuting) nature of the bath points to its relevance for characterizing bath-induced spatial 
correlations, along with their impact on the implementation of quantum technologies.  
By comparing the response of two qubits to different control sequences, for instance, it is now possible in principle to 
discriminate between a classical or a quantum bath with Gaussian statistics. This may have important implications for 
quantum verification and validation protocols and, ultimately, fault-tolerant quantum computing architectures 
in the presence of bath-induced spatial correlations.  While exploring the usefulness of displacement anti-symmetry 
beyond the class of open quantum systems examined here is a natural direction for further investigation, we
anticipate immediate applications of our enhanced DD sequences in the context of multi-qubit noise spectroscopy 
for correlated dephasing environments \cite{cywinski2015,multiNSpaper}.

\section*{Acknowledgements}

It is a pleasure to thank Kaveh Khodjasteh for early contributions,  as well as 
Michael Biercuk and Leigh Norris for valuable discussions on different aspects of this work.
TJG gratefully acknowledges hospitality from the Department of Physics and Astronomy 
at Dartmouth College, where part of his work was performed.  We acknowledge support from 
the US Army Research Office under contracts No. W911NF-11-1-0068 and W911NF-14-1-0682, and 
from the {Constance and Walter Burke Special Project Fund in Quantum Information 
Science}.

\section*{Appendix: Controlled two-qubit dynamics under Gaussian dephasing}

To exemplify the formalism presented in Sec. \ref{sub:exact}, let us consider in detail a two-qubit system undergoing 
stationary zero-mean Gaussian dephasing due to combined classical and quantum bosonic sources. With reference to 
Eq. (\ref{FFprop}), the relevant generalized FFs are
$G^{(1)}_{Z_1}(\omega, T), G^{(1)}_{Z_2}(\omega, T), G^{(1)}_{Z_1 Z_2}(\omega, T)$, and 
$G^{(2)}_{Z_1, Z_2}(\omega, -\omega, T)$.
The interaction-picture two-qubit dynamics is given by 
\beq
\label{dyngaus}
\nonumber \langle\rho(T)  \rangle_{c,q}  =  \hspace*{-5mm}
\sum_{\{ a_1,a_2, b_1, b_2 | a_\ell, b_\ell =0,1 \} } \hspace*{-6mm}
e^{- \chi_{a_1 a_2, b_1 b_2}(T) + i \,\phi_{a_1 a_2, b_1 b_2}(T)} \rho_{a_1 a_2,b_1 b_2} (0) \ket{a_1,a_2} \bra{b_1,b_2},
\eneq
where the decay and phase evolution are determined by the real and imaginary part of the 
noise second-order cumulants, respectively. Specifically, the decay contribution includes {\em all} the classical-noise 
effects [Eq. (\ref{Z1})] plus, from the quantum noise, a contribution that it is identical to what one would find under a 
private-bath assumption [first term in Eq. (\ref{Z2})]:
\begin{align}
\nonumber 
-\chi_{ab}(T) &= \Big\langle \Big( \Delta[|a|,|b|]  \bar{\eta'}_{{1},{2}}(T) + \sum_{{\ell}}  
\Delta[{a}_{{\ell}},{b}_{{\ell}}] \bar{\zeta'}_{{\ell}}(T) \Big)^2   \Big\rangle_c  + 
\Big\langle \Big(\sum_{ {\ell}} \Delta[{a}_{{\ell}},{b}_{{\ell}}]  \bar{{B}}_{{\ell}}(T)\Big)^2 \Big\rangle_q.
\end{align}
Non-commutativity of the bath operators, $[{B}_\ell(t),{B}_{\ell'}(t')] \neq 0$, is responsible for additional 
non-trivial phase evolution in certain off-diagonal density-matrix elements:
\begin{align}
\nonumber 
i \phi_{ab}(T) &= \Delta[|a|,|b|]  \underbrace{(\bar{R}_{12}(T) + \bar{R}_{21}(T))}_{\equiv i \phi^0}  +\, 
\Delta[{a}_{{1}}+{b}_{{2}},{a}_{{2}}+{b}_{{1}}]  \underbrace{\frac{[\bar	{{{B}}}_{1}(T), \bar{{B}}_{2}(T)]}{2}}_{\equiv i \phi^1},
\end{align}
where $| a | \equiv \sum_\ell {a_\ell}$ and we have explicitly identified two distinct contributions: 
(i) $i \phi^0$, resulting from the second-order Magnus term;
(ii)  $i\phi^1$, originating from the partial trace over the quantum bath. 
Thanks to the bosonic algebra, and at variance with $\chi_{ab}(T)$, 
{\em $\phi_{ab}(T)$ depends only on the Hamiltonian} and not on $\rho_B$ (in particular, 
$\phi_{ab}(T)$ is temperature-independent)\footnote{Classical non-Gaussian noise can {\em also} contribute to phase evolution, if odd-order cumulants are non-zero. For classical plus quantum non-Gaussian noise, i.e., when high-order cumulants of $B_\ell(t)$, $\eta_{\ell,\ell'}(t)$ and/or $\zeta_{\ell} (t)$ exist, all factors in Eq.~\eqref{dyn} may contribute to $\phi_{ab}(t)$ and it becomes hard to isolate classical vs quantum effects. }.

The presence of the $\Delta[\cdot,\cdot]$ function in the above equations implies that not all noise sources 
contribute to the evolution of a given density-matrix element. By making the following natural identifications,
$-\chi_{12,\ell} = \langle \bar{\eta'}_{{1},{2}}(T)\bar{\zeta'}_{{\ell}}(T) \rangle$, 
$-\chi_{12,12} = \frac{1}{2} \langle \bar{\eta'}_{{1},{2}}(T)\bar{\eta'}_{{1},{2}}(T) \rangle$, 
$-\chi_{\ell,\ell} = \frac{1}{2}  \langle \bar{\zeta'}_{{\ell}}(T)\bar{\zeta'}_{\ell}(T)\rangle  
+\langle \bar{B}_{{\ell}}(T)^2\rangle_q$, and 
$-\chi_{1,2} =\frac{1}{2} \langle 2  \bar{\zeta'}_{{1}}(T)\bar{\zeta'}_{2}(T)\rangle +
\langle \bar{B}_{{1}}(T)\bar{B}_{2}(T) + \bar{B}_{{2}}(T)\bar{B}_{1}(T)\rangle_q$, the decay pattern has the following form:
$$e^{-\chi_{a,b}(T)} \sim 
\left( \hspace*{-1mm}\begin{array} {cccc}
\cdot & e^{-\chi_{12,12}-\chi_{2,2}-\chi_{12,2}} & e^{-\chi_{12,12}-\chi_{1,1} - \chi_{12,1}} & 
e^{-\chi_{1,1} -\chi_{2,2}-\chi_{1,2}} \\
e^{-\chi_{12,12}-\chi_{2,2}-\chi_{12,2}} & \cdot & e^{-\chi_{1,1}-\chi_{2,2} +\chi_{1,2}} & 
e^{-\chi_{12,12}+\chi_{12,1} - \chi_{1,1}} \\
e^{-\chi_{12,12}-\chi_{12,1}-\chi_{1,1} } & e^{-\chi_{1,1} -\chi_{2,2}+\chi_{1,2}} & \cdot & 
e^{-\chi_{12,12}-\chi_{2,2} + \chi_{12,2}} \\
e^{-\chi_{1,1} -\chi_{2,2} -\chi_{1,2}} & e^{-\chi_{12,12}+\chi_{12,1}-\chi_{1,1} } & 
e^{-\chi_{12,12}-\chi_{2,2} + \chi_{12,2} } & \cdot 
\end{array}\hspace*{-1mm}\right), $$
where $\cdot$ stands for an identity action.
While the diagonal entries (populations) are unaffected by the dephasing noise, off-diagonal terms decay according to different controlled decoherence functions, in the absence of special symmetries\footnote{It is easy to verify that, for 
collective dephasing, qubit-permutation symmetry implies that $\chi_{1,1} = \chi_{2,2} = \chi_{1,2}/2$, in agreement with decoherence-free subspace theory \cite{qec}.}.  
Similarly, the phase evolution is not the same for all the off-diagonal coherence elements. Rather, the diagonal and 
anti-diagonal entries do not exhibit phase evolution, whereas all other elements gain a 
time-dependent phase according to the following pattern:
$$ e^{-i \phi_{a,b}(T)} \sim 
\left(\begin{array}{cccc}
\cdot & e^{ i (\phi^0 - \phi^1)} & e^{ i (\phi^0 + \phi^1)} & \cdot \\
e^{ -i (\phi^0 - \phi^1)} & \cdot & \cdot & e^{ -i (\phi^0 + \phi^1)} \\
e^{ -i (\phi^0 + \phi^1)} & \cdot & \cdot & e^{ -i (\phi^0 - \phi^1)} \\
\cdot & e^{ i (\phi^0 + \phi^1)} & e^{ i (\phi^0 - \phi^1)} & \cdot 
\end{array}\right). $$

A paradigmatic situation is a pure Gaussian two-qubit spin-boson model under DD, 
in which case the classical average in the above expression for $\chi_{ab}(T)$ vanishes and one has
\begin{align}
\nonumber 
- \chi_{ab}(T)
= \sum_{ {\ell}=1,2} \Delta[{a}_{{\ell}},{b}_{{\ell}}]^2  \langle \bar{{B}}_{{\ell}}(T)^2 \rangle_q +
{\Delta[{a}_{2},{b}_{2} ] \Delta[{a}_{{1}},{b}_{{1}}]  \langle \bar{{B}}_{{2}}(T) \bar{{B}}_{{1}}(T) 
+ \bar{{B}}_{{1}}(T) \bar{{B}}_{{2}}(T) \rangle_q ,}
\end{align}
with single- and two-qubit contributions given by
\begin{equation}
\langle \bar{{B}}_{{\ell}}(T)^2 \rangle_q = \hspace*{-0.5mm} - \hspace*{-1mm}
\int_{-\infty}^{\infty}  \frac{d{\omega}}{2\pi}  \, 
G^{(1)}_{Z_{\ell}} (\omega,T) G^{(1)}_{Z_{\ell}} (-\omega,T) S^B_{{{\ell}},{{\ell}}} ({\omega}) 
{ =  - 2 \hspace*{-1mm}\int_0^\infty  \frac{d{\omega}}{2\pi}  \, 
|G^{(1)}_{Z_{\ell}} (\omega,T)|^2  S^{B,+}_{{{\ell}},{{\ell}}} ({\omega}),}  
\label{Bq} 
\end{equation}
\begin{equation}
{  \langle \bar{B}_{2}(T) \bar{B}_{1}(T) + \bar{{B}}_{1}(T) \bar{B}_{2}(T) \rangle_q   =  2
\int_0^\infty  \frac{d{\omega}}{2\pi}  \,  \text{Re} [G^{(1)}_{Z_2} (\omega,T) G^{(1)}_{Z_1} (-\omega,T) 
S^{B,+}_{2,1} ({\omega}) ], }
\label{BBq}
\end{equation}
that is, they are determined by an integral of the overlap 
between a product of FFs, purely dependent on the control, and the noise power spectra 
$S^{B,+}_{\ell,\ell'} (\omega)$.  Thanks to the stationary zero-mean assumptions, note that 
no single FF $G^{(1)}_{Z_\ell} (\omega,t)$  contributes to the reduced dynamics, 
but only $G^{(1)}_{Z_\ell}(\omega,T) G^{(1)}_{Z_\ell}(-\omega,T) = \vert G^{(1)}_{Z_\ell}(\omega,T) \vert^2$.  
Since the latter is an even function of frequency, rewriting the integral in Eq. (\ref{Bq}) as 
one over the non-negative axis makes it clear that only $J(\omega) \coth(\beta \omega/2)$ contributes. 
Similarly, the symmetry properties highlighted in Eq. (\ref{HQspect1}) allow to cast the contribution 
to $\chi_{ab}(T)$ in Eq. (\ref{BBq}) in a form where it is manifestly real.   

The two terms contributing to the controlled phase evolution $i\phi_{ab}(T)$ may likewise be expressed 
as overlap integrals, except that the purely quantum noise spectra, $S^{B,-}_{\ell,\ell'} (\omega)$, which 
arises from bath non-commutativity, is the relevant one in this case.  Explicitly, 
\begin{align}
\nonumber 
i\phi^0 (T)& = -i \int_{-\infty}^\infty \frac{d \omega}{2\pi} {G}^{(2)}_{Z_1,Z_{2}} (\omega,-\omega,T) 
S^{B,-}_{1,2} (\omega) \\
& = - 2i \int_{0}^\infty \frac{d \omega}{2\pi} \, \textrm{Im} [i {G}^{(2)}_{Z_1,Z_{2}} (\omega,-\omega,T) 
S^{B,-}_{1,2} (\omega)],
\label{P0} \\  
\nonumber 
i \phi^1 (T)&= {\frac{1}{2}} \int_{-\infty}^\infty \frac{d \omega}{2\pi} {G}^{(1)}_{Z_1} (\omega,T) 
{G}^{(1)}_{Z_{2}} (-\omega,T) S^{B,-}_{1,2} (\omega) \\
&= i \int_{0}^\infty \frac{d \omega}{2\pi}  \, \textrm{Im} [ {G}^{(1)}_{Z_1} (\omega,T) 
{G}^{(1)}_{Z_{2}} (-\omega,T) S^{B,-}_{1,2} (\omega) ],
\label{P1}
\end{align}
where we have used the anti-symmetry property of $S^{B,-}_{1,2}(-\omega)$, given in Eq. (\ref{HQspect1}), 
together with the relationships 
$$i G^{(2)}_{1, 2}(\omega,-\omega,T) = [i G^{(2)}_{1, 2}(-\omega, \omega,T)]^\ast, \; 
[{G}^{(1)}_{Z_1} (\omega,T) {G}^{(1)}_{Z_{2}} (-\omega,T)]^\ast  = 
{G}^{(1)}_{Z_1} (-\omega,T) {G}^{(1)}_{Z_{2}} (\omega,T), $$ 
which follow directly from the definition.  
Written in the form of Eqs. (\ref{P0})-(\ref{P1}), it is manifest that $i \phi^{0,1}$ are purely imaginary, as expected. 

As anticipated, we remark that the special case of free dynamics may be recovered as a limiting case, 
by letting the control switching functions $y_\ell (t) \equiv 1$ for all $\ell, t$, which yields
\begin{align*}
G^{(1)}_{Z_\ell} (\omega,T) G^{(1)}_{Z_\ell}(-\omega,T)  &= 
G^{(1)}_{Z_1} (\omega,T) G^{(1)}_{Z_2}(-\omega,T)  = 
\frac{ 2[1- \cos (\omega T)]}{\omega^2}, \\
G^{(2)}_{Z_1,Z_2} (\omega,-\omega,T)  &= - \frac{\omega T - \sin (\omega T)}{\omega^2}.
\end{align*}
One may then verify that our expressions recovers existing results obtained for 
both two- and multi-qubit free spin-boson dephasing dynamics, see e.g. \cite{Reina2002,Braun2002}.

\section*{References}
\bibliographystyle{iopart-num}

\providecommand{\newblock}{}

\end{document}